\documentclass[11pt,a4paper]{article}
\pdfoutput=1
\usepackage{jheppub}
\usepackage{rotating}
\usepackage[bb=boondox]{mathalfa}
\usepackage{wrapfig}
\usepackage{comment}
\usepackage{graphicx,graphbox}
\usepackage{amsfonts}
\usepackage{amsmath}
\usepackage{slashed}
\usepackage{amssymb}
\usepackage{latexsym}
\usepackage{multirow}
\usepackage{hyperref}
\usepackage{enumitem}
\hypersetup{colorlinks=true,citecolor=red,linkcolor=magenta,urlcolor=blue}
\usepackage{relsize}
\usepackage{booktabs}
\usepackage{subfigure}
\usepackage{cleveref}
\usepackage{fancyhdr}
\usepackage{float}
\usepackage{mmacells}
\usepackage{graphicx}
\usepackage{microtype}
\usepackage{tabularx}
\usepackage{xcolor}
\usepackage{siunitx,adjustbox,bm}
\usepackage[bottom]{footmisc}
\usepackage{makecell}
\usepackage[normalem]{ulem}

\usepackage{mathtools}
\usepackage[thinc]{esdiff}
\usepackage[euler]{textgreek}
\usepackage{braket}
\usepackage{physics}
\usepackage{xfrac}
\usepackage{soul}
\usepackage{stmaryrd}
\usepackage{pifont}

\newcommand{\nn}{\nonumber}

\renewcommand{\tr}{\text{tr}}

\renewcommand{\L}{\mathcal{L}}
\renewcommand{\c}{\text{c}}

\begin{document}
\allowdisplaybreaks
\title{Renormalization of  Scalar and Fermion  Interacting Field Theory for Arbitrary Loop: Heat-Kernel Approach}
\abstract{We outline a proposal, based on the Heat-Kernel method, to compute 1PI effective action up to any loop order for quantum field theory with scalar and fermion fields. We algebraically extract the divergences associated with the composite operators without explicitly performing any momentum loop integral. We perform this analysis explicitly for one and two-loop cases and pave the way for three-loop as well. Using our prescription we compute the two-loop counter terms for a theory containing higher mass dimensional effective operators that are polynomial in fields for two different cases: (i) real singlet scalar, and (ii) complex fermion-scalar interacting theories. We also discuss how the minimal Heat-Kernel fails to deal with the effective operators involving derivatives. We explicitly compute the one-loop counter terms for such a case within an $O(n)$ symmetric scalar theory employing a non-minimal Heat-Kernel. Our method computes the counter terms of the composite operators directly and is also useful for extracting infrared divergence in massless limits.}	
\author[a]{Upalaparna Banerjee,}
\author[a]{Joydeep Chakrabortty,}
\author[a]{Kaanapuli Ramkumar}

\emailAdd{upalab, joydeep@iitk.ac.in, kaanapuliramkumar08@gmail.com}
\affiliation[a]{Indian Institute of Technology Kanpur, Kalyanpur, Kanpur 208016, Uttar Pradesh, India.}

	
\preprint{}

\maketitle
	
\newpage
\section{Introduction}
In quantum field theory (QFT), the quantum corrections to the tree-level Lagrangian are of great interest. In the case of renormalizable theories, a finite number of counter terms emerge through suitable quantum corrections, and within a perturbative framework, those can be computed order by order. This lays the foundation to estimate the running of the composite operators and their respective couplings. 
This perturbative renormalization technique is not restricted to only relevant and marginal operators and can be extended to renormalize a Lagrangian containing higher mass dimensional effective operators. In such cases, one needs to recall that the effective Lagrangian should be renormalized order by order in the mass dimension of effective operators, and definitely without disrespecting the validity of an effective theory. For example, if an effective action contains up to dimension six effective operators, then in the process of renormalization, further higher dimensional operators, say dimension eight, may emerge which should be ignored. These contributions will be important when we extend the effective Lagrangian beyond dimension six.
Recently, the renormalization techniques in the context of such theories have drawn much attention with an increasing interest in Standard Model Effective Field Theory (SMEFT)~\cite{Henning:2014wua,Brivio:2017vri,Isidori:2023pyp} which has been instrumental in grabbing the footprints of new physics beyond the Standard Model (SM). Relying on collider experimental results, there is expected to be a definite energy gap between the SM and the new physics scales, which solidifies the reliability of the EFT~\cite{Weinberg:1980wa,Georgi:1994qn,Manohar:2018aog,Cohen:2019wxr} parameterizations of BSM scenarios. 

In this context two aspects are important - (i) the emergence of effective operators, either through integrating out a heavy field or computing effective operator bases in a gauge invariant way using information about low energy theory, and (ii) the running of these operators. Several attempts have been made to construct effective operators starting from a UV theory after integrating out heavy fields using the Feynman diagram approach and also the functional method~\cite{Gaillard:1985uh,Cheyette:1987qz,Fuentes-Martin:2016uol,Drozd:2015rsp}. The systematic procedure of operator basis construction within the context of the effective theories for mass dimension six and beyond has been discussed in Refs.~\cite{BUCHMULLER1986621,Grzadkowski:2010es,Lehman:2014jma,Henning:2014wua,Murphy:2020rsh,Li:2020gnx,Li:2022tec,Banerjee:2020jun,Anisha:2019nzx,Banerjee:2019twi,Banerjee:2020bym,Harlander:2023psl,Harlander:2023ozs,Schaaf:2023mpw}. In Refs.~\cite{Banerjee:2023iiv,Chakrabortty:2023yke} the universal effective action has been computed up to dimension eight for degenerate heavy scalar and fermion cases up to one-loop. This method has been generalized further in \cite{Banerjee:2023xak} where one can compute effective operators up to any mass dimension for non-degenerate fields including light-heavy mixing contributions. Besides, there has been development to blend the theoretical aspects of matching and running in the context of EFT suitably with the computational techniques~\cite{Aebischer:2018bkb,Carmona:2021xtq,Criado:2017khh,Celis:2017hod,Bakshi:2018ics,Fuentes-Martin:2022jrf,Fuentes-Martin:2020udw,Cohen:2020qvb} as well, but these are mostly restricted up to one-loop order.

Another important aspect is the computation of the renormalization group evolution (RGE) of these effective operators. The first attempts were made in Refs.~\cite{Jenkins:2013zja,Jenkins:2013wua,Alonso:2013hga} to compute the one-loop RGEs for the SMEFT dimension six operators. Recently, the one-loop RGEs have been computed in Ref.~\cite{Zhang:2023kvw,Chala:2021pll,DasBakshi:2022mwk} for dimension seven and eight operators as well. But based on recent improvements in precision physics, it has been necessitated to look beyond one-loop. For example, in the context of scalar quantum field theory (SQFT), the two-loop RGEs have been computed using functional, algebraic, or geometric approaches in Refs.~\cite{Fuentes-Martin:2023ljp,Jenkins:2023rtg,Jenkins:2023bls}. 

In this article, we propose a Heat-Kernel-based method to renormalize the QFT of scalars and fermions up to any arbitrary loop. This method, minimal and universal, also allows us to add any composite operators with and without derivatives, of arbitrary mass dimensions. The Heat-Kernel allows us to write a two-point Green's function $G(x,y)$ in the basis of the Heat-Kernel coefficients (HKCs) \cite{Jack:1982hf,vonGersdorff:2022kwj,vonGersdorff:2023lle,Hadamard2003,Minakshisundaram:1953xh,DeWitt:1964mxt,Seeley:1969}, which encapsulate the structures of the composite operators. In our earlier papers \cite{Banerjee:2023iiv,Chakrabortty:2023yke,Banerjee:2023xak}, we have computed the HKCs that we have made use of here. The core structures of Green's functions possess algebraic singularities in the limit $d \to 4$. Therefore, the computation of only the distinct vacuum diagrams, at each loop order, is sufficient to note down all the divergences associated with the composite operators present in the Lagrangian; considering multiple Feynman diagrams with external legs for wave-function and coupling renormalizations is not necessary. It is worth mentioning that the divergences are computed as a function of HKCs, thus these results are generic and equally applicable for fermion and scalar QFT including higher-dimensional interactions. 

This article is organized as follows: first, we briefly introduce the computation of non-coincidental HKCs. Then, we discuss the connection between Green's function for scalar fields and the HK and note down the algebraic singularities associated with the component Green's functions. In the following section, we discuss how any generic vacuum diagram can be evaluated employing Green's function and the $n$-point vertex factors derived from the Lagrangian. We explicitly compute the divergent part of the one-loop effective action, i.e., the complete one-loop counter term Lagrangian, and we compute the divergences associated with the topologically distinct two-loop vacuum diagrams including contributions from one-loop corrections of the one-loop counter term. The final two-loop counter term Lagrangian is the cumulative effects of all these diagrams. Following this, we consider a $Z_2$ invariant scalar field Lagrangian containing up to $\phi^8$ operators and employing the divergent structures derived in the previous sections, we compute, explicitly, the complete two-loop counter terms for all the relevant, marginal, and irrelevant operators. We then showcase how we need to go beyond the regular HK method to encapsulate the impact of derivative interactions in the effective operators. We explicitly deal with this case, up to one-loop, considering an $O(n)$ symmetric scalar effective Lagrangian. In the following section, we extend our method for the fermionic fields. First, we discuss how to compute the interacting fermionic Green's function using HK. Then, we demonstrate the extraction of divergent contributions in the presence of both scalar and fermionic fields up to two-loop. We also highlight the source and structure of IR divergences that appear at the two-loop level in the massless limit. In the next section, we pave the direction to compute three-loop and comment on generic loop structures, and in the final section, we conclude.

\section{Heat-Kernel and Scalar Green's Function: Encapsulating Self-interaction}

We start with a UV Lagrangian for a scalar quantum field theory (SQFT); the part containing bilinear terms in fields takes the following form,
\begin{equation}\label{eq:delta}
    \mathcal{L}_{\text{scalar}} = \phi^{\dagger}\,(D^2+M^2+U)\,\phi\;.
\end{equation}
Here, $D_\mu$ is the covariant derivative, $M$ is the mass parameter of the scalar field $\phi$, and $U$ contains the interactions. We define  $ D^2 +  M^2 + U =\Delta$\footnote{Throughout this paper, we use the negative Euclidean metric signature $g_{\mu\nu}=-\delta_{\mu\nu}$.}, which is a second-order elliptic operator, with a positive definite spectrum, for which the Heat-Kernel (HK) is defined as \cite{Kirsten:2001wz,Vassilevich:2003xt,Avramidi:2015ch5}
\begin{equation}\label{eq:heat-kernel}
    K(t,x,y,\Delta) = \sum_n e^{-\Delta t}\,\tilde{\phi}^\dagger_n(x)\,\tilde{\phi}_n(y)\,,
\end{equation}
where $\tilde{\phi}_n$ are the eigenstates of the operator $\Delta$. In the Fourier space, the HK can be written as \cite{Osipov:2021dhc,Osipov:2001bj,Banerjee:2023xak}
\begin{equation}\label{eq:HK}
    K(t,x,y,\Delta) = \int \frac{d^d p}{(4\pi^2t)^{d/2}}e^{\frac{(x-y)^2}{4t}}e^{-M^2 t} e^{p^2}\left[1+\sum_{n=1}^\infty (-1)^n f_n(t,\mathcal A)\right],
\end{equation}
where
\begin{equation}\label{eq:fn}
    f_n(t,\mathcal A)=\int_{0}^{t}ds_1\int_{0}^{s_1}ds_2\cdots\int_{0}^{s_{n-1}}ds_n \,\mathcal A(s_1)\mathcal A(s_2)\cdots \mathcal A(s_n)\;,
\end{equation}
and
\begin{eqnarray}\label{eq:A_mat}
    \mathcal A(t) = e^{M^2 t} (D^2 + 2i\,p\cdot D/\sqrt t + U) e^{-M^2t}\;.
\end{eqnarray}
The Gaussian momentum integral over the $f_n(t,\mathcal A)$ in Eq.~\eqref{eq:HK} generates a polynomial in $t$, and that allows the HK to be  written in the following form
\begin{equation}\label{eq:HK_exp}
    K(t,x,y,\Delta)=\frac{1}{(4\pi t)^{d/2}} e^{\frac{(x-y)^2}{4t}}e^{-M^2 t} \sum_{n=0}^\infty\frac{(-t)^n}{n!}\tilde b_n,
\end{equation}
where $\tilde b_n$ are the Generalized Heat-Kernel Coefficients (g-HKC) which get contribution from $f_m(t,\mathcal A),\ m\leq 2n$\footnote{In case of derivative involved operators, this relation does not hold and $f_m(t,\mathcal{A})$ with $m> 2n$ may also contribute to a specific order of $\tilde b_n$. See App.~\ref{app:gHKC} for more details.}. In the degenerate limit, the g-HKC ($\tilde b_n$) reduces to the usual Heat-Kernel coefficients (HKC) ($b_n(x,y)$) as described in \cite{Banerjee:2023iiv}. One of the important aspects of HKCs is that they capture the information about existing interactions in the Lagrangian.

In general, to compute the scattering cross-sections, one defines a propagator, i.e., a two-point Green's function, for a free theory, and loops are constructed by employing the interactions through vertex factors along with the external states. In this work, we use a Green's function that is defined using full HK, thus it contains all the interactions of the Lagrangian. The advantage of such a definition is that we can only pay attention to the vacuum diagrams of different topologies at each loop order. This allows one to compute the smaller number of diagrams compared to the Feynman diagram approach. Here, each propagator of the vacuum diagrams is represented by a full Green's function that is expanded considering HKCs as the basis. In this expansion, the individual coefficients of the HKCs, i.e., component Green's functions (CGF), encapsulate divergences where all the HKCs are always finite.  Each vacuum topology can be expressed as a polynomial of these CGFs and HKCs. Unlike the other existing methods, the divergences associated with each topology that emerge from the CGF polynomial are computed algebraically, without performing any explicit momentum integrals. 

\begin{figure}[t]
    \centering
    \includegraphics[width=0.3\textwidth]{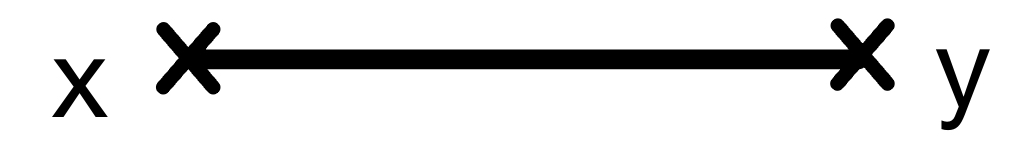}
    \caption{Green's function for interacting theory.}
    \label{fig:propagator}
\end{figure}

The scalar propagator, encapsulating the effects of interactions, (see Fig.~\ref{fig:propagator}) can be defined in terms of the HK as \cite{Jack:1982hf}
\begin{align}\label{eq:propagator}
    G(x,y)=\int_0^\infty dt\ K(t,x,y,\Delta)\,.
\end{align}
We can recast the Green's function in the required form as \cite{Jack:1982hf}, using, Eq.~\eqref{eq:HK_exp} and Eq.~\eqref{eq:propagator}, 
\begin{equation}\label{eq:greens_expansion}
    G(x,y)=\sum_{n=0}^\infty g_n(x,y) \tilde b_n(x,y)\,,
\end{equation}
with the CGFs that can be derived from,
\begin{align}
    g_n(x,y) &= \int_0^\infty dt\, \frac{1}{(4\pi t)^{d/2}} e^{\frac{z^2}{4t}}e^{-M^2 t} \frac{(-t)^n}{n!}\nn\\
    &=\frac{(-1)^n 2^{\frac{d}{2}-n}}{(4\pi)^{\frac{d}{2}}n!} \left(\frac{M}{z}\right)^{\frac{d}{2}-n-1} \mathcal K_{\frac{d}{2}-n-1}(M z),
\end{align}
where $z^2 = (x_\mu-y_\mu)\cdot(x_\mu-y_\mu)$ and $\mathcal K_n(z)$ is the modified Bessel function of the second kind. 
The divergences associated with each diagram can be extracted by noting the algebraic singularities of the CGFs $ g_n(x,y)$ without performing any momentum integrals. Please note that all the g-HKCs are finite in the coincidence limit, i.e., $z \to 0$.

The short distance behaviour (in the limit $z\rightarrow 0$) of the Green's function in $d=4-\epsilon$ dimensions is given from Eq.~\eqref{eq:greens_expansion} as \cite{Jack:1982hf}
\begin{align}\label{eq:propagator_exp}
    &G(x,y) = g_0(x,y)\,\tilde b_0+g_1(x,y)\,\tilde b_1+g_2(x,y)\,\tilde b_2+ \alpha \mathcal{F}(x,y)\,,
    \end{align}
    where the singular structures within  each $ g_n(x,y)$ can be extracted from
 \begin{align}\label{eq:propagator_exp}   
    g_0(x,y)&=\alpha\,\pi^{2-d/2}\bigg[2^{4-d} M^{d-2} \Gamma [1-d/2]-2^{2-d} z^2 M^d \Gamma [-d/2]+\frac{1}{8} M^4 z^{6-d} \Gamma [d/2-3]\nn\\
    &-M^2 z^{4-d} \Gamma [d/2-2]+4 z^{2-d} \Gamma [d/2-1]\bigg]+\mathcal O(z^4)\,,\nn\\
    g_1(x,y)&=\alpha \,\pi^{2-d/2}\bigg[2^{2-d} z^2 M^{d-2} \Gamma [1-d/2]-2^{4-d} M^{d-4} \Gamma [2-d/2]+\frac{1}{4} M^2 z^{6-d} \Gamma [d/2-3]\nn\\
    &-z^{4-d} \Gamma [d/2-2]\bigg]+\mathcal O(z^4)\,,\nn\\
    g_2(x,y)&=\alpha\,\pi^{2-d/2}\bigg[-2^{1-d} z^2 M^{d-4} \Gamma [2-d/2]+2^{3-d} M^{d-6} \Gamma [3-d/2]+\frac{1}{8} z^{6-d} \Gamma [d/2-3]\bigg]\nn\\
    &+\mathcal O(z^4)\,.
\end{align}
Here, $\alpha=1/16\pi^2$ and $\mathcal{F}(x,y)$ contain the higher order HKCs ($\tilde b_n,\ n>2$) and their coefficients do not contribute to the divergences up to two-loop. Some of the terms within $ g_n(x,y)$ may have divergences in the form of poles as they contain gamma functions having zero or negative integer arguments, but their cumulative effects do cancel among themselves making each of the $ g_n(x,y)$ finite. For example, (i) in $g_0(x,y)$, $-\Gamma[d/2-2]$ has $2/\epsilon$ pole that cancels with the $-2/\epsilon$ pole from $\Gamma[1-d/2]$, (ii) in $g_1(x,y)$, poles from $\Gamma[d/2-2]$, and $\Gamma[2-d/2]$ have opposite signs, making the overall propagator finite. 


Another important ingredient that goes in the loop computation is the $n^{th}$  vertex factor ($V^{(n)}(x)$ ) computed from the Lagrangian $\mathcal{L}$
\begin{equation}
    V_{(n)}(x)=\frac{\partial^n\L}{\partial\phi^n}\,.
\end{equation}
Then the general structure of the loop integral is given by
\begin{equation}
    \L_{(L-loop)} = S_f \int d^dx_1\,...\int d^dx_n\, V_{(m_1)}(x_1) ... V_{(m_n)}(x_n) G(x_1,x_1)^{p_1} ...\,G(x_n,x_n)^{p_n} G(x_m,x_n)^{q_{mn}}.
\end{equation}
Here, we assume the diagram consists of vertex factors of $m_1^{th},\cdots,m_n^{th}$ orders in fluctuation at $x_1,\cdots, x_n$, and $p_1,\cdots, p_n$ coincidental propagators at point $x_1, \cdots, x_n$, and on top of that, there can be $q_{mn}$ non-local propagators between points $x_m$ and $x_n$. The generic diagram may have multiple topologically equivalent structures and that information is captured through the symmetry factor $S_f$. We have provided some explicit examples explaining how to compute this symmetry factor in App.~\ref{app:symmetry_factor}.

In addition to these regular distinct diagrams, we will have diagrams that appear for loop order $\geq 2$. These diagrams use the lower order counter term Lagrangian as vertices in other loop diagrams. For instance, in a two-loop case, one such additional diagram will be the one-loop correction of the one-loop counter term Lagrangian. To compute this contribution, we need to evaluate the one-loop counter term vertex factor.  We represent the $L$-loop counter term Lagrangian by $\L_{(L)}$ and the counter term vertex factor for $n$ fluctuations in it is given as
\begin{align}\label{eq:ct_vertex}
    V_{(n)}^{(ct-L)}(x) &= \frac{\partial^n\L_{(L)}}{\partial\phi^n}.
\end{align}

To compute the topologically inequivalent vacuum diagrams it is important to note that for a 1PI $L$-loop  topology with $N$ vertices, and each with $m_i$ fluctuations,
\begin{eqnarray}\label{eq:loop_counting}
    L=\frac{1}{2}\sum_{i=1}^N m_i+1-N.
\end{eqnarray}
Since the minimum number of fluctuations at any vertex for a given loop is 3, we can write 
\begin{eqnarray}\label{eq:loop_constrain}
    \sum_{i=1}^N m_i>3N.
\end{eqnarray}
Now, one can use this information as a starting point to construct all the distinct 1PI vacuum diagrams for any arbitrary $L$-loop with the available m-point interaction terms in the Lagrangian. We also further note that considering suitable coincidence limits starting from a generic ladder diagram can help generate all the inequivalent planar diagrams of that order. 
\section{Renormalization of Composite Operators: Scalar Field-polynomials}

\subsection{One-loop Effective Action: $\mathcal{O}(1/\epsilon)$}

The one-loop effective action, for scalars having tree-level Lagrangian of the form given in Eq.~\eqref{eq:delta}, in terms of the Heat-Kernel is given by \cite{Banerjee:2023iiv},
\begin{equation}
    \L_{1-loop} =c_s \int_0^\infty \frac{dt}{t} K(t,x,x,\Delta).
\end{equation}
As discussed in Ref.~\cite{Banerjee:2023iiv}, the divergent contribution at the one-loop order for the space-time dimension $d=4-\epsilon$ is given by \cite{Banerjee:2023iiv},
\begin{equation}\label{eq:one_loop_divergence}
    \L_{(1)} = \alpha c_s M^{-\epsilon} \left(\Gamma[\epsilon/2-2]M^4\,\tilde b_0 -\Gamma[\epsilon/2-1] M^2\,\tilde b_1 + \frac{1}{2}\,\Gamma[\epsilon/2]\tilde b_2\right),
\end{equation}
where $\tilde b_1$ and $\tilde b_2$ are the g-HKC mentioned in App.~\ref{app:gHKC}, and $c_s=1/2, 1$  for real and complex scalars respectively  \footnote{We continue our discussion with real scalar unless mentioned otherwise.}. In the case of the scalar Lagrangian with a degenerate spectrum and no derivative interactions, the above g-HKCs, when background gauge fields are present, are given by,
\begin{align}\label{eq:degenerate_HKC_minimal}
    \tilde b_0(x,x)=I,\quad \tilde b_1(x,x)=U,\quad \tilde b_2(x,x)=U^2 + \frac{1}{3} (D^2 U) + \frac{1}{6}F_{\mu\nu}F_{\mu\nu},
\end{align}
where $U$ is the interaction matrix defined as $\frac{\delta^2 \mathcal{L}}{\delta \phi^2}$. One should note that though the second term in $\tilde b_2$ is a total derivative and doesn't contribute to the one-loop counter term, it will be important in higher-order loop calculations where the g-HKCs are sandwiched between two or more vertex factors and thus cannot be ignored, as they lead to non-trivial contributions. The one-loop counter term Lagrangian, given by Eq.~\eqref{eq:one_loop_divergence}, will be used to compute the counter term vertex factor for higher-order loop computations. 
\subsection{Two-loop Effective Action: $\mathcal{O}(1/\epsilon+1/\epsilon^2)$}

\begin{figure}[h]
    \centering
    \includegraphics[width=0.7\textwidth]{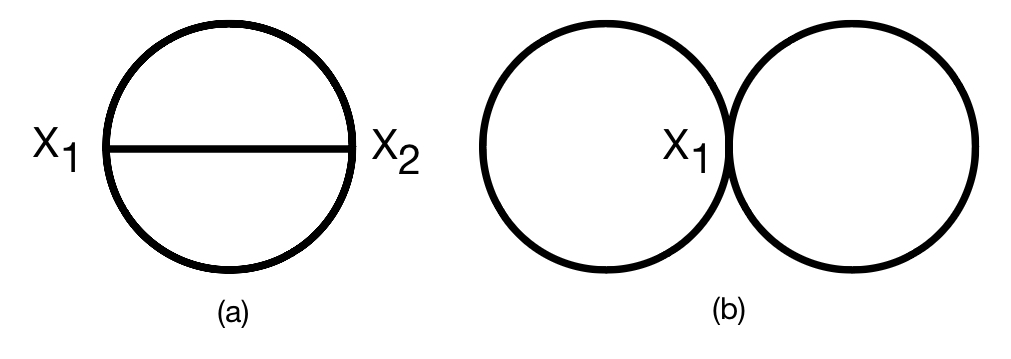}
    \caption{Distinct two-loop vacuum topologies.}
    \label{fig:two_loop}
\end{figure}

In the next loop order, i.e., for two-loop, we find two distinct vacuum diagrams shown in Fig.~\ref{fig:two_loop}.
The divergent contributions from these individual diagrams can be written as
\begin{align}\label{eq:two_loop_divergence}
    \L_{(2)}^{a}& \subset -\frac{1}{12} \int d^dx_1\,d^dx_2\, V_{(3)}(x_1) \,G(x_1,x_2)^3\, V_{(3)}(x_2),\nn\\
    \L_{(2)}^{b}& \subset \frac{1}{8} \int d^dx_1\, V_{(4)}(x_1)\, G(x_1,x_1)^2.
\end{align}
Here, in the case of Fig.~\ref{fig:two_loop} (a), we will use non-coincidence propagators, whereas in Fig.~\ref{fig:two_loop} (b) the propagators are computed at coincidence points. Now, while considering the non-local propagators raised to power $n$ only terms with $|x-y|^{m}\ \text{where, }m=-d,\,-d-2,\,-d-4\,,...$, contribute to the pole. Hence, the specific $g_i(x,y)g_j(x,y)...$ structures that contribute to the pole can be read from App.~\ref{app:formula}. For example, $[G(x,y)]^3$ in $\L_{(2)}^{a}$ contributes to the $\epsilon$ pole through Eqs.~\eqref{eq:g0_2}, \eqref{eq:g0_2_g1_1}, and \eqref{eq:g0_3}.

\begin{figure}[h]
    \centering
    \includegraphics[width=0.22\textwidth]{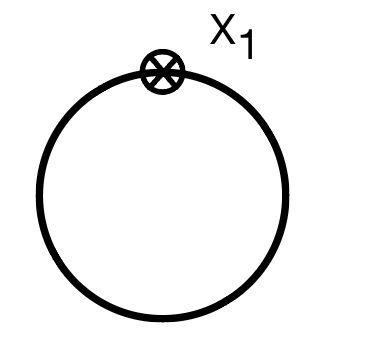}
    \caption{One-loop counter term insertion.}
    \label{fig:one_loop_one_ct}
\end{figure}

In addition to the vacuum diagrams, given in Fig.~\ref{fig:two_loop}, we will also have contributions from a one-loop diagram with a one-loop counter term insertion (see Fig.~\ref{fig:one_loop_one_ct}), which effectively contributes to the two-loop divergences. The crossed circle insertion in Fig.~\ref{fig:one_loop_one_ct} represents the one-loop counter term vertex. The contribution from this diagram can be read as
\begin{align}\label{eq:two_loop_ct_divergence}
    \L_{(2)}^{ct}&\subset \frac{1}{2}\int d^dx_1\, V_{(2)}^{(ct-1)}(x_1)\,G(x_1,x_1),
\end{align}
where the one-loop counter term vertex factor of two fluctuations is computed from Eq.~\eqref{eq:one_loop_divergence} as,
\begin{align}
    V_{(2)}^{(ct-1)}(x) &= \frac{\partial^2\L_{(1)}}{\partial\phi^2}.
\end{align}
Thus, the total two-loop divergent contribution is given by the sum of all these vacuum diagrams (Eqs.~\eqref{eq:two_loop_divergence} and \eqref{eq:two_loop_ct_divergence}) as,
\begin{eqnarray}\label{eq:final_2loop}
    \L_{(2)} = \L_{(2)}^{a}+\L_{(2)}^{b}+\L_{(2)}^{ct}.
\end{eqnarray}
\section{Renormalization of SQFT: Effective field-polynomial operators}

Let us consider a real Scalar Quantum Field Theory described by the following Lagrangian in $d=4$ dimensional  space-time\footnote{We consider the gauge fields in the covariant derivative to be background fields and do not consider their renormalization.},
\begin{equation}
    \L = \frac{1}{2}\phi\,D^2\phi + \frac{1}{2}M^2\phi^2 + \frac{\lambda}{4!}\phi^4 + \frac{c_6}{6!}\phi^6 + \frac{c_8}{8!}\phi^8.
\end{equation}
The strong elliptic operator $\Delta$ for this Lagrangian is identified as  
\[\Delta = \frac{\partial^2 \L}{\partial \phi^2} = D^2+M^2+\frac{\lambda}{2}\phi^2+\frac{c_6}{4!}\phi^4+\frac{c_8}{6!}\phi^6,\]
which defines the Heat-Kernel (see Eq.~\eqref{eq:heat-kernel}) with an interaction matrix $U$ given by,
\[U= \frac{\lambda}{2}\phi^2+\frac{c_6}{4!}\phi^4+\frac{c_8}{6!}\phi^6.\]

Now we can use the generic formalism described in  Eqs.~\eqref{eq:one_loop_divergence}, \eqref{eq:two_loop_divergence}, and \eqref{eq:two_loop_ct_divergence} to calculate the divergent contributions for the one and two-loop cases. Please note that, for this case, we need to evaluate the divergences in the propagator given in Eq.~\eqref{eq:propagator} at the coincidence limit and up to the third power of the non-local limit. The details of this calculation are given in the App.~\ref{app:formula}.

The one-loop divergent contribution, using Eq.~\eqref{eq:one_loop_divergence}, is depicted as
\begin{align}\label{eq:one_loop_divergence_ex}
    \L_{(1)} &=  \frac{\alpha}{2} \left(\Gamma[\epsilon/2-2]M^4-\Gamma[\epsilon/2-1] M^2\,U + \frac{1}{2}\,\Gamma[\epsilon/2]\,U^2\right), \nn\\
    &=  \frac{\alpha}{2} \Bigg(\Gamma[\epsilon/2-2]M^4-\Gamma[\epsilon/2-1] M^2\,\left[\frac{\lambda}{2}\phi^2+\frac{c_6}{4!}\phi^4+\frac{c_8}{6!}\phi^6\right] \nn\\
    & + \frac{1}{2}\,\Gamma[\epsilon/2]\,\left[\frac{\lambda^2}{4}\phi^4+\frac{\lambda c_6}{4!}\phi^6+\left(\frac{c_6^2}{576}+\frac{\lambda c_8}{6!}\right)\phi^8\right]\Bigg).
\end{align}
Here, we ignore the contributions that are suppressed by $\mathcal O (M^{-n})$ with $n > 4$, as we have considered our Lagrangian only up to dimension eight effective operators. They will be important if we wish to renormalize a theory containing operators of further higher mass dimensions. 

To compute the two-loop divergent contributions, we first need to calculate the necessary vertex factors: three and four points. The three-point  and four-point vertex factors are given by
\begin{equation}
    V_{(3)}(x) = \lambda\phi+\frac{c_6}{3!}\phi^3+\frac{c_8}{5!}\phi^5,
\end{equation}
\begin{equation}
    V_{(4)}(x) = \lambda+\frac{c_6}{2}\phi^2+\frac{c_8}{4!}\phi^4,
\end{equation}
respectively.

Now, we can use this information to compute the total two-loop counter terms. The divergent contributions corresponding to the diagram in Fig.~\ref{fig:two_loop} (a) are obtained from the non-local results depicted in Eqs.~\eqref{eq:g0_2}-\eqref{eq:g0_end}, and employing Eq.~\eqref{eq:two_loop_divergence} as
\begin{align}\label{eq:two_loop_divergence_ss}
    \L_{(2)}^{a}& \subset  \frac{1}{12}\int d^dx\,d^dy\, V_{(3)}(x) \left[g_0(x,y)\tilde b_0(x,y) + g_1(x,y)\tilde b_1(x,y)+\alpha \mathcal{F}(x,y)\right]^3 V_{(3)}(y)\nn\\
    & \subset  \frac{1}{12}\int d^dx\,d^dy\, V_{(3)}(x) \big[g_0(x,y)^3\tilde b_0(x,y)^3 + 3g_0(x,y)^2g_1(x,y)\tilde b_1(x,y)\tilde b_0(x,y)\nn\\&
    \quad \quad \quad +3\alpha g_0(x,y)^2\mathcal{F}(x,y)\big] V_{(3)}(y)\nn\\
    &= \frac{1}{12}\alpha^2\, V_{(3)}(x) \bigg[\left[-\frac{1}{2\epsilon}D^2-3M^2\left(\frac{2}{\epsilon^2}+\frac{1}{\epsilon}\right)\right] + 3\left(\frac{2}{\epsilon^2}+\frac{1}{\epsilon}\right)\tilde b_1(x,x)+ \frac{6}{\epsilon}\mathcal{F}(x,x)\bigg] V_{(3)}(x).
\end{align}

In the case of Fig.~\ref{fig:two_loop} (b), the divergent structures emerge from the propagators in the coincidence point as given in Eq.~\eqref{eq:coincidence_pole}. The divergent part is computed (see Eq.~\eqref{eq:two_loop_divergence}) as
\begin{align}\label{eq:two_loop_divergence_inf}
    \L_{(2)}^{b}&\subset \frac{1}{8}\int d^dx\, V_{(4)}(x) \left[g_0(x,x)b_0(x,x) + g_1(x,x)b_1(x,x)+\alpha \mathcal{F}(x,x)\right]^2\nn\\
    &=\frac{1}{8}\alpha^2\,V_{(4)}(x) \left[-\frac{2}{\epsilon}M^2 - \frac{2}{\epsilon}\tilde b_1(x,x)+\mathcal{F}(x,x)\right]^2.    
\end{align}

The counter term vertex factor, computed from one-loop counter term Lagrangian, see Eq.~\eqref{eq:one_loop_divergence_ex}, is given by
\begin{align}
    V_{(2)}^{(ct-1)}(x) &= \frac{\partial^2\L_{(1)}}{\partial\phi^2}\nn\\
    &= \frac{\alpha}{2} \Bigg(-\Gamma[\epsilon/2-1] M^2\,\left[\lambda+\frac{c_6}{2}\phi^2+\frac{c_8}{4!}\phi^4\right] \nn\\
    &\quad\quad\quad\quad + \Gamma[\epsilon/2]\,\left[\frac{3\lambda^2}{2}\phi^2+\frac{5\lambda c_6}{8}\phi^4+\left(\frac{7 c_6^2}{72}+\frac{7\lambda c_8}{90}\right)\phi^6\right]\Bigg).
\end{align}
Using the pole structures depicted in Eq.~\eqref{eq:coincidence_pole} and engaging Eq.~\eqref{eq:two_loop_ct_divergence}, the divergent contribution from the one-loop counter term insertion is given by
\begin{align}\label{eq:two_loop_divergence_ct}
    \L_{(2)}^{ct}&\subset \frac{1}{2}\int d^dx\, V_{(2)}^{(ct-1)}(x) \left[g_0(x,x)b_0(x,x) + g_1(x,x)b_1(x,x)+\alpha \mathcal{F}(x,x)\right]\nn\\
    &=\frac{\alpha}{2}\ V_{(2)}^{(ct-1)}(x)\ \left[-\frac{2}{\epsilon}M^2 - \frac{2}{\epsilon}\tilde b_1(x,x)+\mathcal{F}(x,x)\right]. 
\end{align}

Similar to the earlier case, we drop the terms with the Wilson coefficient of mass dimension $\mathcal O(M^{-n})$ with $n > 4$. Now, combining all the contributions from Eqs.~\eqref{eq:two_loop_divergence_inf},~\eqref{eq:two_loop_divergence_ss}, and ~\eqref{eq:two_loop_divergence_ct}, we find the total two-loop divergent term as  
\begin{align}
    &\phi D^2 \phi\Big|^{div} = \alpha^2\frac{\lambda^2}{24 \epsilon}\,\phi D^2 \phi, \nn\\
    &\phi^2\Big|^{div} = \alpha^2\left[\frac{M^2\lambda^2}{4 \epsilon}-\frac{1}{\epsilon^2}\left(M^2\lambda^2+\frac{M^4c_6}{4}\right)\right]\phi^2,\nn\\
    &\phi^4\Big|^{div}= \alpha^2\left[\frac{1}{\epsilon}\left(\frac{c_6 \lambda  M^2}{12} +\frac{\lambda ^3}{8}\right)-\frac{1}{\epsilon^2}\left(\frac{c_8 M^4}{48}+\frac{11c_6 \lambda  M^2}{24} +\frac{3 \lambda ^3}{8}\right)\right]\phi^4,\nn\\
    &\phi^6\Big|^{div}= \alpha^2\left[\frac{1}{\epsilon}\left(\frac{5 c_6 \lambda ^2}{96}+\frac{c_8 \lambda  M^2}{240} +\frac{c_6^2 M^2}{144}\right)-\frac{1}{\epsilon^2}\left(\frac{3 c_6 \lambda ^2}{16}+\frac{11c_8 \lambda  M^2}{360}+\frac{5 c_6^2 M^2}{144} \right)\right]\phi^6,\nn\\
    &\phi^8\Big|^{div} = \alpha^2\left[\frac{1}{\epsilon}\left( \frac{7 c_8 \lambda ^2}{2880}+\frac{c_6^2 \lambda }{144}\right)-\frac{1}{\epsilon^2}\left( \frac{31 c_8 \lambda ^2}{2880}+\frac{29 c_6^2 \lambda }{1152}\right)\right]\phi^8,\nn\\
    &\phi^3 D^2 \phi\Big|^{div} = \alpha^2\frac{\lambda c_6}{72 \epsilon}\,\phi^3 D^2 \phi,\nn\\
    &\phi^5 D^2 \phi\Big|^{div} = \alpha^2\frac{\lambda c_8}{1440 \epsilon}\,\phi^5 D^2 \phi,\nn\\
    &\phi^3 D^2 \phi^3\Big|^{div} = \alpha^2\frac{ c_6^2}{864 \epsilon}\,\phi^3 D^2 \phi^3.
\end{align}

It is important to note here that, though we do not have any derivative operators in the Lagrangian, they are generated through quantum corrections. We can recast part of the contributions from the last three derivative-involved operators into other composite operators employing the equation of motion given by
\begin{equation}
    D^2\phi = -M^2\phi-\frac{\lambda}{3!}\phi^3-\frac{c_6}{5!}\phi^5.
\end{equation}
Hence, the ($\phi^3 D^2 \phi$) operator can be expressed as
\begin{eqnarray*}
    \phi^3 D^2 \phi = -M^2\phi^4-\frac{\lambda}{3!}\phi^6-\frac{c_6}{5!}\phi^8,
\end{eqnarray*}
while the ($\phi^5 D^2 \phi$) operator can be written as
\begin{eqnarray*}
    \phi^5 D^2 \phi = -M^2\phi^6-\frac{\lambda}{3!}\phi^8,
\end{eqnarray*}
and the ($\phi^3 D^2 \phi^3$) operator can be recast as
\begin{eqnarray*}
    \phi^3 D^2 \phi^3 = -3M^2\phi^6-\frac{\lambda}{2}\phi^8+4\phi^4(D_\mu\phi)^2+2\phi^3(D_\mu\phi)\phi(D_\mu\phi).
\end{eqnarray*}

The final form of the counter terms of all the composite operators at two-loop level are given as 
\begin{align}\label{eq:counter-term-dim8}
    &\phi \,D^2 \phi\Big|^{div} = \alpha^2\frac{\lambda^2}{24 \epsilon}\,\phi \,D^2 \phi,\nn\\
    &\phi^2\Big|^{div} = \alpha^2\left[\frac{M^2\lambda^2}{4 \epsilon}-\frac{1}{\epsilon^2}\left(M^2\lambda^2+\frac{M^4c_6}{4}\right)\right]\phi^2,\nn\\
    &\phi^4\Big|^{div}= \alpha^2\left[\frac{1}{\epsilon}\left(\frac{5c_6 \lambda  M^2}{72} +\frac{\lambda ^3}{8}\right)-\frac{1}{\epsilon^2}\left(\frac{c_8 M^4}{48}+\frac{11c_6 \lambda  M^2}{24} +\frac{3 \lambda ^3}{8}\right)\right]\phi^4,\nn\\
    &\phi^6\Big|^{div} = \alpha^2\left[\frac{1}{\epsilon}\left(\frac{43 c_6 \lambda ^2}{864}-\frac{c_8 \lambda ^3 M^2}{1440}+\frac{ c_8 \lambda  M^2}{240}+\frac{c_6^2 M^2}{288} \right)-\frac{1}{\epsilon^2}\left(\frac{3 c_6 \lambda ^2}{16}+\frac{11c_8 \lambda  M^2}{360}+\frac{5 c_6^2 M^2}{144} \right)\right]\phi^6,\nn\\
    &\phi^8\Big|^{div} = \alpha^2\left[\frac{1}{\epsilon}\left( \frac{c_8 \lambda ^2}{432}+\frac{ c_6^2 \lambda }{160}\right)-\frac{1}{\epsilon^2}\left( \frac{31 c_8 \lambda ^2}{2880}+\frac{29 c_6^2 \lambda }{1152}\right)\right]\phi^8,\nn\\
    &\phi^4 (D_\mu\phi)^2\Big|^{div} = \alpha^2\frac{ c_6^2}{216 \epsilon}\,\phi^4 (D_\mu\phi)^2,\nn\\
    &\phi^3 (D_\mu\phi)\phi(D_\mu\phi)\Big|^{div} = \alpha^2\frac{ c_6^2}{432 \epsilon}\,\phi^3 (D_\mu\phi)\phi(D_\mu\phi).
\end{align}

Here, it is worthy to note the presence of two dimension eight operators involving derivatives: $ \phi^4 (D_\mu\phi)^2$, and  $ \phi^3 (D_\mu\phi)\phi(D_\mu\phi)$ that are absent in the initial Lagrangian but are generated through quantum corrections and are divergent. This implies that we need to add these bosonic operators as well with the $\phi^8$ operator at the dimension eight level, and only then will we get consistent running of all these operators. We also provide the RGEs for these operators up to one-loop, see App.~\ref{eq:app-rges}. Our computed counter terms for this scenario agree with the same, up to dimension six,  given in Refs.~\cite{Fuentes-Martin:2023ljp,Jenkins:2023rtg,Jenkins:2023bls}.

\section{Renormalization of SQFT: With derivative scalar interaction}
\label{sec:derivative_interaction}

We have discussed the computation of counter terms for an SQFT having interactions that do not contain any derivatives other than the usual kinetic term. In this section, we extend the one-loop results for the case where the Lagrangian has derivatively coupled interactions. Let us consider an $O(n)$ scalar field model with the Lagrangian given as
\begin{equation}\label{eq:Lag-deriv-int}
    \L = -\frac{1}{2}(D_\mu \phi)\cdot(D_\mu \phi) +\frac{1}{2}M^2(\phi\cdot\phi)+\frac{\lambda}{4}(\phi\cdot\phi)^2- c_6\ (\phi\cdot\phi)^3 - c_D\ (\phi\cdot\phi)(D_\mu \phi)\cdot(D_\mu \phi).
\end{equation}
We follow the same procedure as in the case of Lagrangian with only polynomial fields and expand the fields $\phi$ field around its classical solutions ($\phi_0$) in linear order in quantum fluctuations as $\phi^k \rightarrow \phi_0^k + \eta^k$ \footnote{We will use $\phi$ instead of $\phi_0$ in the expanded Lagrangian throughout the paper.}. The one-loop 1PI effective action can be obtained from the terms that are quadratic in fluctuations, which are given by
\begin{align}
    \L[\mathcal O(\eta^2)] =&\ \frac{1}{2}\eta\cdot(D^2 \eta) +c_D(\phi\cdot\phi)\eta\cdot(D^2 \eta)-4c_D(\phi\cdot\eta)[(D_\mu\phi)\cdot(D_\mu\eta)]+\frac{1}{2}M^2(\eta\cdot\eta) \nn\\
    & +2c_D[\phi\cdot(D_\mu\phi)][\eta\cdot(D_\mu\eta)]-2c_D(\eta\cdot\eta)(D_\mu\phi)\cdot(D_\mu\phi)\nn\\
    & +\frac{1}{2}\lambda[(\phi\cdot\phi) (\eta\cdot\eta)+2(\phi\cdot\eta)^2] -c_6[3(\phi\cdot\phi)^2 (\eta\cdot\eta) +12(\phi\cdot\phi)(\phi\cdot\eta)^2].
\end{align}
Please note that we are interested in the running of composite operators having mass dimensions up to six. Thus, we neglect terms of $\mathcal{O}(c_6^2)$, $\mathcal{O}(c_D^2)$ and $\mathcal{O}(c_6\,c_D)$ as they are important only at the level of dimension eight operators. 

The one-loop effective action is given by,
\begin{align}\label{eq:non_minimal_oneloop}
    \L_{1-loop}&=\frac{i}{2}\,\tr\ \log\left[\frac{\delta^2 \L}{\delta\eta^i\delta\eta^j}\right]\nn\\
    &=\frac{i}{2}\,\tr\ \log\bigg[ \delta_{ij} D^2 + 2\delta_{ij} c_D\,\phi\cdot\phi\, D^2 -4c_D[(\phi_i)(D_\mu\phi)_j-(D_\mu\phi)_i\phi_j]D_\mu+\delta_{ij}M^2 \nn\\
    &\hspace{1.5cm}  -2c_D\,D_\mu(\phi\cdot(D_\mu\phi))\delta_{ij}-2c_D\,\delta_{ij}(D_\mu\phi)\cdot(D_\mu\phi) +\lambda[(\phi\cdot\phi) \delta_{ij}+2\phi_i\phi_j] \nn\\
    &\hspace{1.5cm}-c_6[6(\phi\cdot\phi)^2 \delta_{ij} +24(\phi\cdot\phi)\phi_i\phi_j] \bigg]\nn\\
    &=\frac{i}{2}\,\tr\ \log\bigg[ \delta_{ij} D^2 + a_{ij} D
    ^2 +(h_\mu)_{ij} D_\mu+\delta_{ij}M^2 + \tilde U_{ij} \bigg],
\end{align}
where,
\begin{align}
     a_{ij} &= 2\,c_D \,(\phi.\phi)\,\delta_{ij},\nn\\
    (h_\mu)_{ij} &= -4c_D[(\phi_i)(D_\mu\phi)_j-(D_\mu\phi)_i\phi_j],\nn\\
    \tilde U_{ij} &= -2c_D\delta_{ij}(D_\mu\phi)\cdot(D_\mu\phi)-2c_D\delta_{ij}D_\mu(\phi\cdot(D_\mu\phi)) +\lambda[(\phi\cdot\phi) \delta_{ij}+2\phi_i\phi_j] \nn\\
    &\quad-c_6[6(\phi\cdot\phi)^2 \delta_{ij} +24(\phi\cdot\phi)\phi_i\phi_j].
\end{align}
Here, we use Greek and Latin indices to represent the Lorentz  and the field spaces, respectively.
Including derivative interaction terms in the Lagrangian leads to a non-minimal second-order elliptic operator whose g-HKCs can be obtained by writing the HK in the Fourier space as described in \cite{Banerjee:2023xak}. The first few g-HKC required for the computation of the one-loop divergent contributions for the above operator is given by,
\begin{align}\label{eq:HKC_non_minimal}
    \tilde b_0(x,x)&=I-2a,\nn\\
    \tilde b_1(x,x)&=\tilde U_{ij} -2(a\,\tilde U)_{ij},\nn\\
    \tilde b_2(x,x)&=\tilde U_{ij}^2 - 2\tilde U_{ij}[D_\mu,h_\mu]_{ij}-2 a_{ij}\,(\tilde U^2)_{ij},
\end{align}
and hence, the one-loop divergent contribution from Eq.~\eqref{eq:one_loop_divergence} is given by,
\begin{align}\label{eq:one_loop_derivative}
    \L_{(1)}&= \frac{1}{2\epsilon}\bigg[4nM^2c_D\,\phi\cdot(D^2\phi)+M^2[2(n+2)\lambda-4n\,M^2c_D](\phi\cdot\phi)\nn\\
    \quad\quad+ & [(n+8)\lambda^2-12(n+4)M^2c_6-8(n+2)M^2\lambda c_D](\phi\cdot\phi)^2\nn\\
    \quad\quad- & [12(n+14)\lambda c_6+4(n+8)\lambda^2 c_D](\phi\cdot\phi)^3 -4(n+2)\lambda c_D(\phi\cdot\phi)(D_\mu\phi)\cdot(D_\mu\phi)\bigg].
\end{align}

\section{Heat-Kernel and Fermion Green's Function: Encapsulating scalar-fermion interaction}

In this section, we extend our counter term computation prescription for the case of fermions. We, first, assume the generic form of fermionic Lagrangian as
\begin{equation}
    \L_{Fermion} = \overline\psi_{i}\left( i \slashed D \delta_{ij} - M_{(f)}^i\delta_{ij} - \Sigma_{ij}\right) \psi_j + \text{H.C.}
\end{equation}
where the ($i,j$) indices represent the fermion generation, the covariant derivative $\slashed D = \gamma_\mu\partial_\mu - i \gamma_\mu A_\mu^i$, $M_{(f)}$ is the diagonal mass matrix, and $\Sigma$ is the interaction matrix. We consider both scalar and pseudo scalar Yukawa interactions. Hence, $\Sigma$ has the general form $\Sigma_{ij} = S_{ij}+ i\gamma_5 R_{ij}$. Considering this form for the Lagrangian, the fermionic propagator can be expressed as
\begin{align}\label{eq:fermion_bosonized}
    G^{f}_{ij}(x,y) &= \frac{1}{i\slashed D \delta_{ij} - M_{(f)}^i\delta_{ij} - \Sigma_{ij}} = \delta_{ik}\ \frac{1}{i\slashed D \delta_{kj} - M_{(f)}^k\delta_{kj} - \Sigma_{kj}}\nn\\
    &= (-i\slashed D \delta_{il} - M_{(f)}^i\delta_{il} -  \Sigma_{il}) \frac{1}{(-i\slashed D \delta_{lk} - M_{(f)}^l\delta_{lk} - \Sigma_{lk})(i\slashed D \delta_{kj}-M_{(f)}^k\delta_{kj} - \Sigma_{kj})}\nn\\
    &= (-i\slashed D \delta_{ik} - M_{(f)}^i \delta_{ik}- \Sigma_{ik})\frac{1}{(\tilde D)^2 \delta_{kj}+ (M_{(f)}^k)^2\delta_{kj} + Y_{kj} + M_{(f)}^k\delta_{kl} \Sigma_{lj} +\Sigma_{kl} M_{(f)}^l\delta_{lj}},
\end{align}
with
\begin{align}
    &Y_{ij} = -\frac{1}{2}\sigma_{\mu\nu}X_{\mu\nu} \delta_{ij} + S_{ik}S_{kj} + 3 R_{ik}R_{kj} +i (\slashed D S_{ij})+i\gamma_5(R_{ik}S_{kj} + S_{ik}R_{kj}),\nn\\
    &\tilde D_\mu^{ij} = D_\mu \delta_{ij} + \gamma^5\gamma_\mu R_{ij},\nn\\
    &\Gamma_{\mu\nu}^{ij} = [\tilde D_{\mu}^{ik},\tilde D_{\mu\nu}^{kj}] = X_{\mu\nu}^i \delta_{ij}- \gamma^5\gamma_{\mu} (D_{\nu}R_{ij}) + \gamma^5\gamma_{\nu} (D_{\mu}R_{ij}) - 2 \sigma_{\mu\nu} R_{ij}R_{jk}.
\end{align}
Here, we have used the notation,\[X_{\mu\nu} = [D_\mu,D_\nu],\quad \sigma_{\mu\nu} = \frac{1}{2}[\gamma_\mu,\gamma_\nu],\quad (D_\mu X) = [D_\mu, X].\]
Now, to employ the HK method to compute the interaction propagator, we have bosonized \cite{Chakrabortty:2023yke} the fermionic propagator in Eq.~\eqref{eq:fermion_bosonized} which leads to the following form of the fermionic Green's function, 
\begin{align}
    G^{f}_{ij}(x,y) &= (-i\slashed D \delta_{ik} - M_{(f)}^i \delta_{ik}- \Sigma_{ik}) \int _0^\infty dt\,  K(t,x,y,\Delta^{(f)}_{kj}).
\end{align}
Here, $\Delta^{(f)}_{kj}= (\tilde D^2)^{kj}+ (M_{(f)}^k)^2\delta_{kj} + U^{(f)}_{kj}$ with $U^{(f)}_{kj} = Y_{kj} + M_{(f)}^k\delta_{kl} \Sigma_{lj} +\Sigma_{kl} M_{(f)}^l\delta_{lj}$.
Now, very similar to the scalar case,  the fermionic Green's function can be written using the HKCs as
\begin{align}\label{eq:fermionic propagator}
     G^{f}_{ij}(x,y) &= (-i\slashed D_x \delta_{ik} - M_{(f)}^i\delta_{ik}- \Sigma_{ik}) \sum_{n=0}^\infty g_n(x,y)\tilde b^{(f)}_n(x,y)_{kj}\nn\\
     &= -i \sum _{n=0}^\infty \left[(\slashed D_x\, g_n(x,y))\tilde b^{(f)}_n(x,y)_{ij} + g_n(x,y)(\slashed D_x\,\tilde b^{(f)}_n(x,y)_{ij})\right]\nn\\
     &\quad\quad\quad -(M_{(f)}^i\delta_{ik}+\Sigma_{ik}) \sum_{n=0}^\infty g_n(x,y)\tilde b^{(f)}_n(x,y)_{kj}.
\end{align}
This form of the interacting fermion propagator can be used to compute the divergent structures.

\section{Composite Operator Renormalization: Fermion-Scalar polynomials}\label{sec:fermion_expample}
In this section, we consider a toy example with interacting massive complex scalar and fermion fields with a few selected dimension six effective operators
\begin{align}
    \L &= i\,\delta_{ij}\,\overline\psi_i(\slashed D\psi_j) - M^i_{(f)}\,\delta_{ij}\, \overline\psi_i\psi_j - y^{(1)}_{ij} \phi\, \overline\psi_i\psi_j  - \frac{y^{(3)}_{ij}}{2} |\phi^\dagger\phi|\, \phi \,\overline\psi_i\psi_j  \nn\\
    &\quad\quad\quad- (D_\mu\phi)^\dagger(D_\mu\phi) + M_s\, \phi^\dagger\phi + \frac{\lambda}{4}|\,\phi^\dagger\phi|^2+ \frac{c_6}{3!3!} \,|\phi^\dagger\phi|^3 + H.C.
\end{align}
We, in the following sections, demonstrate and compute the divergent contributions up to two-loop order. 
\subsection{One-loop effective action: $\mathcal{O}(1/\epsilon)$}\label{sec:fermion_one_loop}

The one-loop divergent structures from the pure fermionic loop can be read off from the one-loop effective action \cite{Chakrabortty:2023yke} (see Eq.~\eqref{eq:one_loop_divergence}) by replacing $U$  in Eq.~\eqref{eq:degenerate_HKC_minimal} with $U^{(f)}_{\rho\sigma}$. Setting $c_s = -1$, the one-loop counter term Lagrangian is given by
\begin{align}\label{eq:fermion_one_loop}
    \L_{(1)}^f &= -\frac{\alpha}{2}\, \tr \,(M_{(f)}^i)^{-\epsilon}\,\bigg( 2\,\Gamma[\epsilon/2-2]\, (M_{(f)}^i)^4 - 2 \,\Gamma[\epsilon/2-1](M_{(f)}^i)^2 U^{(f)}_{ii} \nn\\
    &\quad\quad\quad+ \Gamma[\epsilon/2]\,U^{(f)}_{ij}(U^{(f)}_{ij})^\dagger + \frac{ \Gamma[\epsilon/2]}{6} \Gamma^{ij}_{\mu\nu}(\Gamma^{ij}_{\mu\nu})^\dagger\bigg).
\end{align}
The trace, here, is over the Clifford and internal symmetry spaces.

In the case of the Lagrangian under consideration, the pure fermion loop one-loop counter term, using  Eq.~\eqref{eq:fermion_one_loop}, is given by
\begin{align}
    \L_{(1)}^f&=-\frac{\alpha}{\epsilon}\bigg[10\,M_f^2 {y^{(1)}_{ij}}^\dagger y^{(1)}_{ij}\, (\phi^\dagger \phi) +\big(4\,M_f{y^{(1)}_{ij}}^\dagger y^{(1)}_{ik} y^{(1)}_{kj}\big)(\phi^\dagger\phi)\phi+\big(4\,M_f{y^{(1)}_{ij}}^\dagger{y^{(1)}_{jk}}^\dagger y^{(1)}_{ik}\big)(\phi^\dagger\phi)\phi^\dagger \nn\\
    &\quad\quad+ \big(5\,M_f^2 {y^{(3)}_{ij}}^\dagger y^{(1)}_{ij}+5\,M_f^2 {y^{(1)}_{ij}}^\dagger y^{(3)}_{ij}+ 2\,{y^{(1)}_{ij}}^\dagger y^{(1)}_{ik}{y^{(1)}_{lk}}^\dagger y^{(1)}_{lj}\big)(\phi^\dagger \phi)^2\nn\\
    &\quad\quad + \big( 2\,{y^{(1)}_{ij}}^\dagger {y^{(3)}_{kl}}^\dagger y^{(1)}_{il} y^{(1)}_{kj}+{y^{(1)}_{ij}}^\dagger {y^{(1)}_{kl}}^\dagger y^{(3)}_{il} y^{(1)}_{kj}+{y^{(1)}_{ij}}^\dagger {y^{(1)}_{kl}}^\dagger y^{(1)}_{il} y^{(3)}_{kj}\big)(\phi^\dagger \phi)^3\nn\\
    &\quad\quad +\big(2\,M_f {y^{(1)}_{ij}}^\dagger{y^{(3)}_{jk}}^\dagger y^{(1)}_{ik}+2\,M_f {y^{(1)}_{ij}}^\dagger{y^{(1)}_{jk}}^\dagger y^{(3)}_{ik}+2\,M_f {y^{(1)}_{ij}}^\dagger{y^{(3)}_{ki}}^\dagger y^{(1)}_{kj}\big)(\phi^\dagger\phi)^2\phi^\dagger\nn\\
    &\quad\quad+\big(2\,M_f {y^{(1)}_{ij}}^\dagger y^{(3)}_{kj} y^{(1)}_{ik}+2\,M_f {y^{(1)}_{ij}}^\dagger y^{(1)}_{kj} y^{(3)}_{ik}+2\,M_f {y^{(1)}_{ij}}^\dagger y^{(3)}_{ik} y^{(1)}_{kj}\big)(\phi^\dagger\phi)^2\phi\bigg].
\end{align}

The pure scalar one-loop counter term from Eq.~\eqref{eq:one_loop_divergence} is given by
\begin{align}
    \L_{(1)}^s&=\frac{\alpha}{\epsilon}\bigg[2\,M_s^2\lambda(\phi^\dagger \phi) + \big(2\,\lambda^2+\frac{1}{2}M_s^2c_6\big)(\phi^\dagger \phi)^2+\lambda\,c_6(\phi^\dagger \phi)^3+2\,M_s^2y^{(1)}_{ij}\phi\overline\psi_i\psi_j\nn\\
    &\quad\quad+2\,\lambda\,y^{(3)}_{ij}(\phi^\dagger\phi)\phi\overline\psi_i\psi_+2\,\lambda\,{y^{(3)}_{ij}}^\dagger (\phi^\dagger\phi)\phi^\dagger\overline\psi_j\psi_i\bigg].
\end{align}
\begin{figure}[h]
    \centering
    \includegraphics[width=0.22\textwidth]{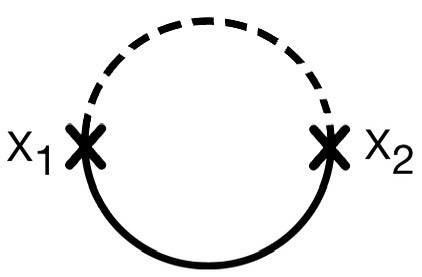}
    \caption{Scalar-fermion mixed one-loop topology. Scalar and fermion Green's functions are depicted by solid and dotted lines respectively.}
    \label{fig:scalar_fermion_mixed}
\end{figure}
We also simultaneously evaluate the additional one-loop topology containing scalar and fermion propagators, i.e., the mixed statistics one-loop contribution (see Fig.~\ref{fig:scalar_fermion_mixed}). The counter term Lagrangian corresponding to this mixed one is given as 
\begin{align}
    \L_{(1)}^{sf}&\subset \int d^dx_1\,d^dx_2\, V^\dagger_{(2)}(x_1)\Big\lvert_{\phi\psi_i} G^f_{ij}(x_1,x_2)G^s(x_1,x_2) V_{(2)}(x_2)\Big\lvert_{\phi^\dagger\overline\psi_j}\nn\\
    &=\frac{\alpha}{\epsilon}\bigg[-2\,{y^{(1)}_{li}}^\dagger y^{(1)}_{lk} y^{(1)}_{kj}\phi\overline\psi_i\psi_j-i\,{y^{(1)}_{ki}}^\dagger y^{(1)}_{kj}\overline\psi_i\slashed D (\psi_j) 
     -i\,{y^{(3)}_{ki}}^\dagger y^{(1)}_{kj}(\phi^\dagger\phi)\overline\psi_i\slashed D(\psi_j) \nn\\
     &-2\,M_f\,{y^{(1)}_{ki}}^\dagger y^{(1)}_{kj}\overline\psi_i\psi_j+\big(-2\,M_f\,{y^{(3)}_{ki}}^\dagger y^{(1)}_{kj}-2\,M_f\,{y^{(1)}_{ki}}^\dagger y^{(3)}_{kj}\big)(\phi^\dagger \phi)\overline\psi_i\psi_j\nn\\
     & -i\,{y^{(1)}_{ki}}^\dagger y^{(3)}_{kj}\overline\psi_i\slashed D(\phi^\dagger\phi\psi_j)
     +\big(2{y^{(3)}_{ki}}^\dagger y^{(1)}_{kl} y^{(1)}_{lj}+{y^{(1)}_{ki}}^\dagger y^{(3)}_{kl} y^{(1)}_{lj}+2{y^{(1)}_{ki}}^\dagger y^{(1)}_{kl} y^{(3)}_{lj}\big)(\phi^\dagger\phi)\phi\overline\psi_i\psi_j\bigg] .
\end{align}
The above mixed-statistics counter term is not the complete contribution to the one-loop order. The topology in Fig.~\ref{fig:scalar_fermion_mixed} can be extended by adding two-point vertices and contributions from those structures must be included. But for the sake of demonstration of our method, this would be sufficient. Computation of such diagrams involves $\epsilon$ pole structures arising from non-trivial space-time functions similar to that described in App.~\ref{app:benz}. Hence, we leave that to our future work.

\subsection{Two-loop effective action: $\mathcal{O}(1/\epsilon+1/\epsilon^2)$}\label{sec:fermion_two_loop}

\begin{figure}[h]
    \centering
    \includegraphics[width=\textwidth]{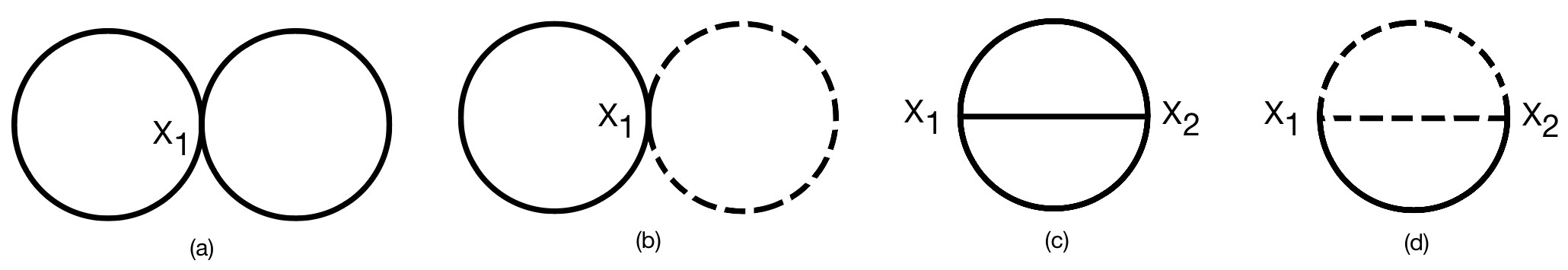}
    \caption{Two-loop topologies that contribute to desired divergent contributions. Here, the dotted lines represent the fermionic propagator and the solid lines represent the scalar propagator.}
    \label{fig:fermion_two_loop}
\end{figure}

\begin{figure}[h]
    \centering
    \includegraphics[width=1\textwidth]{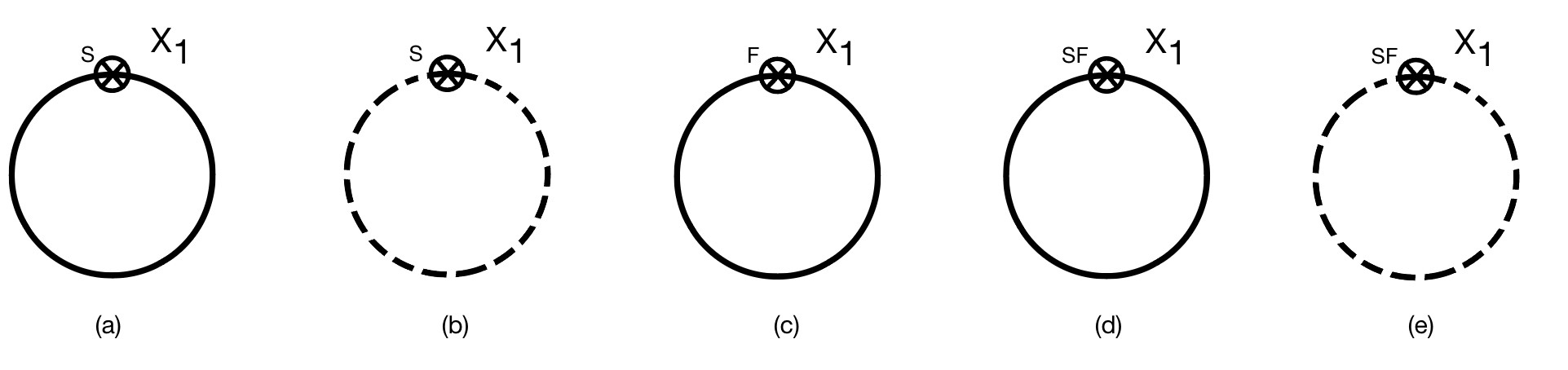}
    \caption{One-loop counter term insertion in one-loop topology. Here scalar and fermion propagators are depicted by solid and dotted lines respectively and the script S/F/SF near the crossed circle denotes scalar/fermion/scalar-fermion one-loop counter term insertion.}
    \label{fig:fermion_oneloop_ct}
\end{figure}

The two-loop counter terms get contribution from pure two-loop structures, as shown in Fig.~\ref{fig:fermion_two_loop}, and also from the one-loop counter term insertions in one-loop as depicted by Fig.~\ref{fig:fermion_oneloop_ct}. 
The different divergent contributions from the individual topologies in Fig.~\ref{fig:fermion_two_loop}, are computed employing the divergence structures from App.~\ref{app:formula}, similar to the scalar case.
The individual contributions are given by 
\begin{align}
    \L_{2}^a &\subset \frac{1}{2}\int d^dx_1\, V_{(4)}^\dagger(x_1)\Big\lvert_{\phi^\dagger\phi\phi^\dagger\phi} G^s(x_1,x_1) (G^s(x_1,x_1))^\dagger,\nn\\
    \L_{2}^b &\subset -\int d^dx_1\, V_{(4)}^\dagger(x_1)\Big\lvert_{\phi^\dagger\phi\overline\psi_i\psi_j} G^s(x_1,x_1)G^f_{ij}(x_1,x_1),\nn\\
    \L_{2}^c &\subset -\frac{1}{2}\int d^dx_1\,d^dx_2\, V^\dagger_{(3)}(x_1)\Big\lvert_{\phi^\dagger\phi\phi} (G^s(x_1,x_2))^3\ V_{(3)}(x_2)\Big\lvert_{\phi\phi^\dagger\phi^\dagger},\nn\\
    \L_{2}^d &\subset \int d^dx_1\,d^dx_2\, V^\dagger_{(3)}(x_1)\Big\lvert_{\phi\overline\psi_i\psi_j} G^f_{jl}(x_1,x_2)(G^f_{ki}(x_1,x_2))^\dagger G^s(x_1,x_2) V_{(3)}(x_2)\Big\lvert_{\phi^\dagger\overline\psi_l\psi_k}.
\end{align}
The necessary vertex factors required for the computation, derived from the example Lagrangian, are 
\begin{align}
    &V_{(4)}(x)\Big\lvert_{\phi^\dagger\phi\phi^\dagger\phi} = \lambda + c_6 \phi^\dagger\phi,\nn\\
    &V_{(4)}(x)\Big\lvert_{\phi^\dagger\phi\overline\psi_i\psi_j} = - y_{ij}^{(3)}\phi,\nn\\
    &V^{(3)}(x)\Big\lvert_{\phi^\dagger\phi\phi}= -y_{ij}^{(3)}\overline\psi_i\psi_j+\lambda \phi^\dagger +\frac{c_6}{2}\phi^\dagger\phi \phi^\dagger,\nn\\
    &V_{(3)}(x)\Big\lvert_{\phi\overline\psi_i\psi_j}= -y^{(1)}_{ij} - y^{(3)}_{ij}\phi^\dagger\phi.
\end{align}
Since we have fields of different characteristics, we tag the vertex factors with their respective fluctuations. Here, the interacting Green's functions are either purely fermionic $(G^f)$ or scalar $(G^s)$. Though the expansion of both the Green's functions is similar in terms of the HKCs, the structures of the HKCs for scalar and fermions cases are different. They are computed using suitable interaction matrices, for the above Lagrangian given as
\begin{align}
    U^{(s)}&=y^{(3)}_{ij}\phi\overline\psi_i\psi_j+\lambda \phi^\dagger\phi + \frac{c_6}{4}(\phi^\dagger\phi)^2,\nn\\
    U^{(f)}_{ij}&=Y_{ij} +2\,M_f^i\Sigma_{ij},\nn\\
    \Sigma_{ij} &= y_{ij}^{(1)}\phi + y_{ij}^{(3)}(\phi^\dagger\phi)\phi,\nn\\
    Y_{ij} &= -\frac{1}{2}\sigma_{\mu\nu}F_{\mu\nu}\delta_{ij} +\Sigma_{ik}\Sigma_{jk}^\dagger+i \slashed D (\Sigma_{ij}),\nn\\
    \Gamma_{\mu\nu}^{ij} &= F_{\mu\nu} \delta_{ij}.
\end{align}

To encapsulate the contributions from the counter term insertion diagrams (see Fig.~\ref{fig:fermion_oneloop_ct}), we employ the one-loop results computed in the previous section. Now, we can calculate the two-loop counter term from individual diagrams as 
\begin{align}
    \L_{(2-ct)}^{(a)} &\subset\int d^dx_1\, {V_{(2)}^{(s,ct-1)}}^\dagger(x_1)\Big\lvert_{\phi^\dagger\phi}\,G^s(x_1,x_1),\nn\\
    \L_{(2-ct)}^{(b)} &\subset -\int d^dx_1\, {V_{(2)}^{(s,ct-1)}}^\dagger(x_1)\Big\lvert_{\overline\psi_j\psi_i}\,G^f_{ij}(x_1,x_1),\nn\\
    \L_{(2-ct)}^{(c)} &\subset\int d^dx_1\, {V_{(2)}^{(f,ct-1)}}^\dagger(x_1)\Big\lvert_{\phi^\dagger\phi}\,G^s(x_1,x_1),\nn\\
    \L_{(2-ct)}^{(d)} &\subset\int d^dx_1\, {V_{(2)}^{(sf,ct-1)}}^\dagger(x_1)\Big\lvert_{\phi^\dagger\phi}\,G^s(x_1,x_1),\nn\\
    \L_{(2-ct)}^{(e)} &\subset-\int d^dx_1\, {V_{(2)}^{(sf,ct-1)}}^\dagger(x_1)\Big\lvert_{\overline\psi_j\psi_i}\,G^f_{ij}(x_1,x_1).\
\end{align}
Here, ${V_{(2)}^{(s,ct-1)}}$ is the counter term vertex factor from pure scalar one-loop, ${V_{(2)}^{(f,ct-1)}}$ is from pure fermion one-loop, and ${V_{(2)}^{(sf,ct-1)}}$ is from the scalar-fermion mixed loop. The counter terms for these individual two-loop equivalent contributions are very rich in structure and lengthy. Thus, we note the explicit functional dependence of the emerged counter terms in App.~\ref{app:fermion_results}, where we showcase only different non-degenerate structures in the coefficients. We also highlight the singled-out IR divergences that are of the form $\frac{1}{\epsilon}\; log(M^2)$ along with the UV divergences of order up to $\epsilon^{-2}$. The complete result can be found in \href{https://github.com/kaanapuliramkumar/Heat-Kernel-Counter-terms}{Github}.
\section{Paving the Path to Higher Loops: Three-loop example}
The prescription that we have formulated and discussed in detail in the previous sections can be generalized for higher-order loops as well. One can generate all the inequivalent vacuum diagrams for any given loop order using Eq.~\eqref{eq:loop_counting}, and these loops are made of higher point ($\geq 3$) vertex factors and coincidental and non-coincidental Green's functions. Now, we can employ a similar strategy to identify the total divergences at each loop order, for both fermion and scalar, as the divergences associated with each loop are effectively replicated from the singular behavior, i.e. the algebraic singularities, of polynomials in $g_n(x,y)$.

\begin{figure}[h]
    \centering
    \includegraphics[width=\textwidth]{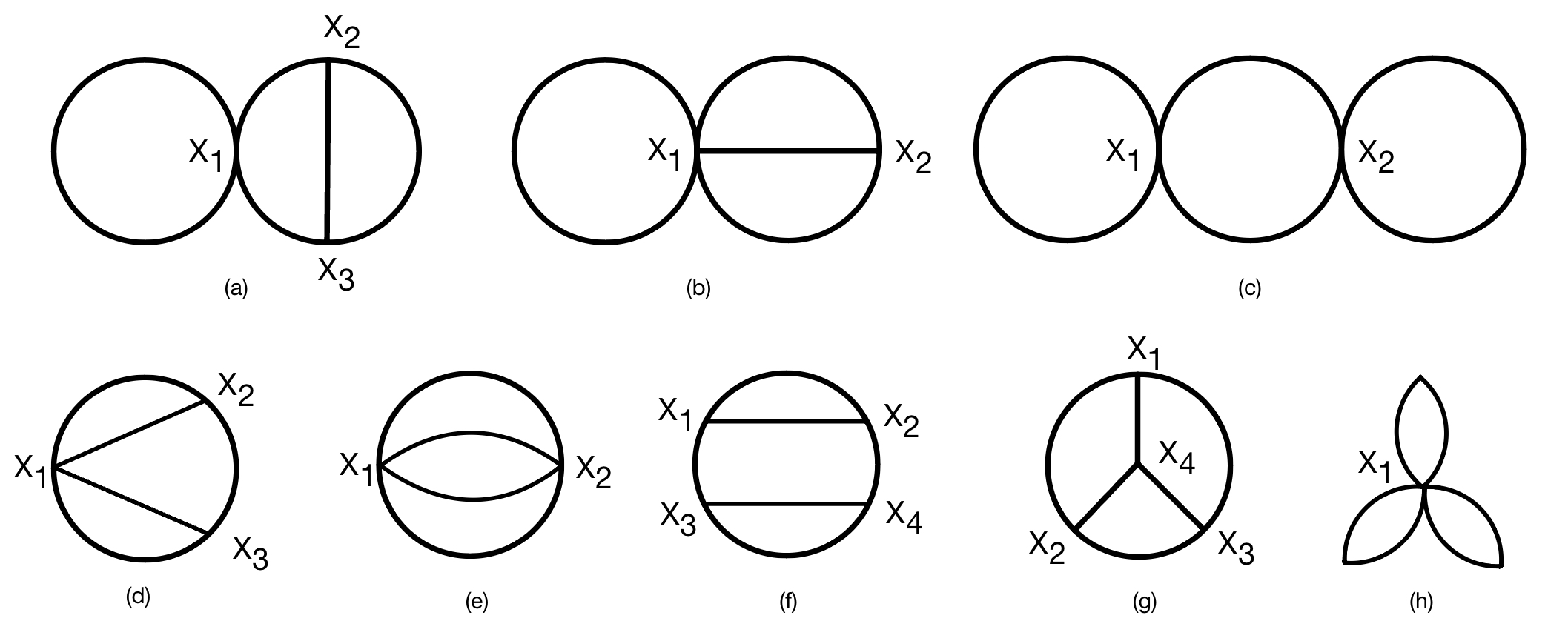}
    \caption{Distinct Three-loop topologies.}
    \label{fig:three_loop}
\end{figure}

Here, we have demonstrated how the knowledge for one and two-loop computations can be used to extrapolate the formalism to compute the same for three-loop vacuum diagrams. The diagrams depicted in Fig.~\ref{fig:three_loop} are the different topologies that contribute to the three-loop order. Following the aforementioned methodologies, the contributions corresponding to each of these topologies can be written in terms of the suitable propagators and the vertex factors as
\begin{align}\label{eq:three_loop_divergence}
    \L_{(3)}^{a}&\subset - \frac{1}{8} \int d^dx_1\,d^dx_2\,d^dx_3\, G(x_1,x_1)\,V_{(4)}(x_1)\,G(x_1,x_2)\,V_{(3)}(x_2)\,G(x_2,x_3)^2\,V_{(3)}(x_3)\,G(x_3,x_1),\nn\\
    \L_{(3)}^{b}&\subset \frac{1}{12}  \int d^dx_1\,d^dx_2 G(x_1,x_1)\,V_{(5)}(x_1)\,G(x_1,x_2)^3\,V_{(3)}(x_2),\nn\\
    \L_{(3)}^{c}&\subset \frac{1}{16}  \int d^dx_1\,d^dx_2 G(x_1,x_1)\,V_{(4)}(x_1)\,G(x_1,x_2)^2\,V_{(4)}(x_2)\,G(x_2,x_2),\nn\\
    \L_{(3)}^{d}&\subset -\frac{1}{8}  \int d^dx_1\,d^dx_2\,d^dx_3\, V_{(4)}(x_1)\,G(x_1,x_2)^2\,V_{(3)}(x_2)\,G(x_2,x_3)\,V_{(3)}(x_3)\,G(x_3,x_1)^2,\nn\\
    \L_{(3)}^{e}&\subset \frac{1}{48}  \int d^dx_1\,d^dx_2\,V_{(4)}(x_1)\,G(x_1,x_2)^4\,V_{(4)}(x_2),\nn\\
    \L_{(3)}^{f}&\subset \frac{1}{16}  \int d^dx_1\,d^dx_2\,d^dx_3\,d^dx_4\, V_{(3)}(x_1)\,G(x_1,x_2)^2\,V_{(3)}(x_2)\,G(x_1,x_4)\,,\nn\\
    & \hspace{2cm} V_{(3)}(x_4)\,G(x_4,x_3)^2\,V_{(3)}(x_3)\,G(x_3,x_1),\nn\\
    \L_{(3)}^{g}&\subset \frac{1}{24}  \int d^dx_1\,d^dx_2\,d^dx_3\,d^dx_4\,V_{(3)}(x_1)\,G(x_1,x_2)\,G(x_1,x_3)\,G(x_1,x_4)\,V_{(3)}(x_4)\, \nn\\&\hspace{2cm} G(x_4,x_2)\,G(x_4,x_3) V_{(3)}(x_2)\,G(x_2,x_3)\,V_{(3)}(x_3),\nn\\
    \L_{(3)}^{h}&\subset -\frac{1}{48}  \int d^dx_1\, V_{(6)}(x_1)\,G(x_1,x_1)^3.
\end{align}
Here, all the diagrams apart from Fig.~\ref{fig:three_loop}(g), referred to as the Benz diagram, are planar. Thus, their singularities can be computed using a similar formalism as in previous cases (see App.~\ref{subsec:singular-noncoincident}).
Due to the non-planar structure of Fig.~\ref{fig:three_loop}(g), obtaining the $\epsilon$ pole structures is non-trivial, unlike the other structures. This is because every loop in the Benz topology is formed between three space-time points, whereas the other structures either have a loop formed at one space-time point, the coincidence point loop, or a loop between two space-time points. This has been further elaborated in App.~\ref{app:benz}. In higher and higher loops, we will encounter different non-planar topologies which require special attention.

\begin{figure}[h]
    \centering
    \includegraphics[width=1\textwidth]{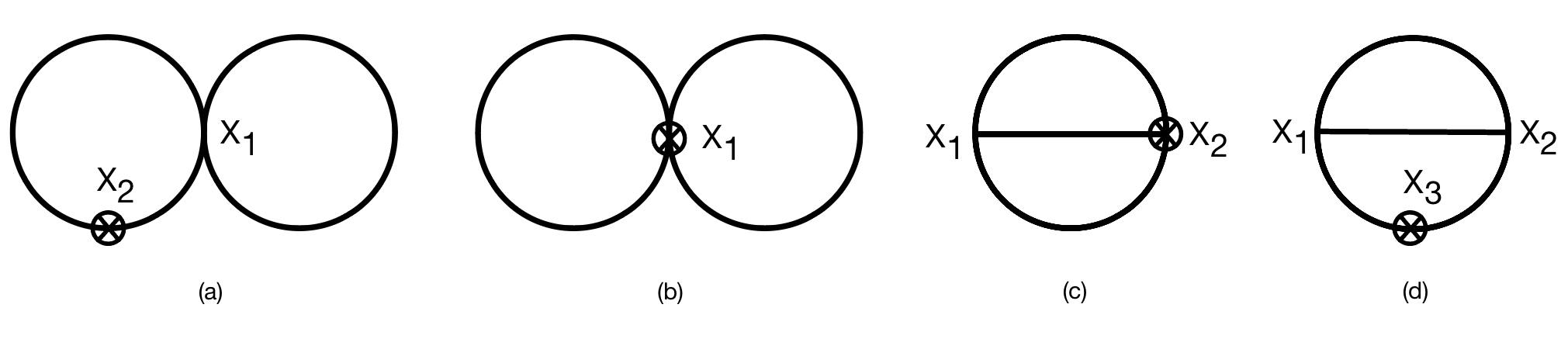}
    \caption{One-loop counter term insertions in two-loop topologies.}
    \label{fig:two_loop_ct}
\end{figure}
\begin{figure}[h]
    \centering
    \includegraphics[width=0.5\textwidth]{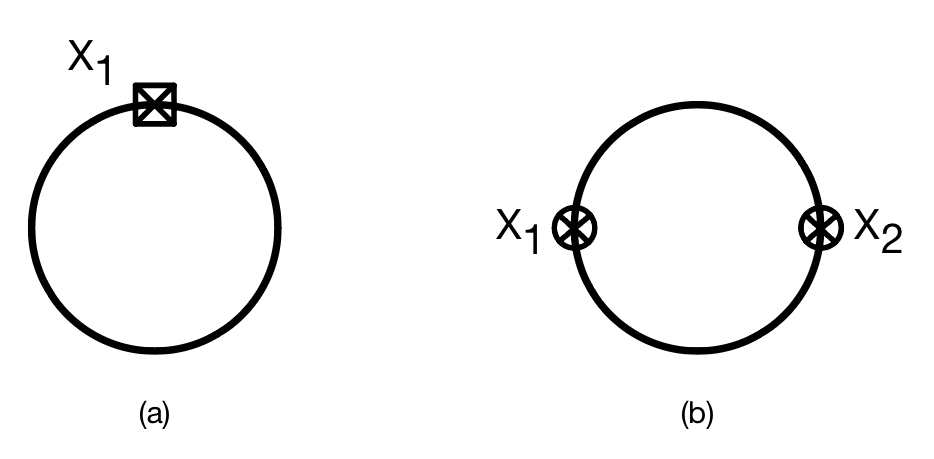}
    \caption{ 2 One-loop and 1  two-loop counter term insertions in one-loop topologies.}
    \label{fig:one_loop_two_ct}
\end{figure}

Similar to previous cases, in three-loop order, we also find contributions from one and two-loop counter term insertions. The one-loop counter term insertion in two-loop diagrams, which is the three-loop equivalent (see Fig.~\ref{fig:two_loop_ct}), can be expressed as
\begin{align}\label{eq:one_loop_ct_two_loop_divergence}
    \L_{(3)}^{a-ct-2}&\subset \frac{1}{16} \int d^dx_1\,d^dx_2\, G(x_1,x_1)V_{(4)}(x_1)\, G(x_1,x_2)^2\,V_{(2)}^{(ct-1)}(x_2),\nn\\
    \L_{(3)}^{b-ct-2}&\subset -\frac{1}{8} \int d^dx_1\, V_{(4)}^{(ct-1)}(x_1)\, G(x_1,x_1)^2,\\
    \L_{(3)}^{c-ct-2}&\subset \frac{1}{12} \int d^dx_1\,d^dx_2\, V_{(3)}(x_1) \,G(x_1,x_2)^3\, V_{(3)}^{(ct-1)}(x_2),\nn\\
    \L_{(3)}^{d-ct-2}&\subset -\frac{1}{24} \int d^dx_1\,d^dx_2\,d^dx_3\, V_{(3)}(x_1) \,G(x_1,x_2)^2\, V_{(3)}(x_2), \,G(x_2,x_3) \,V_{(2)}^{(ct-1)}(x_3)\,G(x_3,x_1).\nn
\end{align}
The additional diagrams consist of two one-loop or one two-loop counter term insertions in one-loop diagrams (see Fig.~\ref{fig:one_loop_two_ct}). Here, the crossed square insertion in Fig.~\ref{fig:one_loop_two_ct} represents the two-loop divergent terms. The individual contributions can be written as  

\begin{align}\label{eq:two_loop_ct_one_loop_divergence}
    \L_{(3)}^{a-ct-1}&\subset -\frac{1}{2} \int d^dx_1\, G(x_1,x_1)\,V_{(2)}^{(ct-2)}(x_1),\nn\\
    \L_{(3)}^{b-ct-1}&\subset\frac{1}{4} \int d^dx_1\, d^dx_2\, \,V_{(2)}^{(ct-1)}(x_1)G(x_1,x_2)^2\,V_{(2)}^{(ct-1)}(x_2).
\end{align}
where the two-loop counter term vertex factor of 2 fluctuations is derived  from Eq.~\eqref{eq:ct_vertex} and Eq.~\eqref{eq:final_2loop} as,
\begin{align}
    V_{(2)}^{(ct-2)}(x) &= \frac{\partial^2\L_{(2)}}{\partial\phi^2}.
\end{align}
Thus, the total three-loop divergent contributions are given by the summation of the derived divergences from all the topologies, i.e., from Eqs.~\eqref{eq:three_loop_divergence}, \eqref{eq:one_loop_ct_two_loop_divergence}, and \eqref{eq:two_loop_ct_one_loop_divergence} as
\begin{align}
    \L_{(3)} =& \L_{(3)}^{a}+\L_{(3)}^{b}+\L_{(3)}^{c}+\L_{(3)}^{d}+\L_{(3)}^{e}+\L_{(3)}^{f}+\L_{(3)}^{g}+\L_{(3)}^{a-ct-2}+\L_{(3)}^{b-ct-2}+\L_{(3)}^{c-ct-2}\nn\\
    &+\L_{(3)}^{d-ct-2}+\L_{(3)}^{a-ct-1}+\L_{(3)}^{b-ct-1}.
\end{align}
Please note that for three-loop computation, we may require the $\epsilon$ and $\epsilon^2$ proportional terms from one and two-loop counter term Lagrangian (which are ignored for their respective cases) to ensure no divergent contributions are missed out. 
\section{Conclusions}

Effective Field Theory computation in the top-down approach, where heavy fields are integrated out from a UV  Lagrangian, is primarily a full theory calculation in an effective way. Thus, it is important to do this precisely, i.e. performing integration out beyond one-loop, if possible. In addition to that, while including the quantum corrections, we can renormalize the effective theory order by order of higher dimensional operator basis. In this method, the effective operators emerge at the matching scale whereas the observable are measured at a much lower scale. Thus, it is important to run down those operators to compute the theoretical predictions for those observable. These running equations are computed by identifying the pole divergences associated with the vacuum diagrams. 

In this paper, we show how a Heat-Kernel-based method can be used to compute these running equations without computing any momentum integral associated with Feynman diagrams or so. Here, we define the interacting Green’s function using HKCs that are finite. Then we construct the distinct vacuum diagrams which consist of these Green’s functions and different vertex factors. We identify the singular behavior in the form of poles employing algebraic identities. We, first, design the set-up for scalar quantum field theory and compute the counter terms for one and two-loops in terms of HKCs assuming polynomial composite operators only. Then, we explicitly compute the counter terms considering a scalar field model adding $\phi^6$, and $\phi^8$ effective operators with the renormalizable Lagrangian. We, then, extend this proposal for effective operators that contain derivatives as well. We highlight the shortcomings of the previous method and discuss how the HK can be used in this case as well by going beyond the minimal scenario. Our method is not restricted to scalar cases only. We can define Green's function for fermion fields also using the notion of bosonization suitably.  We consider a Lagrangian containing scalar and fermion fields and their interactions, encapsulating a few dimension six effective operators. First, we identify the independent vacuum diagrams where apart from pure scalar and fermion Green's functions, we have diagrams that also consist of mixed ones. For this case, we discuss how one can extract the divergent terms up to two-loop order. We also note down the terms that carry signatures of infrared divergences in the limit of scalar and fermion masses to be zero. Then, we discuss how we can generalize this method for any arbitrary loop order, and that too with a discussion on three-loop vacuum diagrams. Our method can be used for any theory where the Lagrangian can be expressed in terms of a second-order elliptic operator. We provide the counter terms as a function of HKCs and the extracted singular structures are independent of the specifics of the models and can be used for any model containing scalar and fermion fields. Here, we can avoid drawing a large number of  Feynman diagrams and computation of divergences through the complicated momentum integrals. The advantage of this method is that one can directly find out the counter terms for the composite operators instead of looking for wave-function and coupling renormalization. In the future, we want to employ this method in the case of SMEFT and also note down the finite part contributions from the higher-order loop as that will help to compute the Wilson coefficients more precisely.

\section*{Acknowledgements}
We thank Diptarka Das, and Nilay Kundu for their useful discussions. We also acknowledge helpful comments from Sabyasachi Chakraborty and Apratim Kaviraj on the manuscript.  
J.C. and U.B. acknowledge the hospitality of HRI, Allahabad, India where part of the research is done.
The work of K.R. is supported by DORD, Indian Institute of Technology Kanpur.
This work is supported by the Core Research Grant (CRG/2023/003200), SERB, India. 

\appendix
\section{Computing the $\beta$-Functions}
\label{app:beta-functions}
We start with the bare Lagrangian including the effective operators $\phi^6$, $\phi^8$ interactions,
\begin{equation}\label{eq:bare-Lag}
    \L = \frac{1}{2}\,(D_\mu\phi_0)\,(D_\mu\phi_0) + \frac{1}{2}M_0^2\,\phi_0^2 + \frac{\lambda_0}{4!}\,\phi_0^4 + \frac{c_{6,0}}{6!}\,\phi_0^6 + \frac{c_{8,0}}{8!}\,\phi_0^8\,.
\end{equation}
In renormalized perturbation theory, the Lagrangian is expressed in terms of the renormalized fields and couplings as
\begin{equation}\label{eq:renorm-Lag}
    \L = \frac{1}{2}\,Z_{\phi}(D_\mu\phi)\,(D_\mu\phi) + \frac{1}{2}\,Z_M M^2\,\phi^2 + \frac{\lambda}{4!}Z_\lambda\,\mu^{\epsilon}\phi^4 + \frac{c_{6}}{6!}\,Z_{c_6}\mu^{2\epsilon}\phi^6 + \frac{c_{8}}{8!}\,Z_{c_8}\mu^{3\epsilon}\phi^8\,.
\end{equation}
Inspecting Eqs.~\eqref{eq:bare-Lag}, and ~\eqref{eq:renorm-Lag}, we find the following relations between the bare and renormalized parameters,
\begin{gather}
    \phi_0 = Z_\phi^{1/2}\phi,\qquad
    M_0 = Z_M^{1/2}\,Z_\phi^{-1/2}M,\qquad
    \lambda_0 = Z_\lambda Z_\phi^{-2} \lambda \mu^{\epsilon},\nn\\
    c_{6,0} = Z_{c_6} Z_\phi^{-3} c_{6}\,\mu^{2\epsilon},\qquad
    c_{8,0} = Z_{c_8} Z_\phi^{-4} c_{8}\,\mu^{3\epsilon}\,,
\end{gather}
where renormalization constants $Z_\phi$, $Z_M$, $Z_\lambda$, $Z_{c_6}$, and $Z_{c_8}$ at one-loop can take the following form when compared with the counter terms given in Eq.~\eqref{eq:one_loop_divergence},
\begin{gather}
    Z_\phi = 1,\qquad Z_M = 1+\cfrac{\alpha\,\lambda}{\epsilon},\qquad Z_\lambda =1+\cfrac{\alpha}{\epsilon}\Bigg[3\lambda+\cfrac{M^2c_6}{\lambda}\Bigg],\nn\\
    Z_{c_6} = 1+ \cfrac{\alpha}{\epsilon}\Bigg[15\lambda+\cfrac{c_8M^2}{c_6}\Bigg],\qquad Z_{c_8}=1+\cfrac{\alpha}{\epsilon}\Bigg[28\lambda+\cfrac{35c_6^2}{c_8}\Bigg]\,.
\end{gather}
Since bare terms are scale ($\mu$) independent, we can write  $\mu\cfrac{dC_i}{d\mu}=0$ ($C_i=M_0,\lambda_0,c_{6,0},c_{8,0}$), and that  leads to
\begin{gather}\label{eq:betaeqn}
    \beta_\lambda \Bigg[\cfrac{M^2\alpha}{\epsilon}\Bigg]+\beta_{M^2}\Bigg[1+\cfrac{\alpha\lambda}{\epsilon}\Bigg] =0\,,\nn\\
    \beta_{\lambda} \Bigg[\cfrac{1}{\lambda}+\cfrac{6\,\alpha}{\epsilon}\Bigg] + \beta_{M^2} \Bigg[\cfrac{\alpha}{\epsilon}\cfrac{c_6}{\lambda}\Bigg] +\beta_{c_6} \Bigg[\cfrac{\alpha}{\epsilon}\cfrac{M^2}{\lambda}\Bigg] +\alpha\Bigg[3\lambda+\cfrac{M^2c_6}{\lambda}\Bigg] +\epsilon= 0\,,\nn\\
    \beta_\lambda\Bigg[\cfrac{15\alpha}{\epsilon}\Bigg] +\beta_{M^2} \Bigg[\cfrac{\alpha}{\epsilon}\cfrac{c_8}{c_6}\Bigg] +\beta_{c_6}\Bigg[\cfrac{1}{c_6}+\cfrac{\alpha}{\epsilon}\cfrac{15\lambda}{c_6}\Bigg] + \beta_{c_8}\Bigg[\cfrac{\alpha}{\epsilon}\cfrac{M^2}{c_6}\Bigg]+\alpha\Bigg[30\lambda+\cfrac{2c_8M^2}{c_6}\Bigg]+2\epsilon =0\,,\nn\\
    \beta_\lambda \Bigg[\cfrac{28\alpha}{\epsilon}\Bigg]+\beta_{c_6}\Bigg[\cfrac{\alpha}{\epsilon}\cfrac{70c_6}{c_8}\Bigg]+\beta_{c_8}\Bigg[\cfrac{1}{c_8}+\cfrac{\alpha}{\epsilon}\cfrac{28\lambda}{c_8}\Bigg]+\alpha\Bigg[84\lambda+\cfrac{c_6^2}{c_8}\Bigg]+3\epsilon=0\,.
    \label{eq:betaeqn}
\end{gather}
Solving the set of equations simultaneously,  depicted in Eq.~\eqref{eq:betaeqn}, we note down the $\beta$-functions for respective couplings of composite operators as
\begin{eqnarray}\label{eq:app-rges}
    \beta_{M^2} &=& \alpha M^2\lambda\,,\nn\\
    \beta_\lambda &=& \alpha (3\lambda^2+M^2c_6)\,,\nn\\
    \beta_{c_6} &=& \alpha (15\,c_6\lambda+M^2c_8)\,,\nn\\
    \beta_{c_8} &=& \alpha (35\,c_6^2+28\,c_8\lambda)\,.
\end{eqnarray}

\section{Symmetry Factor of Vacuum Topologies}\label{app:symmetry_factor}

In this section, we illustrate how one can find the distinct 1PI topologies and the symmetry factor, associated with each such diagram, at any given loop order. 

We have noted from Eqs.~\eqref{eq:loop_counting}, and ~\eqref{eq:loop_constrain} that at a given loop order $L$, the number of available vertices is given by $ N\leq 2(L-1)$. For example,  a three-loop vacuum diagram can contain a maximum of four vertices. The type of vertex is determined by  the total number of allowed fluctuations, and how we can distribute them among the different vertices satisfying Eq.~\eqref{eq:loop_constrain}. 
\begin{figure}[h]
    \centering
    \includegraphics[width=\textwidth]{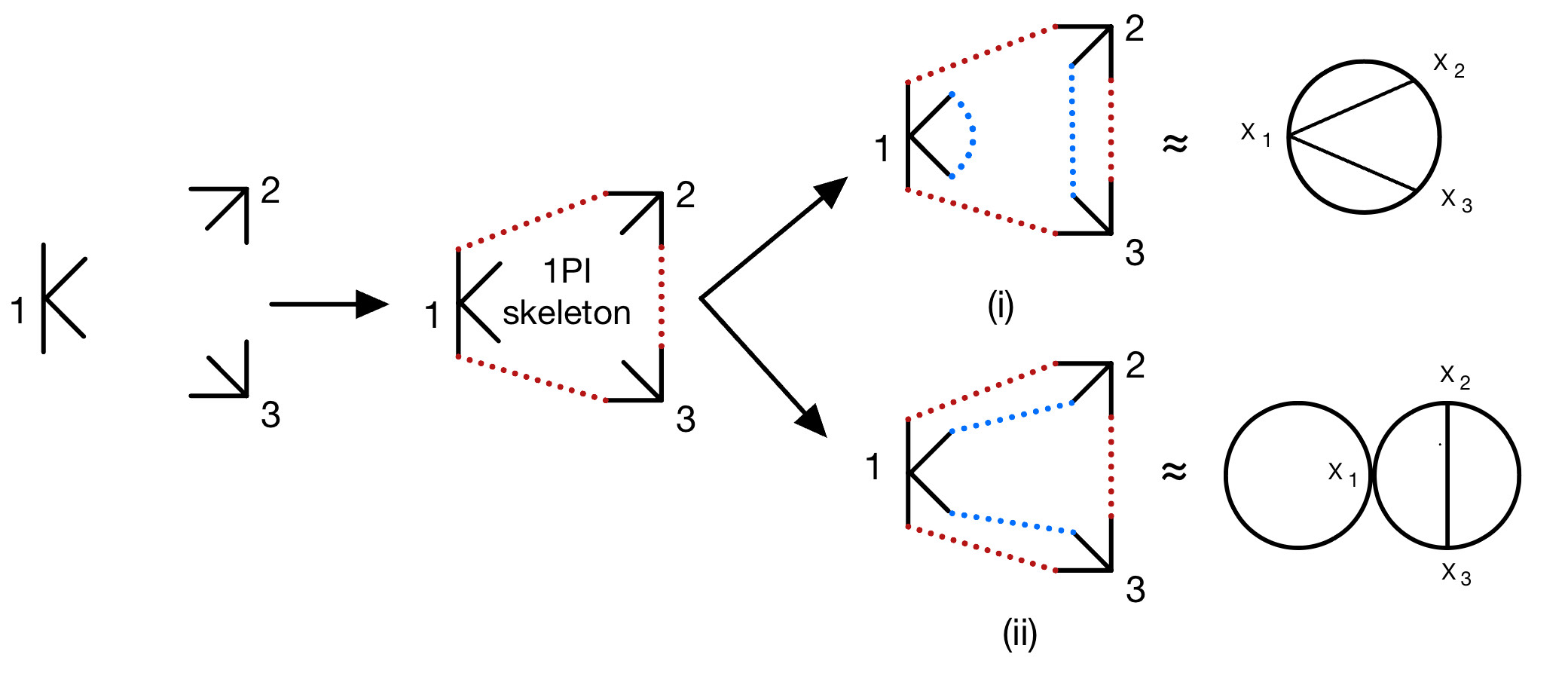}
    \caption{1PI Three-loop topologies with three vertices.}
    \label{fig:1PI_topology_n3}
\end{figure}
\begin{figure}[h]
    \centering
    \includegraphics[width=\textwidth]{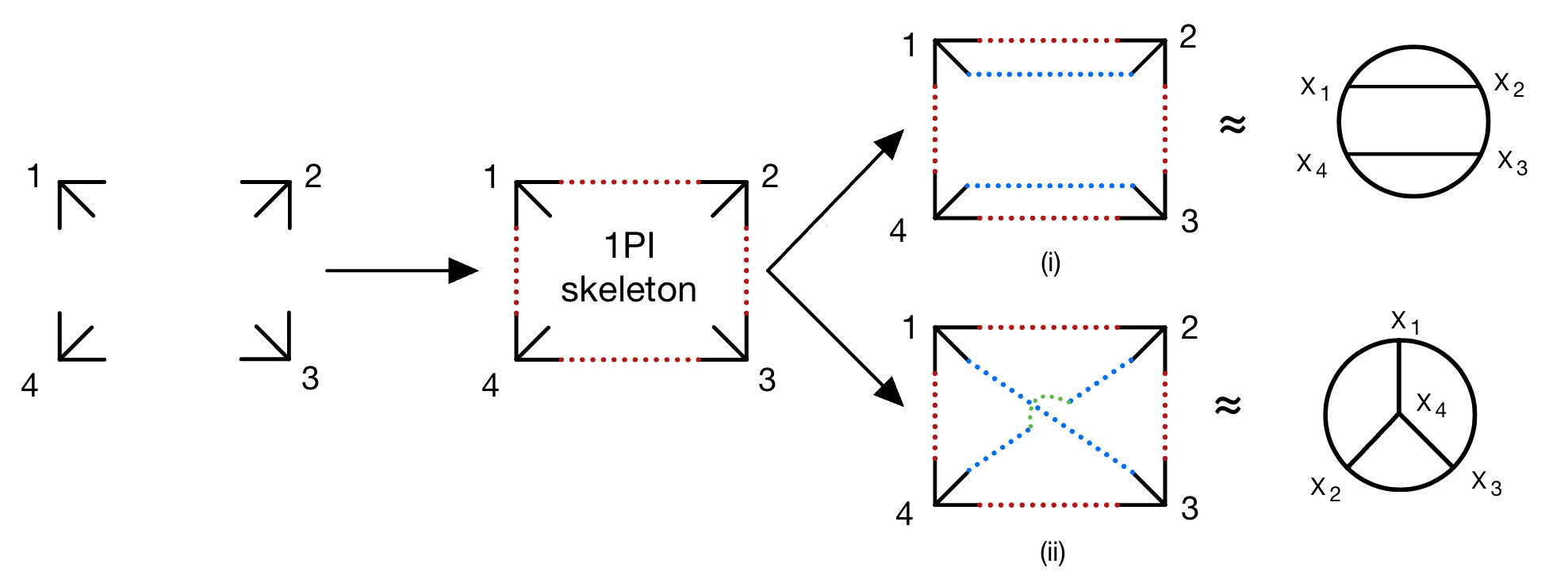}
    \caption{1PI Three-loop topologies with four vertices.}
    \label{fig:1PI_topology_n4}
\end{figure}
To elaborate this point, we consider the case of (i) $L=3,\quad N=3$, this gives $\sum m_i = 10$ that can be divided among three vertices as $4+3+3$, see Fig.~\ref{fig:1PI_topology_n3}, and for (ii) $L=3,\quad N=4$, we have $\sum m_i = 12$ that can be divided among four vertices as $3+3+3+3$, see Fig.~\ref{fig:1PI_topology_n4}. Now, to construct the topologies, as a first step, one must focus on 1PI topologies only. Hence, each of the vertices is connected to two other vertices and  forms a closed loop. Once the 1PI skeleton is obtained, the remaining free  legs, coming out of the vertices, are contracted in distinct possible ways.
Now, we are ready to compute the symmetry factor ($S_f$) for a given topology. For illustration, we consider the following example topology, see Fig.~\ref{fig:1PI_topology_n4}(ii), and with the help of Fig.~\ref{fig:symmetry_factor} we demonstrate our strategy:
\begin{figure}[h]
    \centering
    \includegraphics[width=0.4\textwidth]{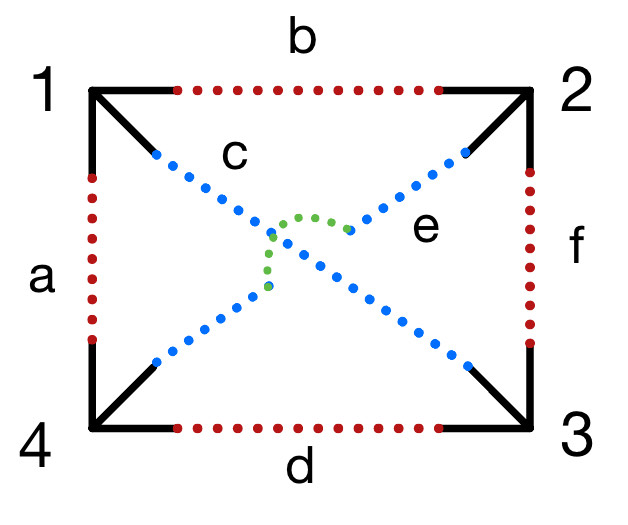}
    \caption{Illustrating symmetry factor computation.}
    \label{fig:symmetry_factor}
\end{figure}
\begin{itemize}
    \item To start with the very first propagator contraction, vertex 1 can be made to connect to any of the other vertices (2, 3, or 4), and there are nine possible ways to make the contractions.
    
    \item Once the first contraction is made, let us call it `$a$', the second contraction (`$b$') can be performed to the other two  vertices, apart from 4, i.e., 2 or 3. We can do that in six ways.
    
    \item Similarly, the third contraction (`$c$') from vertex 1 is made to the remaining vertex 3, and there are three ways to achieve this.
    
    \item Now all the free legs attached with vertex 1 are contracted. We will repeat the similar contraction procedure with all other vertices systematically.
  
    \item Completing the task we have counting for different contractions.  At this point, we need to focus on the identical ones and note down the over-counting factor. 
    
    \item In the case of a real scalar field, all the vertices and associated fluctuations are identical.  Hence, we divide the symmetry factor by $4!$,  and $3!$ to take care of the effect of identical  vertices, and  fluctuations at each of the vertices respectively.
    
\end{itemize}
Hence the symmetry factor for the considered topology, see  Fig.~\ref{fig:1PI_topology_n4}(ii),  is given by
\begin{equation*}
    S_f = \frac{9\times6\times3\times4\times2}{4!\times3!\times3!\times3!\times3!}=\frac{1}{4!}.
\end{equation*}

\section{Obtaining the Generalized Heat-Kernel Coefficients}\label{app:gHKC}
Here we will consider the case of obtaining the g-HKC up to $\tilde b_2$. We start from Eq.~\eqref{eq:HK} and expand the summation over $f_n(t,\mathcal A)$ up to $n=4$, which is enough to get the g-HKC up to $\tilde b_2$, to obtain \cite{Banerjee:2023xak}
\begin{equation}
    K(t,x,y,\Delta) = \int \frac{d^d p}{(4\pi^2t)^{d/2}}e^{\frac{(x-y)^2}{4t}}e^{-M^2 t} e^{p^2}\left[1-f_1(t,\mathcal A)+f_2(t,\mathcal A)-f_3(t,\mathcal A)+f_4(t,\mathcal A)\right],
\end{equation}
where $f_n(t,\mathcal A)$ is given in Eq.~\eqref{eq:fn}. For a theory with $m$ particles, the mass term $M$ and $\mathcal A$ are $m\cross m$ matrices. Performing the Gaussian momentum integral in the above expression, we obtain an expression that is a polynomial in $t$. Comparing the coefficients of different orders of $t$ with \cite{Banerjee:2023xak}
\begin{equation}\label{eq:HK_exp_expanded}
    K(t,x,y,\Delta)=\frac{1}{(4\pi t)^{d/2}} e^{\frac{(x-y)^2}{4t}}e^{-M^2 t} \left(\tilde b_0(x,y)-t\,\tilde b_1(x,y)+\frac{t^2}{2}\tilde b_2(x,y)+\mathcal{O}(t^3)\right),
\end{equation}
the g-HKCs can be identified. As we have truncated the expansion up to $f_4(t,\mathcal A)$, the coefficients of the higher powers of $t$ ($>t^3$) in the above equation do not contain the complete information of the higher g-HKCs.

For the case of derivatively coupled interactions, discussed in Sec.~\ref{sec:derivative_interaction}, employing $D_\mu\rightarrow D_\mu + i\,p_\mu $ in Eq.~\eqref{eq:non_minimal_oneloop},  and removing the Gaussian in momentum,  the $\mathcal A$ matrix takes the following form
\begin{equation}
    \mathcal A= (1+a)D^2 + h_\mu D_\mu - a\,p^2 + \frac{2}{\sqrt t}(1+a)p.D + \frac{1}{\sqrt t}b.p +\tilde U.
\end{equation}
For the example in Sec.~\ref{sec:derivative_interaction}, $a$ is of the order $\mathcal O(1/\Lambda^2)$. Hence truncating at $\mathcal O (1/\Lambda^2)$, to obtain g-HKC up to $\tilde b_2$, one has to consider the series expansion till
$f_5(t,\mathcal A)$.

\section{Algebraic Singularities in Green's Function Polynomial}\label{app:formula}
In evaluating the loop integral, we encounter products of propagators in various combinations. The following relations are useful in evaluating the loop integrals.
\subsection{Singularity at coincidence point}\label{app:formula_coincidence}
To find the pole structures in the coincidence limit, we analytically continue from $d<2$ dimensions and set $z=0$ in Eq.~\eqref{eq:propagator_exp}. This gives,
\begin{align}\label{eq:coincidence_pole}
    g_0(x,x)&= \alpha \Gamma[1-d/2]M^{d-2},\nn\\
    g_1(x,x)&= -\alpha \Gamma[2-d/2]M^{d-4}.
\end{align}

\subsection{Singularity at non-coincidence point}\label{subsec:singular-noncoincident}
In the short distance limit $z\rightarrow 0$, for the non-local propagator, the $1/z^{2a}$ for $a\geq2$ can be series expanded to obtain $1/\epsilon$ pole from its Fourier transform as \cite{gel2016generalized,Jack:1982hf},
\begin{align}
    \int d^d z\, \frac{1}{|z|^{2a}}e^{ik\cdot z} &= \pi^{d/2}\frac{\Gamma[d/2-a]}{\Gamma[a]}\left[\left(\frac{-k^2}{4^{a-d/2}}\right)^{a-d/2}\right]\nn\\
    &= \frac{\pi^{d/2}}{4^{a-d/2}}\frac{\Gamma[d/2-a]}{\Gamma[a]}\int d^d z\,e^{ik\cdot z}(D^2)^n\,\int d^d k'\,(-k'^2)^\xi e^{-ik'\cdot z},
\end{align}
where $a-d/2=n+\xi$ where $n$ is a non-negative integer and $\xi\rightarrow 0$. For instance if $a=3+\epsilon$ and $d=4-\epsilon$ with $\epsilon\rightarrow 0$, then $n=1$ and $\xi = 3\epsilon/2$. The Gamma function in the numerator of the above expression has simple poles ($1/\xi$) for $a\geq2$ and the integral over $k'$ in the above equation is $\delta^d(z)$ in the leading order. Hence we get \cite{gel2016generalized},
\begin{equation}\label{eq:pole_expansion}
    \frac{1}{z^{2a}} = \frac{\pi^{d/2}}{4^{a-d/2}}\frac{\Gamma[d/2-a]}{\Gamma[a]}(D^2)^n\,\delta^d(z) + \mathcal{O}(\xi^0).
\end{equation}
For the case of the fermion propagator, one also has derivatives acting on the $g_i$'s, hence the following expression will be useful in obtaining the $\epsilon$ pole structures for the fermion propagator \cite{gel2016generalized},
\begin{align}
    \frac{z_\mu}{z^{2a+2}} &= \frac{-1}{2a}\frac{\pi^{d/2}}{4^{a-d/2}}\frac{\Gamma[d/2-a]}{\Gamma[a]}D_\mu(D^2)^n\,\delta^d(z) + \mathcal{O}(\xi^0),\nn\\
    \frac{z_\mu z_\nu}{z^{2a+2}} &= \frac{\pi^{d/2}}{4^{a-d/2}}\frac{\Gamma[d/2-a]}{\Gamma[a+1]}\left[\frac{\delta_{\mu\nu}D^2}{2}+(d/2-a)D_\mu D_\nu\right]\,(D^2)^{n-1}\,\delta^d(z) + \mathcal{O}(\xi^0).
\end{align}
For computation of the two-loop results, the above expansion up to $\mathcal{O}(1/\xi)$ is sufficient. Using these expansions we obtain the pole structure for different combinations of $g_i(x,y)$'s as follows \cite{gel2016generalized} 
\begin{align}
    &g_0(x,y)^2 = \alpha\frac{2^{\epsilon } \pi ^{\epsilon /2} \Gamma \left[1-\frac{\epsilon }{2}\right]^2 \Gamma \left[\frac{\epsilon }{2}\right]}{\Gamma [2-\epsilon ]} \delta(z),\label{eq:g0_2}\\
    &g_0(x,y)^2 g_1(x,y) = -\alpha^2(4 \pi )^{\epsilon } \Gamma \left[1-\frac{\epsilon }{2}\right]^2 \left( \frac{\Gamma \left[-\frac{\epsilon }{2}\right] \Gamma [\epsilon]}{\Gamma \left[2-\frac{3 \epsilon }{2}\right] }+M^{-\epsilon }\frac{\Gamma \left[\frac{\epsilon }{2}\right]^2}{ \Gamma [2-\epsilon ]}\right)\delta (z),\label{eq:g0_2_g1_1}\\
    &g_0(x,y)^3 = \alpha^2(4 \pi )^{\epsilon }  \Gamma \left[1-\frac{\epsilon }{2}\right]^2 \bigg(\frac{ \Gamma \left[1-\frac{\epsilon }{2}\right] \Gamma [\epsilon -1]}{ \Gamma \left[3-\frac{3 \epsilon }{2}\right] }D^2-3M^2  \frac{\Gamma \left[-\frac{\epsilon }{2}\right] \Gamma [\epsilon ]}{\Gamma \left[2-\frac{3 \epsilon }{2}\right]}\nn\\
    &\hspace{3cm}+3M^{2-\epsilon} \frac{ \Gamma \left[\frac{\epsilon }{2}\right] \Gamma \left[\frac{\epsilon }{2}-1\right]}{ \Gamma [2-\epsilon ]}\bigg)\delta(z),\label{eq:g0_3}\\
    &g_0(x,y)(D_\mu\, g_0(x,y)) = \alpha\frac{2^{\epsilon } \pi ^{\epsilon /2} \Gamma \left[1-\frac{\epsilon }{2}\right]^2 \Gamma \left[\frac{\epsilon }{2}\right]}{2\,\Gamma [2-\epsilon ]}D_\mu \delta(z),\\
    &g_0(x,y)^2 (D_\mu\, g_0(x,y)) = \frac{\alpha^2}{3}(4 \pi )^{\epsilon }  \Gamma \left[1-\frac{\epsilon }{2}\right]^2 \bigg(\frac{ \Gamma \left[1-\frac{\epsilon }{2}\right] \Gamma [\epsilon -1]}{ \Gamma \left[3-\frac{3 \epsilon }{2}\right] }D^2-3M^2  \frac{\Gamma \left[-\frac{\epsilon }{2}\right] \Gamma [\epsilon ]}{\Gamma \left[2-\frac{3 \epsilon }{2}\right]}\nn\\
    &\hspace{3cm}+3M^{2-\epsilon} \frac{ \Gamma \left[\frac{\epsilon }{2}\right] \Gamma \left[\frac{\epsilon }{2}-1\right]}{ \Gamma [2-\epsilon ]}\bigg)D_\mu\delta(z),\\
    &g_0(x,y)(D_\mu\, g_0(x,y))g_1(x,y)= -\alpha^2 (4 \pi )^{\epsilon } \Gamma \left[1-\frac{\epsilon }{2}\right] \Gamma \left[2-\frac{\epsilon }{2}\right] \bigg(M^{-\epsilon }\frac{ \Gamma \left[\frac{\epsilon }{2}\right]^2}{\Gamma [3-\epsilon ]}\nn\\
    &\hspace{3cm}+\frac{\Gamma \left[-\frac{\epsilon }{2}\right] \Gamma [\epsilon ]}{\Gamma \left[3-\frac{3 \epsilon }{2}\right]}\bigg)D_\mu\delta(z),\\
    &g_0(x,y)^2(D_\mu\, g_1(x,y)) = -\alpha^2(4 \pi )^{\epsilon } \frac{\Gamma \left[1-\frac{\epsilon }{2}\right]^3 \Gamma [\epsilon]}{\Gamma \left[3-\frac{3 \epsilon }{2}\right]}D_\mu\delta(z),\\
    &g_0(x,y)(D_\mu\, g_0(x,y))(D_\nu\, g_0(x,y))=-(4 \pi )^{\epsilon }\alpha^2\bigg[M^{4-\epsilon } \Gamma \left[2-\frac{\epsilon }{2}\right] \Gamma \left[1-\frac{\epsilon }{2}\right] \Gamma \left[\frac{\epsilon }{2}\right] \nn\\
    &\hspace{3cm} \left(\frac{\Gamma \left[2-\frac{\epsilon }{2}\right] \Gamma \left[\frac{\epsilon }{2}-2\right]}{2 \Gamma [3-\epsilon] \Gamma \left[1-\frac{\epsilon }{2}\right]}-\frac{\Gamma \left[\frac{\epsilon }{2}-2\right]}{\Gamma [3-\epsilon]}+\frac{\Gamma \left[\frac{\epsilon }{2}-1\right]}{\Gamma [3-\epsilon]}\right) \delta _{\mu \nu }\nn\\
    &\hspace{1cm}+M^4\,\frac{\Gamma \left[2-\frac{\epsilon }{2}\right] \Gamma \left[1-\frac{\epsilon }{2}\right] \Gamma [\epsilon]}{2 \Gamma \left[3-\frac{3 \epsilon }{2}\right]} \left(-\frac{\Gamma \left[1-\frac{\epsilon }{2}\right]^2}{\Gamma \left[2-\frac{\epsilon }{2}\right]}-3 \Gamma \left[-\frac{\epsilon }{2}\right]-\frac{\Gamma \left[-\frac{\epsilon }{2}-1\right] \Gamma \left[2-\frac{\epsilon }{2}\right]}{2 \Gamma \left[1-\frac{\epsilon }{2}\right]}\right) \delta _{\mu \nu }\nn\\
    &\hspace{1cm}-M^2 \Gamma \left[2-\frac{\epsilon }{2}\right]^2 \bigg(\frac{D^2 \Gamma \left[1-\frac{\epsilon }{2}\right] \Gamma [\epsilon -1]}{M^2 \Gamma \left[5-\frac{3 \epsilon }{2}\right]}+\frac{M^{-\epsilon } \Gamma \left[\frac{\epsilon }{2}\right] \Gamma \left[\frac{\epsilon }{2}-1\right]}{\Gamma [4-\epsilon ]}\nn\\
    &\hspace{3cm}-\frac{\Gamma [\epsilon] \Gamma \left[-\frac{\epsilon }{2}\right]}{\Gamma \left[4-\frac{3 \epsilon }{2}\right]}-\frac{2 \Gamma \left[1-\frac{\epsilon }{2}\right]^2 \Gamma [\epsilon]}{\Gamma \left[2-\frac{\epsilon }{2}\right] \Gamma \left[4-\frac{3 \epsilon }{2}\right]}\bigg)D_{\mu } D_{\nu } \nn\\
    &\hspace{1cm}-\frac{M^2}{2}\Gamma \left[2-\frac{\epsilon }{2}\right]^2  \bigg(\frac{D^2 \Gamma \left[1-\frac{\epsilon }{2}\right] \Gamma [\epsilon -2]}{M^2 \Gamma \left[5-\frac{3 \epsilon }{2}\right]}+\frac{M^{-\epsilon } \Gamma \left[\frac{\epsilon }{2}-1\right]^2}{\Gamma [4-\epsilon ]}\nn\\
    &\hspace{3cm}-\frac{2 \Gamma \left[1-\frac{\epsilon }{2}\right]^2 \Gamma [\epsilon -1]}{\Gamma \left[4-\frac{3 \epsilon }{2}\right] \Gamma \left[2-\frac{\epsilon }{2}\right]}-\frac{\Gamma [\epsilon -1] \Gamma \left[-\frac{\epsilon }{2}\right]}{\Gamma \left[4-\frac{3 \epsilon }{2}\right]}\bigg)\delta _{\mu \nu }D^2\bigg]\delta(z),\\
    &g_1(x,y)(D_\mu\,g_0(x,y))(D_\nu\,g_0(x,y)) = \alpha^2 (4 \pi )^{\epsilon } \Gamma \left[2-\frac{\epsilon }{2}\right]^2\bigg[\nn\\
    &\hspace{2cm}-\left(D^2 \left[\frac{M^{-\epsilon } \Gamma \left[\frac{\epsilon }{2}\right] \Gamma \left[\frac{\epsilon }{2}-1\right]}{2 \Gamma (4-\epsilon )}+\frac{\Gamma \left[-\frac{\epsilon }{2}\right] \Gamma (\epsilon -1)}{2 \Gamma \left[4-\frac{3 \epsilon }{2}\right]}\right] \delta _{\mu\nu }\right)\nn\\
    &\hspace{2cm}+M^2 \bigg(M^{-\epsilon }\frac{ \left[\Gamma \left[1-\frac{\epsilon }{2}\right] \Gamma \left[\frac{\epsilon }{2}\right]^2\right]}{\Gamma [3-\epsilon ] \Gamma \left[2-\frac{\epsilon }{2}\right]}+M^{-\epsilon }\frac{ \left[\Gamma \left[\frac{\epsilon }{2}-1\right] \Gamma \left[\frac{\epsilon }{2}\right]\right]}{2 \Gamma [3-\epsilon ]}\nn\\
    &\hspace{3cm}+\frac{\Gamma \left[-\frac{\epsilon }{2}\right] \Gamma \left[1-\frac{\epsilon }{2}\right] \Gamma [\epsilon ]}{\Gamma \left[3-\frac{3 \epsilon }{2}\right] \Gamma \left[2-\frac{\epsilon }{2}\right]}+\frac{\Gamma \left[-\frac{\epsilon }{2}-1\right] \Gamma [\epsilon ]}{2 \Gamma \left[3-\frac{3 \epsilon }{2}\right]}\bigg) \delta _{\mu\nu }\nn\\
    &\hspace{2cm}- \left(\frac{M^{-\epsilon } \Gamma \left[\frac{\epsilon }{2}\right]^2}{\Gamma (4-\epsilon )}+\frac{\Gamma \left[-\frac{\epsilon }{2}\right] \Gamma [\epsilon ]}{\Gamma \left[4-\frac{3 \epsilon }{2}\right]}\right)D_{\mu } D_{\nu }\bigg]\delta(z),\\
    &g_0(x,y)(D_\mu\,g_0(x,y))(D_\nu\,g_1(x,y)) = \alpha^2(4 \pi )^{\epsilon } \Gamma \left[1-\frac{\epsilon }{2}\right]^2 \Gamma \left[2-\frac{\epsilon }{2}\right]\bigg[\bigg(-\frac{D^2 \Gamma (\epsilon -1)}{2 M^2 \Gamma \left[4-\frac{3 \epsilon }{2}\right]}\nn\\
    &\hspace{3cm}-\frac{M^{-\epsilon } \Gamma \left[\frac{\epsilon }{2}-1\right] \Gamma \left[\frac{\epsilon }{2}\right]}{\Gamma \left[1-\frac{\epsilon }{2}\right] \Gamma [3-\epsilon ]}+\frac{\Gamma \left[-\frac{\epsilon }{2}\right] \Gamma [\epsilon ]}{\Gamma \left[3-\frac{3 \epsilon }{2}\right] \Gamma \left[1-\frac{\epsilon }{2}\right]}\nn\\
    &\hspace{3cm} +\frac{\Gamma \left[1-\frac{\epsilon }{2}\right] \Gamma [\epsilon ]}{2 \Gamma \left[3-\frac{3 \epsilon }{2}\right] \Gamma \left[2-\frac{\epsilon }{2}\right]}\bigg)M^2 \delta _{\mu\nu }-\frac{\Gamma [\epsilon ]}{\Gamma \left[4-\frac{3 \epsilon }{2}\right]}D_{\mu } D_{\nu }\bigg]\delta(z),\\
    &g_0(x,y)(D_\mu\,g_1(x,y))(D_\nu\,g_1(x,y)) =\alpha^2(4 \pi )^{\epsilon }\frac{\Gamma \left[1-\frac{\epsilon }{2}\right]^3 \Gamma [\epsilon ] }{2\ \Gamma \left[3-\frac{3 \epsilon }{2}\right]}\delta _{\mu\nu }\,\delta(z),\\
    &g_1(x,y)(D_\mu\,g_0(x,y))(D_\nu\,g_1(x,y)) =\alpha^2(4 \pi )^{\epsilon }\Gamma \left[1-\frac{\epsilon }{2}\right] \Gamma \left[2-\frac{\epsilon }{2}\right]\nn\\
    &\hspace{3cm}\bigg(\frac{M^{-\epsilon } \Gamma \left[\frac{\epsilon }{2}\right]^2}{2 \Gamma [3-\epsilon ]}+\frac{\Gamma \left[-\frac{\epsilon }{2}\right] \Gamma [\epsilon ]}{2 \Gamma \left[3-\frac{3 \epsilon }{2}\right]}\bigg)\delta_{\mu\nu}\,\delta(z),\\
    &g_0(x,y)(D_\mu\,g_2(x,y))(D_\nu\,g_0(x,y)) =\alpha^2(4 \pi )^{\epsilon }\Gamma \left[1-\frac{\epsilon }{2}\right] \Gamma \left[2-\frac{\epsilon }{2}\right]\nn\\
    &\hspace{3cm}\bigg(\frac{M^{-\epsilon } \Gamma \left[\frac{\epsilon }{2}\right]^2}{4 \Gamma [3-\epsilon ]}+\frac{\Gamma \left[-\frac{\epsilon }{2}\right] \Gamma [\epsilon ]}{4 \Gamma \left[3-\frac{3 \epsilon }{2}\right]}\bigg)\delta_{\mu\nu}\,\delta(z),\\
    &g_2(x,y)(D_\mu\,g_0(x,y))(D_\nu\,g_0(x,y)) =\alpha^2(4 \pi )^{\epsilon }\Gamma \left[2-\frac{\epsilon }{2}\right]^2\nn\\
     &\hspace{3cm}\bigg(\frac{\Gamma \left[-\frac{\epsilon }{2}-1\right] \Gamma [\epsilon ]}{4 \Gamma \left[3-\frac{3 \epsilon }{2}\right]}-\frac{M^{-\epsilon } \Gamma \left[\frac{\epsilon }{2}\right]^2}{4 \Gamma [3-\epsilon ]}\bigg)\delta_{\mu\nu}\,\delta(z).\label{eq:g0_end}
\end{align}

\subsection{Singularity of Gamma functions}
The following Taylor expansion of gamma functions are useful in the computation of the divergent terms of the loops.
\begin{align}
    &\Gamma\left[\pm \frac{\epsilon}{2}\right] = \pm \frac{2}{\epsilon}-\gamma \pm \frac{\epsilon}{24}(6\gamma^2+\pi^2),\\
    &\Gamma\left[\pm \frac{\epsilon}{2}-1\right] = \mp \frac{2}{\epsilon}+(\gamma-1) \pm \frac{\epsilon}{24}(-12+12\gamma-6\gamma^2-\pi^2),\\
    &\Gamma\left[ \pm \frac{\epsilon}{2}-2\right] = \pm \frac{1}{\epsilon}+\frac{1}{4}(3-2\gamma) \pm \frac{\epsilon}{48}(21-18\gamma+6\gamma^2\pi^2),\\
    &\Gamma\left[ \pm \frac{\epsilon}{2}-3\right] = \mp \frac{1}{3\epsilon}+\frac{1}{36}(6\gamma-11) \pm \frac{\epsilon}{432}(-85+66\gamma-18\gamma^2-3\pi^2).
\end{align}
\subsection{Singularity associated with non-planar diagrams at non-coincidence point}\label{app:benz}

As demonstrated in App.~\ref{subsec:singular-noncoincident}, the divergent contribution from structures such as $1/|x_1-x_2|^{2a}$, where $a\geq2$ can be obtained from the expansion given in Eq.~\eqref{eq:pole_expansion}. These are formed when a loop is formed between two distinct space-time by two or more propagators. The divergent contribution from loops formed at a single space-time point can also be calculated trivially by analytically continuing from $d<2$ dimensions as explained in App.~\ref{app:formula_coincidence}. In the case of three-loop Benz topology (Fig.~\ref{fig:three_loop} (g)), each of the loops is formed between three distinct space-time points. The space-time dependence of the propagator now has the form,\[\frac{1}{|x_1-x_2|^2|x_2-x_3|^2|x_3-x_1|^2},\] whose $\epsilon$ pole structure can not be extracted through the expansion in Eq.~\eqref{eq:pole_expansion} at $d=4-\epsilon$ dimensions. These kinds of non-planar structures, where a loop is formed between more than two space-time points start appearing at three and higher loop order and the evaluation of the $\epsilon$ pole structure has to be dealt with separately.

\section{Two-loop Counter Terms: Interacting fermion-scalar fields}\label{app:fermion_results}
Here we provide the functional dependence of the two-loop counter term Wilson coefficients on the different coupling constants for the Lagrangian considered in Sec. \ref{sec:fermion_expample}.

Contributions from pure two-loop topologies are given by,
\begin{align}
    \L_{2}^a &\subset \frac{1}{2}\int d^dx_1\, V_{(4)}^\dagger(x_1)\Big\lvert_{\phi^\dagger\phi\phi^\dagger\phi} G^s(x_1,x_1) (G^s(x_1,x_1))^\dagger\nn\\
    &=\alpha^2 \bigg[\left(\frac{1}{\epsilon}\left\{2 M_s^2\lambda^2 +2c_6 M_s^4\right\}+\frac{1}{\epsilon^2}\left\{4 M_s^2\lambda^2 +2 c_6 M_s^4\right\}-\frac{\log M_s^2}{\epsilon}\left\{4 M_s^2\lambda^2 +2 c_6 M_s^4\right\}\right)\phi^\dagger\phi\nn\\
    &\quad\quad+ \left(\frac{5}{2\epsilon} M_s^2\lambda c_6 +\frac{1}{\epsilon^2}\left\{5 M_s^2\lambda c_6 +2 \lambda^3\right\}-\frac{\log M_s^2}{\epsilon}\left\{5 M_s^2\lambda c_6 +2 \lambda^3\right\}\right)(\phi^\dagger\phi)^2\nn\\
    &\quad\quad + \left(\frac{3}{\epsilon^2}\lambda^2 c_6-\frac{3\log M_s^2}{\epsilon}\lambda^2 c_6\right)(\phi^\dagger\phi)^3 + \bigg(\frac{1}{\epsilon} M_s^2 \lambda y^{(3)}_{ij} +\frac{2}{\epsilon^2} M_s^2 \lambda y^{(3)}_{ij}\nn\\
    &\quad\quad -\frac{2\log M_s^2}{\epsilon} M_s^2 \lambda y^{(3)}_{ij}\bigg)\phi\overline\psi_i\psi_j
    + \left(\frac{1}{\epsilon} M_s^2 \lambda y^{(3)}_{ij} +\frac{2}{\epsilon^2} M_s^2 \lambda {y^{(3)}_{ij}}^\dagger-\frac{2\log M_s^2}{\epsilon} M_s^2 \lambda {y^{(3)}_{ij}}^\dagger\right)\phi^\dagger\overline\psi_j\psi_i \nn\\
    &\quad\quad+\left(\frac{2\lambda^2y^{(3)}_{ij}}{\epsilon^2}-\frac{2\log M_s^2}{\epsilon}\lambda^2y^{(3)}_{ij}\right)(\phi^\dagger\phi)\phi\overline\psi_i\psi_j\nn\\
    &\quad\quad+\left(\frac{2\lambda^2{y^{(3)}_{ij}}^\dagger}{\epsilon^2}-\frac{2\log M_s^2}{\epsilon}\lambda^2{y^{(3)}_{ij}}^\dagger\right)(\phi^\dagger\phi)\phi^\dagger\overline\psi_j\psi_i\bigg],
\end{align}
\begin{align}
    \L_{2}^b &\subset -\int d^dx_1\, V_{(4)}^\dagger(x_1)\Big\lvert_{\phi^\dagger\phi\overline\psi_i\psi_j} G^s(x_1,x_1)G^f_{ij}(x_1,x_1)\nn\\
    &=\alpha^2\bigg[\bigg(\frac{1}{\epsilon}\left\{8M_f^2M_s^2{y^{(3)}_{ij}}^\dagger y^{(1)}_{ij}\right\}+\frac{1}{\epsilon^2}\left\{12M_f^2M_s^2y^{(1)}_{ik}{y^{(3)}_{ik}}^\dagger\right\}\nn\\
    &\quad\quad\quad\quad-\frac{\log M_s^2}{\epsilon}\left\{6\,M_f^2M_s^2y^{(1)}_{ik}{y^{(3)}_{ik}}^\dagger\right\}-\frac{\log M_f^2}{\epsilon}\left\{6\,M_f^2M_s^2y^{(1)}_{ik}{y^{(3)}_{ik}}^\dagger\right\}\bigg)\phi^\dagger\phi\nn\\
    &\quad\quad+\bigg(\frac{1}{\epsilon}\left\{2M_f^2\lambda{y^{(3)}_{ij}}^\dagger y^{(1)}_{ij}+2M_s^2{y^{(1)}_{il}}^\dagger{y^{(3)}_{ji}}^\dagger y^{(1)}_{jk}y^{(1)}_{kl}\right\}\nn\\
    &\quad\quad\quad\quad\quad+\frac{1}{\epsilon^2}\left\{12M_f^2\lambda y^{(1)}_{ik}{y^{(3)}_{ik}}^\dagger+2M_s^2{y^{(1)}_{il}}^\dagger{y^{(3)}_{ji}}^\dagger y^{(1)}_{jk}y^{(1)}_{kl}\right\}\nn\\
    &\quad\quad\quad\quad\quad-\frac{\log M_s^2}{\epsilon}\left\{6\,M_f^2\lambda y^{(1)}_{ik}{y^{(3)}_{ik}}^\dagger+2M_s^2{y^{(1)}_{il}}^\dagger{y^{(3)}_{ji}}^\dagger y^{(1)}_{jk}y^{(1)}_{kl}\right\}\nn\\
    &\quad\quad\quad\quad\quad-\frac{\log M_f^2}{\epsilon}\left\{6M_f^2\lambda y^{(1)}_{ik}{y^{(3)}_{ik}}^\dagger+2M_s^2{y^{(1)}_{il}}^\dagger{y^{(3)}_{ji}}^\dagger y^{(1)}_{jk}y^{(1)}_{kl}\right\}\bigg)(\phi^\dagger\phi)^2\nn\\
    &\quad\quad+\bigg(\frac{1}{\epsilon}\left\{4M_f M_s^2{y^{(3)}_{ji}}^\dagger y^{(1)}_{jk}y^{(1)}_{ki}\right\}+\frac{1}{\epsilon^2}\left\{8M_f M_s^2 {y^{(3)}_{ji}}^\dagger y^{(1)}_{jk}y^{(1)}_{ki}\right\}\nn\\
    &\quad\quad\quad\quad\quad-\frac{\log M_s^2}{\epsilon}\left\{4M_f M_s^2 {y^{(3)}_{ji}}^\dagger y^{(1)}_{jk}y^{(1)}_{ki}\right\}-\frac{\log M_f^2}{\epsilon}\left\{4M_f M_s^2 {y^{(3)}_{ji}}^\dagger y^{(1)}_{jk}y^{(1)}_{ki}\right\}\bigg)(\phi^\dagger\phi)\phi \nn\\
    &\quad\quad + \bigg(\frac{1}{\epsilon}\left\{2M_f M_s^2{y^{(1)}_{ji}}^\dagger {y^{(3)}_{kj}}^\dagger y^{(1)}_{ki}\right\}+\frac{1}{\epsilon^2}\left\{4M_f M_s^2 {y^{(1)}_{ji}}^\dagger {y^{(3)}_{kj}}^\dagger y^{(1)}_{ki}\right\}\nn\\
    &\quad\quad\quad\quad\quad-\frac{\log M_s^2}{\epsilon}\left\{2M_f M_s^2 {y^{(1)}_{ji}}^\dagger {y^{(3)}_{kj}}^\dagger y^{(1)}_{ki}\right\}-\frac{\log M_f^2}{\epsilon}\left\{2M_f M_s^2 {y^{(1)}_{ji}}^\dagger {y^{(3)}_{kj}}^\dagger y^{(1)}_{ki}\right\}\bigg)(\phi^\dagger\phi)\phi^\dagger\nn\\
    &\quad\quad +\bigg(\frac{1}{\epsilon^2}\left\{4 M_f \lambda {y^{(1)}_{il}}^\dagger{y^{(3)}_{ki}}^\dagger y^{(1)}_{kl}\right\}-\frac{\log M_s^2}{\epsilon}\left\{2 M_f \lambda {y^{(1)}_{il}}^\dagger{y^{(3)}_{ki}}^\dagger y^{(1)}_{kl}\right\}\nn\\
    &\quad\quad\quad\quad\quad-\frac{\log M_f^2}{\epsilon}\left\{2 M_f \lambda {y^{(1)}_{il}}^\dagger{y^{(3)}_{ki}}^\dagger y^{(1)}_{kl}\right\}\bigg)(\phi^\dagger\phi)^2\phi^\dagger+\bigg(\frac{1}{\epsilon^2}\left\{8 M_f \lambda {y^{(3)}_{il}}^\dagger{y^{(1)}_{ik}} y^{(1)}_{kl}\right\}\nn\\
    &\quad\quad\quad\quad\quad-\frac{\log M_s^2}{\epsilon}\left\{4 M_f \lambda {y^{(3)}_{il}}^\dagger{y^{(1)}_{ik}} y^{(1)}_{kl}\right\}-\frac{\log M_f^2}{\epsilon}\left\{4 M_f \lambda {y^{(3)}_{il}}^\dagger{y^{(1)}_{ik}} y^{(1)}_{kl}\right\}\bigg)(\phi^\dagger\phi)^2\phi\nn\\
    &\quad\quad+\bigg(\frac{1}{\epsilon^2}\left\{4\lambda{y^{(1)}_{il}}^\dagger{y^{(3)}_{ji}}^\dagger y^{(1)}_{jk}y^{(1)}_{kl}\right\}-\frac{\log M_s^2}{\epsilon}\left\{2\lambda{y^{(1)}_{il}}^\dagger{y^{(3)}_{ji}}^\dagger y^{(1)}_{jk}y^{(1)}_{kl}\right\}\nn\\
    &\quad\quad\quad\quad\quad-\frac{\log M_f^2}{\epsilon}\left\{2\lambda{y^{(1)}_{il}}^\dagger{y^{(3)}_{ji}}^\dagger y^{(1)}_{jk}y^{(1)}_{kl}\right\}\bigg)(\phi^\dagger\phi)^3\bigg],
\end{align}
\begin{align}
    \L_{2}^c &\subset -\frac{1}{2}\int d^dx_1\,d^dx_2\, V^\dagger_{(3)}(x_1)\Big\lvert_{\phi^\dagger\phi\phi} (G^s(x_1,x_2))^3\ V_{(3)}(x_2)\Big\lvert_{\phi\phi^\dagger\phi^\dagger},\nn\\
    &=\alpha^2\bigg[\left(\frac{1}{\epsilon^2}\left\{ 3M_s^2\lambda^2\right\}+\frac{1}{2 \epsilon}\left\{9M_s^2\lambda^2\right\}-\frac{\log M_s^2}{\epsilon}\left\{3\,M_s^2\lambda^2 \right\}\right)(\phi^\dagger \phi)\nn\\
    &\quad\quad+\left(\frac{1}{\epsilon^2}\left\{3\lambda^3+3M_s^2\lambda c_6\right\}+\frac{1}{2\epsilon}\left\{3\lambda^3+9M_s^2\lambda c_6\right\}-\frac{\log M_s^2}{\epsilon}\left\{3\,\lambda^3+3\,M_s^2\lambda c_6\right\}\right)(\phi^\dagger \phi)^2\nn\\
    &\quad\quad+\left(\frac{1}{4\epsilon^2}\left\{15 \lambda^2 c_6\right\}+\frac{1}{8\epsilon}\left\{15\lambda^2 c_6\right\}-\frac{\log M_s^2}{4\epsilon}\left\{15\lambda^2 c_6\right\}\right)(\phi^\dagger \phi)^3\nn\\
    &\quad\quad+\left(\frac{1}{\epsilon^2}\left\{4\lambda^2 {y^{(3)}_{ij}}^\dagger\right\}+\frac{1}{\epsilon}\left\{2\lambda^2 {y^{(3)}_{ij}}^\dagger\right\}-\frac{\log M_s^2}{\epsilon}\left\{4\lambda^2 {y^{(3)}_{ij}}^\dagger\right\}\right)(\phi^\dagger \phi)\phi^\dagger \overline\psi_j\psi_i\nn\\
    &\quad\quad+\left(\frac{1}{\epsilon^2}\left\{5\lambda^2 y^{(3)}_{ij}\right\}+\frac{1}{2\epsilon}\left\{5\lambda^2 y^{(3)}_{ij}\right\}-\frac{\log M_s^2}{\epsilon}\left\{5\lambda^2 y^{(3)}_{ij}\right\}\right)(\phi^\dagger \phi)\phi \overline\psi_i\psi_j\nn\\
    &\quad\quad+\left(\frac{1}{\epsilon^2}\left\{3M_s^2\lambda y^{(3)}_{ij}\right\}+\frac{1}{2\epsilon}\left\{9M_s^2\lambda y^{(3)}_{ij}\right\}-\frac{\log M_s^2}{\epsilon}\left\{3M_s^2\lambda y^{(3)}_{ij}\right\}\right)\phi\overline\psi_i\psi_j\nn\\
    &\quad\quad+\left(\frac{1}{\epsilon^2}\left\{3M_s^2\lambda {y^{(3)}_{ij}}^\dagger\right\}+\frac{1}{2\epsilon}\left\{9M_s^2\lambda {y^{(3)}_{ij}}^\dagger\right\}-\frac{\log M_s^2}{\epsilon}\left\{3M_s^2\lambda {y^{(3)}_{ij}}^\dagger\right\}\right)\phi^\dagger\overline\psi_j\psi_i\nn\\
    &\quad\quad+\bigg(\frac{1}{8\epsilon}\bigg\{2\lambda^2\phi^\dagger D^2(\phi)+3\lambda c_6 (\phi^\dagger\phi)\phi^\dagger D^2(\phi)+\lambda c_6 (\phi^\dagger\phi)\phi D^2(\phi^\dagger)\nn\\
    &\quad\quad\quad\quad\quad\quad+2\lambda Y^{(3)}_{ij}\overline\psi_i\psi_j D^2(\phi)+2\lambda {Y^{(3)}_{ij}}^\dagger\overline\psi_j\psi_i D^2(\phi^\dagger)\bigg\}\bigg)\bigg],
\end{align}
\begin{align}
    \L_{2}^d &\subset \int d^dx_1\,d^dx_2\, V^\dagger_{(3)}(x_1)\Big\lvert_{\phi\overline\psi_i\psi_j} G^f_{jl}(x_1,x_2)(G^f_{ki}(x_1,x_2))^\dagger G^s(x_1,x_2) V_{(3)}(x_2)\Big\lvert_{\phi^\dagger\overline\psi_l\psi_k}\nn\\
    &=\alpha^2\bigg[\bigg(\frac{1}{\epsilon^2}\bigg\{-3M_f^2\,\lambda\,{y^{(1)}_{ij}}^\dagger y^{(1)}_{ij}+\frac{1}{2}M_s^2\,\lambda\,{y^{(1)}_{ij}}^\dagger y^{(1)}_{ij}-\frac{9}{2}M_f^4\,{y^{(3)}_{ij}}^\dagger y^{(1)}_{ij}+\frac{1}{4}M_s^4\,{y^{(3)}_{ij}}^\dagger y^{(1)}_{ij}\nn\\
    &\quad\quad\quad\quad-3M_f^2M_s^2\,{y^{(3)}_{ij}}^\dagger y^{(1)}_{ij}-\frac{7}{2}M_f^2\,{y^{(1)}_{ij}}^\dagger y^{(1)}_{il}y^{(1)}_{kj}{y^{(1)}_{kl}}^\dagger-\frac{1}{2}M_s^2\,{y^{(1)}_{ij}}^\dagger y^{(1)}_{il}y^{(1)}_{kj}{y^{(1)}_{kl}}^\dagger+\cdot\cdot\cdot\bigg\}\nn\\
    &\quad\quad\quad\quad+\frac{1}{\epsilon}\bigg\{-\frac{5}{4}M_f^2\,\lambda\,{y^{(1)}_{ij}}^\dagger y^{(1)}_{ij}+\frac{5}{8}M_s^2\,\lambda\,{y^{(1)}_{ij}}^\dagger y^{(1)}_{ij}-\frac{57}{8}M_f^4\,{y^{(3)}_{ij}}^\dagger y^{(1)}_{ij}+\frac{7}{16}M_s^4\,{y^{(3)}_{ij}}^\dagger y^{(1)}_{ij}\nn\\
    &\quad\quad\quad\quad-\frac{17}{4}M_f^2M_s^2\,{y^{(3)}_{ij}}^\dagger y^{(1)}_{ij}-\frac{17}{8}M_f^2\,{y^{(1)}_{ij}}^\dagger y^{(1)}_{il}y^{(1)}_{kj}{y^{(1)}_{kl}}^\dagger-\frac{5}{8}M_s^2\,{y^{(1)}_{ij}}^\dagger y^{(1)}_{il}y^{(1)}_{kj}{y^{(1)}_{kl}}^\dagger+\cdot\cdot\cdot\bigg\}\nn\\
    &\quad\quad\quad\quad+\frac{\log M_s^2}{\epsilon}\bigg\{-\frac{1}{4}M_s^4\,{y^{(3)}_{ij}}^\dagger y^{(1)}_{ij}+3M_f^2M_s^2\,{y^{(3)}_{ij}}^\dagger y^{(1)}_{ij}+3M_f^2\,\lambda\,{y^{(1)}_{ij}}^\dagger y^{(1)}_{ij}-\frac{1}{2}M_s^2\,\lambda\,{y^{(1)}_{ij}}^\dagger y^{(1)}_{ij}\nn\\
    &\quad\quad\quad\quad+\frac{1}{2}M_s^2\,{y^{(1)}_{ij}}^\dagger y^{(1)}_{il}y^{(1)}_{kj}{y^{(1)}_{kl}}^\dagger+\cdot\cdot\cdot\bigg\}+\frac{\log M_f^2}{\epsilon}\bigg\{\frac{9}{2}M_f^4\,{y^{(3)}_{ij}}^\dagger y^{(1)}_{ij}+\frac{7}{2}M_f^2\,{y^{(1)}_{ij}}^\dagger y^{(1)}_{il}y^{(1)}_{kj}{y^{(1)}_{kl}}^\dagger\nn\\
    &\quad\quad\quad\quad+\cdot\cdot\cdot\bigg\}\bigg)(\phi^\dagger\phi)\nn\\
    &\quad\quad+\bigg(\frac{1}{\epsilon^2}\bigg\{-\frac{1}{4}{y^{(1)}_{ij}}^\dagger {y^{(1)}_{kl}}^\dagger y^{(1)}_{il}{y^{(1)}_{mn}}^\dagger y^{(1)}_{kn} y^{(1)}_{mj}-\frac{7}{4} M_f^2\,{y^{(1)}_{ij}}^\dagger y^{(1)}_{il}y^{(3)}_{kj}{y^{(1)}_{kl}}^\dagger-\frac{1}{4} M_s^2\,{y^{(1)}_{ij}}^\dagger y^{(1)}_{il}y^{(3)}_{lk}{y^{(1)}_{jk}}^\dagger \nn\\
    &\quad\quad\quad\quad +\frac{1}{4}\lambda^2\,{y^{(1)}_{ij}}^\dagger y^{(1)}_{ij}-\frac{1}{2}\lambda\,{y^{(1)}_{ij}}^\dagger y^{(1)}_{il}y^{(1)}_{kj}{y^{(1)}_{kl}}^\dagger-3M_f^2\,\lambda {y^{(3)}_{ij}}^\dagger y^{(1)}_{ij}+\frac{1}{2}M_s^2\,\lambda {y^{(3)}_{ij}}^\dagger y^{(1)}_{ij}\nn\\
    &\quad\quad\quad\quad-\frac{3}{4}M_f^2\,c_6\, {y^{(1)}_{ij}}^\dagger y^{(1)}_{ij}+\frac{1}{8}M_s^2\,c_6\, {y^{(1)}_{ij}}^\dagger y^{(1)}_{ij}+\cdot\cdot\cdot\bigg\}+\frac{1}{\epsilon}\bigg\{-\frac{1}{16}{y^{(1)}_{ij}}^\dagger {y^{(1)}_{kl}}^\dagger y^{(1)}_{il}{y^{(1)}_{mn}}^\dagger y^{(1)}_{mj} y^{(1)}_{kn}\nn\\
    &\quad\quad\quad\quad-\frac{17}{8} M_f^2\,{y^{(1)}_{ij}}^\dagger y^{(3)}_{il}y^{(1)}_{kj}{y^{(1)}_{kl}}^\dagger-\frac{5}{8} M_s^2\,{y^{(1)}_{ij}}^\dagger y^{(3)}_{il}y^{(1)}_{kj}{y^{(1)}_{kl}}^\dagger +\frac{1}{16}\lambda^2\,{y^{(1)}_{ij}}^\dagger y^{(1)}_{ij}-\frac{1}{8}\lambda\,{y^{(1)}_{ij}}^\dagger y^{(1)}_{il}y^{(1)}_{kj}{y^{(1)}_{kl}}^\dagger\nn\\
    &\quad\quad\quad\quad-\frac{5}{4}M_f^2\,\lambda {y^{(3)}_{ij}}^\dagger y^{(1)}_{ij}+\frac{5}{8}M_s^2\,\lambda {y^{(3)}_{ij}}^\dagger y^{(1)}_{ij}-\frac{5}{16}M_f^2\,c_6\, {y^{(1)}_{ij}}^\dagger y^{(1)}_{ij}+\frac{5}{32}M_s^2\,c_6\, {y^{(1)}_{ij}}^\dagger y^{(1)}_{ij}+\cdot\cdot\cdot\bigg\}\nn\\
    &\quad\quad\quad\quad+\frac{\log M_s^2}{\epsilon}\bigg\{\frac{1}{4} M_s^2\,{y^{(1)}_{ij}}^\dagger y^{(1)}_{il}y^{(3)}_{kj}{y^{(1)}_{kl}}^\dagger -\frac{1}{4}\lambda^2\,{y^{(1)}_{ij}}^\dagger y^{(1)}_{ij}+\frac{1}{2}\lambda\,{y^{(1)}_{ij}}^\dagger y^{(1)}_{kj}y^{(1)}_{il}{y^{(1)}_{kl}}^\dagger\nn\\
    &\quad\quad\quad\quad+3M_f^2\,\lambda{y^{(3)}_{ij}}^\dagger y^{(1)}_{ij}-\frac{1}{2}M_s^2\,\lambda {y^{(3)}_{ij}}^\dagger y^{(1)}_{ij}+\frac{3}{4}M_f^2\,c_6\, {y^{(1)}_{ij}}^\dagger y^{(1)}_{ij}-\frac{1}{8}M_s^2\,c_6\, {y^{(1)}_{ij}}^\dagger y^{(1)}_{ij}+\cdot\cdot\cdot\bigg\}\nn\\
    &\quad\quad\quad\quad+\frac{\log M_f^2}{\epsilon}\bigg\{\frac{1}{4}{y^{(1)}_{ij}}^\dagger {y^{(1)}_{kl}}^\dagger y^{(1)}_{il}{y^{(1)}_{mn}}^\dagger y^{(1)}_{kn} y^{(1)}_{mj}+\frac{7}{4} M_f^2\,{y^{(1)}_{ij}}^\dagger y^{(1)}_{il}y^{(3)}_{kj}{y^{(1)}_{kl}}^\dagger+\cdot\cdot\cdot\bigg\}\bigg)(\phi^\dagger\phi)^2\nn\\
    &\quad\quad+\bigg(\frac{1}{\epsilon^2}\bigg\{-\frac{1}{4}{y^{(3)}_{mn}}^\dagger {y^{(1)}_{ij}}^\dagger y^{(1)}_{mj}{y^{(1)}_{kl}}^\dagger y^{(1)}_{kn} y^{(1)}_{il}-\frac{1}{2}\lambda\, {y^{(1)}_{ij}}^\dagger y^{(1)}_{kj}{y^{(1)}_{kl}}^\dagger y^{(1)}_{il} +\frac{1}{4}\lambda^2\, {y^{(3)}_{ij}}^\dagger y^{(1)}_{ij}\nn\\
    &\quad\quad\quad\quad +\frac{1}{8}\lambda\, c_6 \,{y^{(1)}_{ij}}^\dagger y^{(1)}_{ij}-\frac{1}{8} c_6 {y^{(1)}_{ij}}^\dagger y^{(1)}_{il}{y^{(1)}_{kl}}^\dagger y^{(1)}_{kj}+\cdot\cdot\cdot\bigg\}+\frac{1}{\epsilon}\bigg\{\frac{1}{16}\lambda^2\, {y^{(3)}_{ij}}^\dagger y^{(1)}_{ij}\nn\\
    &\quad\quad\quad\quad-\frac{1}{4}{y^{(3)}_{mk}}^\dagger {y^{(1)}_{ij}}^\dagger y^{(1)}_{in}{y^{(1)}_{kl}}^\dagger y^{(1)}_{mj} y^{(1)}_{nl}-\frac{1}{8}\lambda\, {y^{(1)}_{ik}}^\dagger y^{(1)}_{kj}{y^{(1)}_{kl}}^\dagger y^{(3)}_{il} +\frac{1}{32}\lambda\, c_6 \,{y^{(1)}_{ij}}^\dagger y^{(1)}_{ij}\nn\\
    &\quad\quad\quad\quad-\frac{1}{32} c_6 {y^{(1)}_{ij}}^\dagger y^{(1)}_{il}{y^{(1)}_{kl}}^\dagger y^{(1)}_{kj}+\cdot\cdot\cdot\bigg\} +\frac{\log M_s^2}{\epsilon}\bigg\{\frac{3}{4}\lambda\, {y^{(3)}_{kl}}^\dagger y^{(1)}_{kj}{y^{(1)}_{ij}}^\dagger y^{(1)}_{il} \nn\\
    &\quad\quad\quad\quad-\frac{1}{4}\lambda\, c_6 \,{y^{(1)}_{ij}}^\dagger y^{(1)}_{ij} +\frac{1}{4} c_6 {y^{(1)}_{ij}}^\dagger y^{(1)}_{il}{y^{(1)}_{kl}}^\dagger y^{(1)}_{kj}-\frac{1}{2}\lambda^2\,{y^{(3)}_{ij}}^\dagger y^{(1)}_{ij}+\cdot\cdot\cdot\bigg\} \nn\\
    &\quad\quad\quad\quad+\frac{\log M_f^2}{\epsilon}\bigg\{ {y^{(1)}_{jk}}^\dagger {y^{(1)}_{ij}}^\dagger y^{(1)}_{im}{y^{(1)}_{kl}}^\dagger y^{(1)}_{mn} y^{(1)}_{nl}+\cdot\cdot\cdot\bigg\}\bigg)(\phi^\dagger\phi)^3\nn\\
    &\quad\quad + \bigg(\frac{1}{\epsilon^2}\bigg\{-2M_f{y^{(1)}_{ij}}^\dagger {y^{(1)}_{kl}}^\dagger y^{(1)}_{im} y^{(1)}_{jk} y^{(1)}_{ml}-9M_f^3{y^{(1)}_{ij}}^\dagger  y^{(3)}_{ik} y^{(1)}_{kj}-3M_f\lambda{y^{(1)}_{ij}}^\dagger  y^{(1)}_{ik} y^{(1)}_{kj}\nn\\
    &\quad\quad\quad\quad-3M_fM_s^2{y^{(1)}_{ij}}^\dagger  y^{(3)}_{ik} y^{(1)}_{kj}+\cdot\cdot\cdot\bigg\}+\frac{1}{\epsilon}\bigg\{-\frac{1}{8}M_f{y^{(1)}_{ij}}^\dagger {y^{(1)}_{kl}}^\dagger y^{(1)}_{im} y^{(1)}_{jk} y^{(1)}_{ml}\nn\\
    &\quad\quad\quad\quad-\frac{39}{4}M_f^3{y^{(1)}_{ij}}^\dagger  y^{(3)}_{ik} y^{(1)}_{kj}-\frac{5}{4}M_f\lambda{y^{(1)}_{ij}}^\dagger  y^{(1)}_{ik} y^{(1)}_{kj}-\frac{17}{4}M_f M_s^2{y^{(1)}_{ij}}^\dagger  y^{(3)}_{ik} y^{(1)}_{kj}+\cdot\cdot\cdot\bigg\}\nn\\
    &\quad\quad\quad\quad+\frac{\log M_s^2}{\epsilon}\bigg\{3M_f\lambda{y^{(1)}_{ij}}^\dagger  y^{(1)}_{ik} y^{(1)}_{kj}+3M_fM_s^2{y^{(1)}_{ij}}^\dagger  y^{(3)}_{ik} y^{(1)}_{kj}+\cdot\cdot\cdot\bigg\}\nn\\
    &\quad\quad\quad\quad+\frac{\log M_f^2}{\epsilon}\bigg\{2M_f{y^{(1)}_{ij}}^\dagger {y^{(1)}_{kl}}^\dagger y^{(1)}_{im} y^{(1)}_{jk} y^{(1)}_{ml} +9M_f^3{y^{(1)}_{ij}}^\dagger  y^{(3)}_{ik} y^{(1)}_{kj}+\cdot\cdot\cdot\bigg\}\bigg)(\phi^\dagger\phi)\phi \nn\\
    &\quad\quad + \bigg(\frac{1}{\epsilon^2}\bigg\{-2M_f{y^{(1)}_{ij}}^\dagger {y^{(1)}_{kl}}^\dagger y^{(1)}_{im} {y^{(1)}_{kj}}^\dagger y^{(1)}_{ml}-9M_f^3{y^{(3)}_{ij}}^\dagger  {y^{(1)}_{ki}}^\dagger y^{(1)}_{kj}-3M_f\lambda{y^{(1)}_{ij}}^\dagger  {y^{(1)}_{ki}}^\dagger y^{(1)}_{kj}\nn\\
    &\quad\quad\quad\quad-3M_fM_s^2{y^{(1)}_{ij}}^\dagger  {y^{(3)}_{ki}}^\dagger y^{(1)}_{kj}+\cdot\cdot\cdot\bigg\}+\frac{1}{\epsilon}\bigg\{-\frac{9}{8}M_f{y^{(1)}_{ij}}^\dagger {y^{(1)}_{kl}}^\dagger y^{(1)}_{im} {y^{(1)}_{kj}}^\dagger y^{(1)}_{ml}\nn\\
    &\quad\quad\quad\quad-\frac{39}{4}M_f^3{y^{(1)}_{ij}}^\dagger  {y^{(1)}_{ki}}^\dagger y^{(3)}_{kj}-\frac{5}{4}M_f\lambda{y^{(3)}_{ij}}^\dagger  {y^{(1)}_{ki}}^\dagger y^{(1)}_{kj}-\frac{17}{4}M_f M_s^2{y^{(3)}_{ij}}^\dagger  {y^{(1)}_{ki}}^\dagger y^{(1)}_{kj}+\cdot\cdot\cdot\bigg\}\nn\\
    &\quad\quad\quad\quad+\frac{\log M_s^2}{\epsilon}\bigg\{3M_f\lambda{y^{(1)}_{ij}}^\dagger  {y^{(1)}_{ki}}^\dagger y^{(1)}_{kj}+3M_fM_s^2{y^{(1)}_{ij}}^\dagger  {y^{(3)}_{ki}}^\dagger y^{(1)}_{kj}+\cdot\cdot\cdot\bigg\}\nn\\
    &\quad\quad\quad\quad+\frac{\log M_f^2}{\epsilon}\bigg\{\frac{5}{2}M_f{y^{(1)}_{ij}}^\dagger {y^{(1)}_{kl}}^\dagger {y^{(1)}_{mi}}^\dagger y^{(1)}_{jk} y^{(1)}_{ml} +9M_f^3{y^{(1)}_{ij}}^\dagger  y^{(3)}_{ik} {y^{(1)}_{jk}}^\dagger+\cdot\cdot\cdot\bigg\}\bigg)(\phi^\dagger\phi)\phi^\dagger \nn\\
    &\quad\quad+\bigg(\frac{1}{\epsilon^2}\bigg\{-\frac{3}{4}M_f\,c_6\,{y^{(1)}_{ij}}^\dagger y^{(1)}_{ik} y^{(1)}_{kj}-3M_f\,\lambda\,{y^{(1)}_{ij}}^\dagger  y^{(3)}_{ik} y^{(1)}_{kj}-2M_f{y^{(3)}_{lk}}^\dagger y^{(1)}_{lm}{y^{(1)}_{ki}}^\dagger {y^{(1)}_{ij}}^\dagger y^{(1)}_{mj}+\cdot\cdot\cdot\bigg\}\nn\\
    &\quad\quad\quad\quad+\frac{1}{\epsilon}\bigg\{-\frac{5}{16}M_f\,c_6\,{y^{(1)}_{ij}}^\dagger y^{(1)}_{ik} y^{(1)}_{kj}-\frac{5}{4}M_f\,\lambda\,{y^{(1)}_{ij}}^\dagger  y^{(3)}_{ik} y^{(1)}_{kj}-\frac{9}{8}M_f{y^{(3)}_{lk}}^\dagger y^{(1)}_{lm}{y^{(1)}_{ki}}^\dagger {y^{(1)}_{ij}}^\dagger y^{(1)}_{mj}+\cdot\cdot\cdot\bigg\}\nn\\
    &\quad\quad\quad\quad+\frac{\log M_s^2}{\epsilon}\bigg\{\frac{3}{4}M_f\,c_6\,{y^{(1)}_{ij}}^\dagger y^{(1)}_{ik} y^{(1)}_{kj}+3M_f\,\lambda\,{y^{(1)}_{ij}}^\dagger  y^{(3)}_{ik} y^{(1)}_{kj}+\cdot\cdot\cdot\bigg\}\nn\\
    &\quad\quad\quad\quad+\frac{2\log M_f^2}{\epsilon}M_f{y^{(3)}_{lk}}^\dagger y^{(1)}_{lm}{y^{(1)}_{ki}}^\dagger {y^{(1)}_{ij}}^\dagger y^{(1)}_{mj}+\cdot\cdot\cdot\bigg)(\phi^\dagger\phi)^2\phi^\dagger \nn\\
    &\quad\quad+\bigg(\frac{1}{\epsilon^2}\bigg\{-\frac{3}{4}M_f\,c_6\,{y^{(1)}_{ij}}^\dagger {y^{(1)}_{ki}}^\dagger y^{(1)}_{kj}-3M_f\,\lambda\,{y^{(1)}_{ij}}^\dagger  y^{(3)}_{ik} {y^{(1)}_{jk}}^\dagger-2M_f{y^{(3)}_{mn}}^\dagger y^{(1)}_{mk}{y^{(1)}_{ik}}^\dagger {y^{(1)}_{nj}}^\dagger y^{(1)}_{ij}+\cdot\cdot\cdot\bigg\}\nn\\
    &\quad\quad\quad\quad+\frac{1}{\epsilon}\bigg\{-\frac{5}{16}M_f\,c_6\,{y^{(1)}_{ij}}^\dagger {y^{(1)}_{ki}}^\dagger y^{(1)}_{kj}-\frac{5}{4}M_f\,\lambda\,{y^{(1)}_{ij}}^\dagger  y^{(3)}_{ik} {y^{(1)}_{jk}}^\dagger-\frac{1}{8}M_f{y^{(3)}_{mn}}^\dagger y^{(1)}_{mk}{y^{(1)}_{ik}}^\dagger {y^{(1)}_{nj}}^\dagger y^{(1)}_{ij}+\cdot\cdot\cdot\bigg\}\nn\\
    &\quad\quad\quad\quad+\frac{\log M_s^2}{\epsilon}\bigg\{\frac{3}{4}M_f\,c_6\,{y^{(1)}_{ij}}^\dagger {y^{(1)}_{ki}}^\dagger y^{(1)}_{kj}\nn\\
    &\quad\quad\quad\quad+3M_f\,\lambda\,{y^{(1)}_{ij}}^\dagger  y^{(3)}_{ik} {y^{(1)}_{jk}}^\dagger+\cdot\cdot\cdot\bigg\}+\frac{2\log M_f^2}{\epsilon}M_f{y^{(3)}_{mn}}^\dagger y^{(1)}_{mk}{y^{(1)}_{ik}}^\dagger {y^{(1)}_{nj}}^\dagger y^{(1)}_{ij}+\cdot\cdot\cdot\bigg)(\phi^\dagger\phi)^2\phi \nn\\
    &\quad\quad+\bigg(\frac{1}{\epsilon^2}\bigg\{-3M_fM_s^2 {y^{(1)}_{ij}}^\dagger y^{(1)}_{ik} y^{(1)}_{kj}-9M_f^3 {y^{(1)}_{ij}}^\dagger y^{(1)}_{ik} y^{(1)}_{kj}+\cdot\cdot\cdot\bigg\}\nn\\
    &\quad\quad\quad\quad+\frac{1}{\epsilon}\bigg\{-\frac{17}{4}M_fM_s^2 {y^{(1)}_{ij}}^\dagger y^{(1)}_{ik} y^{(1)}_{kj}-\frac{39}{4}M_f^3 {y^{(1)}_{ij}}^\dagger y^{(1)}_{ik} y^{(1)}_{kj}+\cdot\cdot\cdot\bigg\}\nn\\
    &\quad\quad\quad\quad+\frac{3\log M_s^2}{\epsilon}M_fM_s^2 {y^{(1)}_{ij}}^\dagger y^{(1)}_{ik} y^{(1)}_{kj}+\frac{9\log M_f^2}{\epsilon}M_f^3 {y^{(1)}_{ij}}^\dagger y^{(1)}_{ik} y^{(1)}_{kj}+\cdot\cdot\cdot\bigg)\phi\nn\\
    &\quad\quad+\bigg(\frac{1}{\epsilon^2}\bigg\{-3M_fM_s^2 {y^{(1)}_{ij}}^\dagger {y^{(1)}_{ki}}^\dagger y^{(1)}_{kj}-9M_f^3 {y^{(1)}_{ij}}^\dagger {y^{(1)}_{ki}}^\dagger y^{(1)}_{kj}+\cdot\cdot\cdot\bigg\}\nn\\
    &\quad\quad\quad\quad+\frac{1}{\epsilon}\bigg\{-\frac{17}{4}M_fM_s^2 {y^{(1)}_{ij}}^\dagger {y^{(1)}_{ki}}^\dagger y^{(1)}_{kj}-\frac{39}{4}M_f^3 {y^{(1)}_{ij}}^\dagger {y^{(1)}_{ki}}^\dagger y^{(1)}_{kj}+\cdot\cdot\cdot\bigg\}\nn\\
    &\quad\quad\quad\quad+\frac{3\log M_s^2}{\epsilon}M_fM_s^2 {y^{(1)}_{ij}}^\dagger {y^{(1)}_{ki}}^\dagger y^{(1)}_{kj}+\frac{9\log M_f^2}{\epsilon}M_f^3 {y^{(1)}_{ij}}^\dagger {y^{(1)}_{ki}}^\dagger y^{(1)}_{kj}+\cdot\cdot\cdot\bigg)\phi^\dagger\nn\\
    &\quad\quad+\bigg(\frac{1}{2\epsilon^2}\bigg\{-3M_f^2\,y^{(3)}_{ij}{y^{(1)}_{kl}}^\dagger y^{(1)}_{kl}+\frac{1}{2}M_s^2\,y^{(3)}_{ij}{y^{(1)}_{kl}}^\dagger y^{(1)}_{kl}+ \cdot\cdot\cdot\bigg\}\nn\\
    &\quad\quad\quad\quad+\frac{1}{\epsilon}\bigg\{-\frac{5}{4}M_f^2\,y^{(3)}_{ij}{y^{(1)}_{kl}}^\dagger y^{(1)}_{kl}+\frac{5}{8}M_s^2\,y^{(3)}_{ij}{y^{(1)}_{kl}}^\dagger y^{(1)}_{kl}+ \cdot\cdot\cdot\bigg\}\nn\\
    &\quad\quad\quad\quad-\frac{\log M_s^2}{\epsilon}\bigg\{-3M_f^2\,y^{(3)}_{ij}{y^{(1)}_{kl}}^\dagger y^{(1)}_{kl}+\frac{1}{2}M_s^2\,y^{(3)}_{ij}{y^{(1)}_{kl}}^\dagger y^{(1)}_{kl}+ \cdot\cdot\cdot\bigg\}\bigg)\phi\overline\psi_i\psi_j\nn\\
    &\quad\quad+\bigg(\frac{1}{2\epsilon^2}\bigg\{\frac{1}{4}\lambda\,y^{(3)}_{ij}{y^{(1)}_{kl}}^\dagger y^{(1)}_{kl}-\frac{1}{2}y^{(3)}_{ij}{y^{(1)}_{mn}}^\dagger y^{(1)}_{mk}{y^{(1)}_{nl}}^\dagger y^{(1)}_{kl} + \cdot\cdot\cdot\bigg\}\nn\\
    &\quad\quad\quad\quad +\frac{1}{\epsilon}\bigg\{\frac{1}{16}\lambda\,y^{(3)}_{ij}{y^{(1)}_{kl}}^\dagger y^{(1)}_{kl}-\frac{1}{8}y^{(3)}_{ij}{y^{(1)}_{mn}}^\dagger y^{(1)}_{mk}{y^{(1)}_{nl}}^\dagger y^{(1)}_{kl}  + \cdot\cdot\cdot\bigg\}\nn\\
    &\quad\quad\quad\quad-\frac{\log M_s^2}{\epsilon}\bigg\{-\frac{1}{4}\lambda\,y^{(3)}_{ij}{y^{(1)}_{kl}}^\dagger y^{(1)}_{kl}+\frac{1}{2}y^{(3)}_{ij}{y^{(1)}_{mn}}^\dagger y^{(1)}_{mk}{y^{(1)}_{nl}}^\dagger y^{(1)}_{kl} + \cdot\cdot\cdot\bigg\}\bigg)(\phi^\dagger\phi)\phi\overline\psi_i\psi_j\nn\\
    &\quad\quad + \bigg(-\frac{3}{\epsilon^2}M_f y^{(3)}_{ij}{y^{(1)}_{mn}}^\dagger {y^{(1)}_{lm}}^\dagger y^{(1)}_{ln}-\frac{5}{4\epsilon}M_f y^{(3)}_{ij}{y^{(1)}_{mn}}^\dagger {y^{(1)}_{lm}}^\dagger y^{(1)}_{ln}\nn\\
    &\quad\quad\quad\quad+\frac{3\log M_s^2}{\epsilon}M_f y^{(3)}_{ij}{y^{(1)}_{mn}}^\dagger {y^{(1)}_{lm}}^\dagger y^{(1)}_{ln}\bigg)(\phi^\dagger\phi)\overline\psi_i\psi_j \nn\\
    &\quad\quad + \bigg(\frac{1}{4\epsilon^2}\lambda\, {y^{(3)}_{ji}}^\dagger {y^{(1)}_{mn}}^\dagger y^{(1)}_{mn}+\frac{1}{16\epsilon}\lambda\, {y^{(3)}_{ji}}^\dagger {y^{(1)}_{mn}}^\dagger  y^{(1)}_{mn}-\frac{\log M_s^2}{4\epsilon}\lambda {y^{(3)}_{ij}}^\dagger {y^{(1)}_{mn}}^\dagger y^{(1)}_{mn}\bigg)(\phi^\dagger\phi)\phi^\dagger\overline\psi_i\psi_j \nn\\
    &\quad\quad + \bigg(\frac{1}{4\epsilon^2}\lambda\, y^{(3)}_{ij}{y^{(1)}_{mn}}^\dagger y^{(1)}_{mn}+\frac{1}{16\epsilon}\lambda\, y^{(3)}_{ij}{y^{(1)}_{mn}}^\dagger  y^{(1)}_{mn}-\frac{\log M_s^2}{4\epsilon}\lambda y^{(3)}_{ij}{y^{(1)}_{mn}}^\dagger y^{(1)}_{mn}\bigg)(\phi^\dagger\phi)\phi\overline\psi_i\psi_j \nn\\
    &\quad\quad + \bigg(-\frac{3}{\epsilon^2}M_f\, y^{(3)}_{ij}{y^{(1)}_{mn}}^\dagger y^{(1)}_{ml}y^{(1)}_{ln}-\frac{5}{4\epsilon}M_f\, y^{(3)}_{ij}{y^{(1)}_{mn}}^\dagger y^{(1)}_{ml}y^{(1)}_{ln}\nn\\
    &\quad\quad\quad\quad+\frac{3\log M_s^2}{\epsilon}M_f\, y^{(3)}_{ij}{y^{(1)}_{mn}}^\dagger y^{(1)}_{ml}y^{(1)}_{ln}\bigg)(\phi)^2\overline\psi_i\psi_j \nn\\
    &\quad\quad + \bigg(-\frac{4}{\epsilon^2}M_f^2\, {y^{(1)}_{ij}}^\dagger {y^{(1)}_{jk}}^\dagger {y^{(1)}_{kl}}^\dagger y^{(1)}_{il}-2M_f^2\, {y^{(1)}_{ij}}^\dagger {y^{(1)}_{jk}}^\dagger {y^{(1)}_{kl}}^\dagger y^{(1)}_{il}\nn\\
    &\quad\quad\quad\quad+\frac{4\log M_s^2}{\epsilon}M_f\, M_f^2\, {y^{(1)}_{ij}}^\dagger {y^{(1)}_{jk}}^\dagger {y^{(1)}_{kl}}^\dagger y^{(1)}_{il}\bigg)(\phi^\dagger)^2 \nn\\
    &\quad\quad + \bigg(-\frac{4}{\epsilon^2}M_f^2\, {y^{(1)}_{ij}}^\dagger y^{(1)}_{kj} y^{(1)}_{kl} y^{(1)}_{il}-2M_f^2\, {y^{(1)}_{ji}}^\dagger y^{(1)}_{jk} y^{(1)}_{kl} y^{(1)}_{li}\nn\\
    &\quad\quad\quad\quad+\frac{4\log M_s^2}{\epsilon}M_f\, M_f^2\, {y^{(1)}_{ji}}^\dagger y^{(1)}_{jk} y^{(1)}_{kl} y^{(1)}_{li}\bigg)(\phi)^2 -\frac{1}{2\epsilon} {y^{(1)}_{ij}}^\dagger {y^{(1)}_{jk}}^\dagger y^{(1)}_{kl} y^{(1)}_{il} \phi D^2(\phi^\dagger)\nn\\
    &\quad\quad+\bigg(\frac{1}{\epsilon^2}\bigg\{-\frac{5}{12}{y^{(1)}_{mn}}^\dagger y^{(1)}_{mk}{y^{(1)}_{nl}}^\dagger y^{(3)}_{kl} +\frac{5}{36}\lambda\,{y^{(3)}_{ij}}^\dagger y^{(1)}_{ij} + \cdot\cdot\cdot \bigg\}\nn\\
    &\quad\quad\quad\quad+\frac{1}{\epsilon}\bigg\{-\frac{11}{144}{y^{(1)}_{mn}}^\dagger y^{(1)}_{mk}{y^{(1)}_{nl}}^\dagger y^{(3)}_{kl} +\frac{7}{432}\lambda\,{y^{(3)}_{ij}}^\dagger y^{(1)}_{ij} + \cdot\cdot\cdot\bigg\}\nn\\
    &\quad\quad\quad\quad+\frac{\log M_s^2}{\epsilon}\bigg\{-\frac{5}{36}\lambda\,{y^{(3)}_{ij}}^\dagger y^{(1)}_{ij}+\cdot\cdot\cdot\bigg\}\nn\\
    &\quad\quad\quad\quad+\frac{\log M_f^2}{\epsilon}\bigg\{\frac{5}{12}{y^{(1)}_{mn}}^\dagger y^{(1)}_{mk}{y^{(1)}_{nl}}^\dagger y^{(3)}_{kl} +\cdot\cdot\cdot\bigg\}\bigg)(\phi^\dagger\phi)D^2(\phi^\dagger\phi)\nn\\
    &\quad\quad+\bigg(-\frac{5}{6\epsilon^2}y^{(1)}_{mk}{y^{(1)}_{nk}}^\dagger y^{(3)}_{nm}-\frac{11}{72\epsilon}y^{(1)}_{mk}{y^{(1)}_{nk}}^\dagger y^{(3)}_{nm}+\frac{5\log M_f^2}{6\epsilon}y^{(1)}_{mk}{y^{(1)}_{nk}}^\dagger y^{(3)}_{nm} +\cdot\cdot\cdot\bigg)(\phi^\dagger\phi)D^2(\phi)\nn\\
    &\quad\quad+\bigg(-\frac{5}{6\epsilon^2}y^{(1)}_{mk}{y^{(3)}_{nk}}^\dagger y^{(1)}_{nm}-\frac{11}{72\epsilon}y^{(1)}_{mk}{y^{(2)}_{nk}}^\dagger y^{(1)}_{nm}+\frac{5\log M_f^2}{6\epsilon}y^{(1)}_{mk}{y^{(3)}_{nk}}^\dagger y^{(1)}_{nm} +\cdot\cdot\cdot\bigg)(\phi^\dagger\phi)D^2(\phi^\dagger)\bigg].
\end{align}

Contributions from one-loop counter term insertion in one-loop topology are given by,
\begin{align}
    \L_{(2-ct)}^{(a)} &\subset\int d^dx_1\, {V_{(2)}^{(s-1)}}^\dagger(x_1)\Big\lvert_{\phi^\dagger\phi}\,G^s(x_1,x_1)\nn\\
    &= \alpha^2\bigg[\bigg(\frac{1}{\epsilon^2}\bigg\{-20M_s^2\lambda^2-4M_s^4c_6\bigg\}-\frac{1}{\epsilon}\bigg\{10M_s^2\lambda^2+4M_s^4c_6\bigg\}\nn\\
    &\quad\quad\quad\quad+\frac{\log M_s^2}{\epsilon}\bigg\{20M_s^2\lambda^2+4M_s^4c_6\bigg\}\bigg)(\phi^\dagger\phi)\nn\\
    &\quad\quad+\bigg(\frac{1}{\epsilon^2}\bigg\{-16\lambda^3-23M_s^2\,\lambda\,c_6\bigg\}+\frac{1}{2\epsilon}\bigg\{-23M_s^2\,\lambda\,c_6\bigg\}\nn\\
    &\quad\quad\quad\quad+\frac{\log M_s^2}{\epsilon}\bigg\{16\lambda^3+23M_s^2\,\lambda\,c_6\bigg\}\bigg)(\phi^\dagger\phi)^2\nn\\
    &\quad\quad+\bigg(\frac{1}{\epsilon^2}\bigg\{-22\lambda^2\,c_6\bigg\}+\frac{\log M_s^2}{\epsilon}\bigg\{22\lambda^2\,c_6\bigg\}\bigg)(\phi^\dagger\phi)^3\nn\\
    &\quad\quad+\bigg(\frac{1}{\epsilon^2}\bigg\{-24\lambda^2\,y^{(3)}_{ij}\bigg\}+\frac{\log M_s^2}{\epsilon}\bigg\{24\lambda^2\,y^{(3)}_{ij}\bigg\}\bigg)(\phi^\dagger\phi)\phi\overline\psi_i\psi_j\nn\\
    &\quad\quad+\bigg(\frac{1}{\epsilon^2}\bigg\{-8M_s^2\,\lambda\,{y^{(3)}_{ij}}^\dagger\bigg\}+\frac{\log M_s^2}{\epsilon}\bigg\{8M_s^2\,\lambda\,{y^{(3)}_{ij}}^\dagger\bigg\}\bigg)(\phi^\dagger\phi)\phi^\dagger\overline\psi_i\psi_j\nn\\
    &\quad\quad+\bigg(\frac{1}{\epsilon^2}\bigg\{-8M_s^2\,\lambda\,{y^{(3)}_{ij}}^\dagger\bigg\}+\frac{1}{\epsilon}\bigg\{-4M_s^2\,\lambda\,{y^{(3)}_{ij}}^\dagger\bigg\}+\frac{\log M_s^2}{\epsilon}\bigg\{8M_s^2\,\lambda\,{y^{(3)}_{ij}}^\dagger\bigg\}\bigg)\phi^\dagger\overline\psi_i\psi_j\nn\\
    &\quad\quad+\bigg(\frac{1}{\epsilon^2}\bigg\{-12M_s^2\,\lambda\,{y^{(3)}_{ij}}^\dagger\bigg\}+\frac{1}{\epsilon}\bigg\{-6M_s^2\,\lambda\,{y^{(3)}_{ij}}^\dagger\bigg\}+\frac{\log M_s^2}{\epsilon}\bigg\{12M_s^2\,\lambda\,y^{(3)}_{ij}\bigg\}\bigg)\phi\overline\psi_i\psi_j\bigg],
\end{align}
\begin{align}
    \L_{(2-ct)}^{(b)} &\subset -\int d^dx_1\, {V_{(2)}^{(s-1)}}^\dagger(x_1)\Big\lvert_{\overline\psi_j\psi_i}\,G^f_{ij}(x_1,x_1)\nn\\
    &=\alpha^2\bigg[\bigg(\frac{-12}{\epsilon^2}M_s^2M_f^2{y^{(1)}_{ij}}^\dagger y^{(3)}_{ij}-\frac{8}{\epsilon}M_s^2M_f^2{y^{(1)}_{ij}}^\dagger y^{(3)}_{ij}-\frac{6\log M_f^2}{\epsilon}M_s^2M_f^2{y^{(1)}_{ij}}^\dagger y^{(3)}_{ij}\nn\\
    &\quad\quad-\frac{6\log M_f}{\epsilon^2}M_s^2M_f^2{y^{(1)}_{ij}}^\dagger y^{(3)}_{ij}\bigg)(\phi^\dagger\phi) +\bigg(-\frac{4}{\epsilon^2}\bigg\{3M_f^2\lambda\,{y^{(1)}_{ij}}^\dagger y^{(3)}_{ij}+M_s^2{y^{(1)}_{ij}}^\dagger y^{(3)}_{ik}{y^{(1)}_{lk}}^\dagger y^{(1)}_{lj}\bigg\}\nn\\
    &\quad\quad\quad\quad-\frac{2}{\epsilon}\bigg\{M_f^2\lambda\,{y^{(1)}_{ij}}^\dagger y^{(3)}_{ij}+M_s^2{y^{(1)}_{ij}}^\dagger y^{(3)}_{ik}{y^{(1)}_{lk}}^\dagger y^{(1)}_{lj}\bigg\}\nn\\
    &\quad\quad\quad\quad+\frac{\log M_s^2}{\epsilon}\bigg\{6M_f^2\lambda\,{y^{(1)}_{ij}}^\dagger y^{(3)}_{ij}+2M_s^2{y^{(1)}_{ij}}^\dagger y^{(3)}_{ik}{y^{(1)}_{lk}}^\dagger y^{(1)}_{lj}\bigg\}\nn\\
     &\quad\quad\quad\quad+\frac{\log M_f^2}{\epsilon}\bigg\{6M_f^2\lambda\,{y^{(1)}_{ij}}^\dagger y^{(3)}_{ij}+2M_s^2{y^{(1)}_{ij}}^\dagger y^{(3)}_{ik}{y^{(1)}_{lk}}^\dagger y^{(1)}_{lj}\bigg\}\bigg)(\phi^\dagger\phi)^2\nn\\
    &\quad\quad +\bigg(\frac{-4}{\epsilon^2}c_6\,{y^{(1)}_{ij}}^\dagger y^{(1)}_{ik}{y^{(1)}_{lk}}^\dagger y^{(3)}_{lj}+\frac{2\log M_f^2}{\epsilon}c_6\,{y^{(1)}_{ij}}^\dagger y^{(1)}_{ik}{y^{(1)}_{lk}}^\dagger y^{(1)}_{lj} \nn\\
    &\quad\quad\quad\quad+\frac{2\log M_s^2}{\epsilon}\c_6\,{y^{(1)}_{ij}}^\dagger y^{(1)}_{ik}{y^{(1)}_{lk}}^\dagger y^{(1)}_{lj} \bigg)(\phi^\dagger\phi)^3\nn\\
    &\quad\quad +\bigg(\frac{4}{\epsilon^2}\bigg\{-2{y^{(1)}_{ij}}^\dagger y^{(3)}_{ik}{y^{(1)}_{jk}}^\dagger-{y^{(1)}_{ij}}^\dagger y^{(1)}_{ik}{y^{(3)}_{jk}}^\dagger\bigg\}\nn\\
    &\quad\quad\quad\quad+\frac{2\log M_f^2}{\epsilon}\bigg\{2{y^{(1)}_{ij}}^\dagger y^{(3)}_{ik}{y^{(1)}_{jk}}^\dagger+{y^{(1)}_{ij}}^\dagger y^{(1)}_{ik}{y^{(3)}_{jk}}^\dagger\bigg\}\nn\\
    &\quad\quad\quad\quad+\frac{2\log M_s^2}{\epsilon}\bigg\{2{y^{(1)}_{ij}}^\dagger y^{(3)}_{ik}{y^{(1)}_{jk}}^\dagger+{y^{(1)}_{ij}}^\dagger y^{(1)}_{ik}{y^{(3)}_{jk}}^\dagger\bigg\}\bigg)(\phi^\dagger\phi)^2\phi^\dagger\nn\\
    &\quad\quad +\bigg(-\frac{4}{\epsilon^2}{y^{(1)}_{ij}}^\dagger y^{(3)}_{ik}y^{(1)}_{kj}+\frac{2\log M_f^2}{\epsilon}{y^{(1)}_{ij}}^\dagger y^{(3)}_{ik}y^{(1)}_{kj}+\frac{2\log M_s^2}{\epsilon}{y^{(1)}_{ij}}^\dagger y^{(3)}_{ik}y^{(1)}_{kj}\bigg)(\phi^\dagger\phi)^2\phi\nn\\
    &\quad\quad +\bigg(-\frac{8}{\epsilon^2}M_fM_s^2\,y^{(3)}_{ij}{y^{(1)}_{kl}}^\dagger y^{(1)}_{kl}-\frac{4}{\epsilon^2}M_fM_s^2\,y^{(3)}_{ij}{y^{(1)}_{kl}}^\dagger y^{(1)}_{kl}\nn\\
    &\quad\quad\quad\quad+\frac{4\log M_f^2}{\epsilon}M_fM_s^2\,y^{(3)}_{ij}{y^{(1)}_{kl}}^\dagger y^{(1)}_{kl}+\frac{4\log M_s^2}{\epsilon}M_fM_s^2\,y^{(3)}_{ij}{y^{(1)}_{kl}}^\dagger y^{(1)}_{kl}\bigg)(\phi^\dagger\phi)\phi\nn\\
    &\quad\quad +\bigg(-\frac{4}{\epsilon^2}M_fM_s^2{y^{(1)}_{ij}}^\dagger y^{(3)}_{ik}{y^{(1)}_{jk}}^\dagger-\frac{2}{\epsilon}M_fM_s^2{y^{(1)}_{ij}}^\dagger y^{(3)}_{ik}{y^{(1)}_{jk}}^\dagger\nn\\
    &\quad\quad\quad\quad+\frac{2\log M_f^2}{\epsilon}M_fM_s^2{y^{(1)}_{ij}}^\dagger y^{(3)}_{ik}{y^{(1)}_{jk}}^\dagger+\frac{2\log M_s^2}{\epsilon}M_fM_s^2{y^{(1)}_{ij}}^\dagger y^{(3)}_{ik}{y^{(1)}_{jk}}^\dagger\bigg)(\phi^\dagger\phi)\phi^\dagger\bigg],
\end{align}
\begin{align}
    \L_{(2-ct)}^{(c)} &\subset\int d^dx_1\, {V_{(2)}^{(f-1)}}^\dagger(x_1)\Big\lvert_{\phi^\dagger\phi}\,G^s(x_1,x_1)\nn\\
    &= \alpha^2\bigg[ \bigg(\frac{1}{\epsilon^2}\bigg\{20M_f^2\lambda{y^{(1)}_{ij}}^\dagger y^{(1)}_{ij}+16M_s^2\lambda\,{y^{(1)}_{ij}}^\dagger{y^{(1)}_{kl}}^\dagger y^{(1)}_{il}y^{(1)}_{kj}+40\,M_f^2M_s^2{y^{(3)}_{ij}}^\dagger y^{(1)}_{ij}+\cdot\cdot\cdot\bigg\}\nn\\
    &\quad\quad\quad\quad + \frac{1}{\epsilon}\bigg\{2M_f^2\lambda{y^{(1)}_{ij}}^\dagger y^{(1)}_{ij}+8M_s^2\lambda\,{y^{(1)}_{ij}}^\dagger{y^{(1)}_{kl}}^\dagger y^{(1)}_{il}y^{(1)}_{kj}+24\,M_f^2M_s^2{y^{(3)}_{ij}}^\dagger y^{(1)}_{ij}+\cdot\cdot\cdot\bigg\}\nn\\
    &\quad\quad\quad\quad - \frac{\log M_s^2}{\epsilon}\bigg\{10M_f^2\lambda{y^{(1)}_{ij}}^\dagger y^{(1)}_{ij}+8M_s^2\lambda\,{y^{(1)}_{ij}}^\dagger{y^{(1)}_{kl}}^\dagger y^{(1)}_{il}y^{(1)}_{kj}+20\,M_f^2M_s^2{y^{(3)}_{ij}}^\dagger y^{(1)}_{ij}+\cdot\cdot\cdot\bigg\}\nn\\
    &\quad\quad\quad\quad - \frac{\log M_f^2}{\epsilon}\bigg\{10M_f^2\lambda{y^{(1)}_{ij}}^\dagger y^{(1)}_{ij}+8M_s^2\lambda\,{y^{(1)}_{ij}}^\dagger{y^{(1)}_{kl}}^\dagger y^{(1)}_{il}y^{(1)}_{kj}\nn\\
    &\quad\quad\quad\quad+20\,M_f^2M_s^2{y^{(3)}_{ij}}^\dagger y^{(1)}_{ij}+\cdot\cdot\cdot\bigg\}\bigg)(\phi^\dagger\phi)  \nn\\
    &\quad\quad+\bigg(\frac{16}{\epsilon^2}M_f^2M_s^2{y^{(3)}_{ij}}^\dagger y^{(1)}_{ij}+\frac{8}{\epsilon}M_f^2M_s^2{y^{(3)}_{ij}}^\dagger y^{(1)}_{ij}-\frac{8\log M_s^2}{\epsilon}M_f^2M_s^2{y^{(3)}_{ij}}^\dagger y^{(1)}_{ij}\nn\\
    &\quad\quad\quad\quad-\frac{8\log M_f^2}{\epsilon}M_f^2M_s^2{y^{(3)}_{ij}}^\dagger y^{(1)}_{ij}\bigg)\phi^\dagger\nn\\
    &\quad\quad+\bigg(\frac{16}{\epsilon^2}M_f^2M_s^2{y^{(1)}_{ij}}^\dagger y^{(3)}_{ij}+\frac{8}{\epsilon}M_f^2M_s^2{y^{(1)}_{ij}}^\dagger y^{(3)}_{ij}-\frac{8\log M_s^2}{\epsilon}M_f^2M_s^2{y^{(1)}_{ij}}^\dagger y^{(3)}_{ij}\nn\\
    &\quad\quad\quad\quad-\frac{8\log M_f^2}{\epsilon}M_f^2M_s^2{y^{(1)}_{ij}}^\dagger y^{(3)}_{ij}\bigg)\phi\nn\\
    &\quad\quad+\bigg(\frac{20}{\epsilon^2}M_f^2y^{(3)}_{ij}{y^{(1)}_{kl}}^\dagger y^{(1)}_{kl}+\frac{2}{\epsilon}M_f^2y^{(3)}_{ij}{y^{(1)}_{kl}}^\dagger y^{(1)}_{kl}\nn\\
    &\quad\quad\quad\quad-\frac{10\log M_s^2}{\epsilon}M_f^2y^{(3)}_{ij}{y^{(1)}_{kl}}^\dagger y^{(1)}_{kl}-\frac{10\log M_f^2}{\epsilon}M_f^2y^{(3)}_{ij}{y^{(1)}_{kl}}^\dagger y^{(1)}_{kl}\bigg)\phi\overline\psi_i\psi_j\nn\\
    &\quad\quad+\bigg(\frac{1}{\epsilon^2}\bigg\{16M_f\,\lambda\,{y^{(1)}_{ij}}^\dagger y^{(1)}_{ik} y^{(1)}_{kj} +24\,M_fM_s^2y^{(1)}_{ij}{y^{(1)}_{kj}}^\dagger y^{(3)}_{ki}+\cdot\cdot\cdot\bigg\}\nn\\
    &\quad\quad\quad\quad+\frac{1}{\epsilon}\bigg\{12M_fM_s^2y^{(1)}_{ij}{y^{(1)}_{kj}}^\dagger y^{(3)}_{ki}+\cdot\cdot\cdot\bigg\}\nn\\
    &\quad\quad\quad\quad-\frac{\log M_f^2}{\epsilon}\bigg\{8M_f\,\lambda\,{y^{(1)}_{ij}}^\dagger y^{(1)}_{ik} y^{(1)}_{kj} +12\,M_fM_s^2y^{(1)}_{ij}{y^{(1)}_{kj}}^\dagger y^{(3)}_{ki}+\cdot\cdot\cdot\bigg\}\nn\\
    &\quad\quad\quad\quad-\frac{\log M_s^2}{\epsilon}\bigg\{8M_f\,\lambda\,{y^{(1)}_{ij}}^\dagger y^{(1)}_{ik} y^{(1)}_{kj} +12\,M_fM_s^2y^{(1)}_{ij}{y^{(1)}_{kj}}^\dagger y^{(3)}_{ki}+\cdot\cdot\cdot\bigg\}\bigg)(\phi^\dagger\phi)\phi\nn\\
    &\quad\quad+\bigg(\frac{1}{\epsilon^2}\big\{16M_f\,\lambda\,{y^{(1)}_{ij}}^\dagger {y^{(1)}_{ki}}^\dagger y^{(1)}_{kj} +24\,M_fM_s^2y^{(1)}_{ij}{y^{(1)}_{kj}}^\dagger y^{(3)}_{ki}+\cdot\cdot\cdot\bigg\}\nn\\
    &\quad\quad\quad\quad+\frac{1}{\epsilon}\bigg\{12M_fM_s^2y^{(1)}_{ij}{y^{(1)}_{kj}}^\dagger y^{(3)}_{ki}+\cdot\cdot\cdot\bigg\}\nn\\
    &\quad\quad\quad\quad-\frac{\log M_f^2}{\epsilon}\big\{8M_f\,\lambda\,{y^{(1)}_{ij}}^\dagger {y^{(1)}_{ki}}^\dagger y^{(1)}_{kj} +12\,M_fM_s^2y^{(1)}_{ij}{y^{(1)}_{kj}}^\dagger y^{(3)}_{ki}+\cdot\cdot\cdot\bigg\}\nn\\
    &\quad\quad\quad\quad-\frac{\log M_s^2}{\epsilon}\big\{8M_f\,\lambda\,{y^{(1)}_{ij}}^\dagger {y^{(1)}_{ki}}^\dagger y^{(1)}_{kj} +12\,M_fM_s^2y^{(1)}_{ij}{y^{(1)}_{kj}}^\dagger y^{(3)}_{ki}+\cdot\cdot\cdot\bigg\}\bigg)(\phi^\dagger\phi)\phi^\dagger\nn\\
    &\quad\quad+\bigg(\frac{16}{\epsilon^2}{y^{(1)}_{mn}}^\dagger y^{(1)}_{mk}{y^{(1)}_{lk}}^\dagger y^{(1)}_{ln}y^{(3)}_{ij}-\frac{8\log M_f^2}{\epsilon^2}{y^{(1)}_{mn}}^\dagger y^{(1)}_{mk}{y^{(1)}_{lk}}^\dagger y^{(1)}_{ln}y^{(3)}_{ij}\nn\\
    &\quad\quad\quad\quad-\frac{8\log M_s^2}{\epsilon^2}{y^{(1)}_{mn}}^\dagger y^{(1)}_{mk}{y^{(1)}_{lk}}^\dagger y^{(1)}_{ln}y^{(3)}_{ij}\bigg)(\phi^\dagger\phi)\phi\overline\psi_i\psi_j\nn\\
    &\quad\quad+\bigg(\frac{1}{\epsilon^2}\bigg\{5M_f^2\c_6\,{y^{(1)}_{ij}}^\dagger y^{(1)}_{ij} + 40\,M_f^2\,\lambda\,{y^{(3)}_{ij}}^\dagger y^{(1)}_{ij} + 16\,\lambda\,{y^{(1)}_{ij}}^\dagger y^{(1)}_{ik}{y^{(3)}_{lk}}^\dagger y^{(1)}_{lj} \nn\\
    &\quad\quad\quad\quad+ 36M_s^2{y^{(1)}_{kj}}^\dagger y^{(1)}_{kl}{y^{(1)}_{nl}}^\dagger y^{(3)}_{nj}+\cdot\cdot\cdot\bigg\}\nn\\
    &\quad\quad\quad\quad+\frac{1}{\epsilon}\bigg\{\frac{1}{2}M_f^2\c_6\,{y^{(1)}_{ij}}^\dagger y^{(1)}_{ij} + 4\,M_f^2\,\lambda\,{y^{(3)}_{ij}}^\dagger y^{(1)}_{ij}  + 18 M_s^2{y^{(1)}_{kj}}^\dagger y^{(1)}_{kl}{y^{(1)}_{nl}}^\dagger y^{(3)}_{nj}+\cdot\cdot\cdot\bigg\}\nn\\
    &\quad\quad\quad\quad-\frac{\log M_f^2}{\epsilon}\bigg\{\frac{5}{2}M_f^2\c_6\,{y^{(1)}_{ij}}^\dagger y^{(1)}_{ij} + 20\,M_f^2\,\lambda\,{y^{(3)}_{ij}}^\dagger y^{(1)}_{ij} + 8\,\lambda\,{y^{(1)}_{ij}}^\dagger y^{(1)}_{ik}{y^{(3)}_{lk}}^\dagger y^{(1)}_{lj} \nn\\
    &\quad\quad\quad\quad+ 18M_s^2{y^{(1)}_{kj}}^\dagger y^{(1)}_{kl}{y^{(1)}_{nl}}^\dagger y^{(3)}_{nj}+\cdot\cdot\cdot\bigg\}\nn\\
    &\quad\quad\quad\quad-\frac{\log M_s^2}{\epsilon}\bigg\{\frac{5}{2}M_f^2\c_6\,{y^{(1)}_{ij}}^\dagger y^{(1)}_{ij} + 20\,M_f^2\,\lambda\,{y^{(3)}_{ij}}^\dagger y^{(1)}_{ij} + 8\,\lambda\,{y^{(1)}_{ij}}^\dagger y^{(1)}_{ik}{y^{(3)}_{lk}}^\dagger y^{(1)}_{lj} \nn\\
    &\quad\quad\quad\quad+ 18M_s^2{y^{(1)}_{kj}}^\dagger y^{(1)}_{kl}{y^{(1)}_{nl}}^\dagger y^{(3)}_{nj}+\cdot\cdot\cdot\bigg\}\bigg)(\phi^\dagger\phi)^2\nn\\
    &\quad\quad+\bigg(\frac{1}{\epsilon^2}\bigg\{24M_f\,\lambda\,y^{(1)}_{ij}{y^{(1)}_{kj}}^\dagger {y^{(3)}_{ik}}^\dagger+4M_f\,c_6\,y^{(1)}_{ij}{y^{(1)}_{kj}}^\dagger {y^{(1)}_{ik}}^\dagger+\cdot\cdot\cdot\bigg\}\nn\\
    &\quad\quad\quad\quad-\frac{\log M_s^2}{\epsilon}\bigg\{12M_f\,\lambda\,y^{(1)}_{ij}{y^{(1)}_{kj}}^\dagger {y^{(3)}_{ik}}^\dagger+2M_f\,c_6\,y^{(1)}_{ij}{y^{(1)}_{kj}}^\dagger {y^{(1)}_{ik}}^\dagger+\cdot\cdot\cdot\bigg\}\nn\\
    &\quad\quad\quad\quad-\frac{\log M_f^2}{\epsilon}\bigg\{12M_f\,\lambda\,y^{(1)}_{ij}{y^{(1)}_{kj}}^\dagger {y^{(3)}_{ik}}^\dagger+2M_f\,c_6\,y^{(1)}_{ij}{y^{(1)}_{kj}}^\dagger {y^{(1)}_{ik}}^\dagger+\cdot\cdot\cdot\bigg\}\bigg)(\phi^\dagger\phi)^2\phi^\dagger\nn\\
    &\quad\quad+\bigg(\frac{1}{\epsilon^2}\bigg\{24M_f\,\lambda\,y^{(1)}_{ij}{y^{(1)}_{kj}}^\dagger y^{(3)}_{ki}+4M_f\,c_6\,y^{(1)}_{ij}{y^{(1)}_{kj}}^\dagger y^{(1)}_{ki}+\cdot\cdot\cdot\bigg\}\nn\\
    &\quad\quad\quad\quad-\frac{\log M_s^2}{\epsilon}\bigg\{12M_f\,\lambda\,y^{(1)}_{ij}{y^{(1)}_{kj}}^\dagger y^{(3)}_{ki}+2M_f\,c_6\,y^{(1)}_{ij}{y^{(1)}_{kj}}^\dagger y^{(1)}_{ki}+\cdot\cdot\cdot\bigg\}\nn\\
    &\quad\quad\quad\quad-\frac{\log M_f^2}{\epsilon}\bigg\{12M_f\,\lambda\,y^{(1)}_{ij}{y^{(1)}_{kj}}^\dagger y^{(3)}_{ki}+2M_f\,c_6\,y^{(1)}_{ij}{y^{(1)}_{kj}}^\dagger y^{(1)}_{ki}+\cdot\cdot\cdot\bigg\}\bigg)(\phi^\dagger\phi)^2\phi\nn\\
    &\quad\quad+\bigg(\frac{1}{\epsilon^2}\bigg\{18\lambda\,{y^{(1)}_{kj}}^\dagger y^{(1)}_{kl}{y^{(1)}_{nl}}^\dagger y^{(3)}_{nj}+4\,c_6\,{y^{(1)}_{kj}}^\dagger y^{(1)}_{kl}{y^{(1)}_{nl}}^\dagger y^{(1)}_{nj}+\cdot\cdot\cdot\bigg\}\nn\\
    &\quad\quad\quad\quad-\frac{\log M_s^2}{\epsilon}\bigg\{9\lambda\,{y^{(1)}_{kj}}^\dagger y^{(1)}_{kl}{y^{(1)}_{nl}}^\dagger y^{(3)}_{nj}+2\,c_6\,{y^{(1)}_{kj}}^\dagger y^{(1)}_{kl}{y^{(1)}_{nl}}^\dagger y^{(1)}_{nj}+\cdot\cdot\cdot\bigg\}\nn\\
    &\quad\quad\quad\quad-\frac{\log M_f^2}{\epsilon}\bigg\{9\lambda\,{y^{(1)}_{kj}}^\dagger y^{(1)}_{kl}{y^{(1)}_{nl}}^\dagger y^{(3)}_{nj}+2\,c_6\,{y^{(1)}_{kj}}^\dagger y^{(1)}_{kl}{y^{(1)}_{nl}}^\dagger y^{(1)}_{nj}+\cdot\cdot\cdot\bigg\}\bigg)(\phi^\dagger\phi)^3\bigg].
\end{align}

For each of the above topologies computed, there is also a conjugate contribution that we are not explicitly mentioning here. As mentioned in Sec. \ref{sec:fermion_one_loop} since the one-loop scalar-fermion mixed contributions are incomplete, we have not provided, here, the explicit results from this counter term insertion in one-loop but the computation of this contribution is also similar to what is done for the other cases. 

\newpage
\bibliographystyle{jhep}
\bibliography{ref.bib}
\end{document}